\documentclass[10pt,a4paper]{article}
\pdfoutput=1
\usepackage[dvipsnames]{xcolor}
\usepackage{amsmath,fourier,amssymb,amsthm,graphicx,epstopdf,tikz,hyperref,
	color,cases,float,mathdots,cancel,MnSymbol}
	\usepackage{accents}
\usetikzlibrary{shapes,snakes}
\usepackage{chngcntr}
\counterwithout{figure}{section}
\usepackage[T1]{fontenc}
\allowdisplaybreaks
\numberwithin{equation}{section}

\newtheorem{innercustomthm}{Theorem}
\newenvironment{customthm}[1]
{\renewcommand\theinnercustomthm{#1}\innercustomthm}
{\endinnercustomthm}

\theoremstyle{plain}
\newtheorem {hypo}{\bf\hspace{-\parindent}Hypothesis}[section]

\newtheorem {prop}[hypo]{Proposition}
\newtheorem {lemma}[hypo]{Lemma}
\newtheorem {theo}[hypo]{Theorem}
\newtheorem {defin}[hypo]{Definition}
\newtheorem {cor}[hypo]{Corollary}

\theoremstyle{remark}
\newtheorem {rmk}[hypo]{Remark}
\newtheorem {eg}[hypo]{Example} 
\newcommand{\pf}{\begin{bpf}}

	\newcommand{\pfms}{\begin{bpfms}}
		\newcommand{\epf}{\end{bpf}\hfill$\square$\vspace{0.1cm}}
	\newcommand{\epfms}{\end{bpfms}\hfill$\square$\\ }
\newcommand\ben{\begin{equation*}}
\newcommand\ebn{\end{equation*}}
\newcommand\beq{\begin{equation}}
\newcommand\eeq{\end{equation}}
\newcommand\ds{\displaystyle}
\newcommand\lb{\left(}
\newcommand\rb{\right)} 

\newcommand\Cb{\mathbb{C}} 

\newcommand\Pb{\mathbb{P}}

\newcommand\PIIeq{P$_\text{II}$ }

\newcommand{\hG}{\hat{G}}
\newcommand{\Li}{\operatorname{Li}_2}

\begin{document}
\noindent 
   \LARGE
   \noindent
   \textbf{Painlevé II connection problem revisited}
   \normalsize
	\vspace{0.5cm}\\
	\textit{ 
	    N. Iorgov$\,^{a,b}$,
		O. Lisovyy$\,^{c}$,
        N. Tamarin$\,^{d}$,
		Yu. Zhuravlov$\,^{a,b}$}
	\vspace{0.2cm}\\
	$^a$  Bogolyubov Institute for Theoretical Physics, 14B Metrolohichna st., 03143 Kyiv, Ukraine \\
        $^b$  Kyiv Academic University, 36 Vernadsky Ave., 03142 Kyiv, Ukraine          \\
	$^c$  Institut Denis-Poisson, Universit\'e de Tours, CNRS, Parc de Grandmont,
	37200 Tours, France\\
    	$^d$  CAMGSD, Departamento de Matem\`atica, Instituto Superior Técnico,
Universidade de Lisboa, \\ 1049-001 Lisboa, Portugal

    \begin{abstract}
    We study the inhomogeneous Painlevé II equation for generic monodromy data.
    Using Kapaev's asymptotic formulas, the parameters of Zak transform representations of the Painlevé II tau function on six special rays $\arg t=\frac{\pi p}{3}$ ($p=0,\ldots,5$) in the complex time plane are expressed in terms of Stokes parameters of the associated linear system. We then solve the connection problem for the tau function between these special rays, and provide analytical and numerical checks of our result.
    \end{abstract}

   \section{Introduction}
  A characterization of the space of all solutions of a nonlinear PDE is in general a difficult task. Even inside the class of integrable equations such as the Korteweg-De Vries (KdV) equation
  \beq\label{KdV}
 u_t-6uu_x+u_{xxx}=0,
 \eeq
   one has different types of solutions: rapidly decaying solutions arising in the context of inverse scattering, finite-gap solutions, rational solutions that appear in the Sato Grassmannian formalism, Kontsevich-Witten solution encoding intersection numbers on moduli spaces of curves, etc. 
   
   For ODEs, and in particular for Painlevé equations that are obtained from integrable PDEs under symmetry reductions, the problem of classifying all solutions becomes more tractable. The main example relevant for this work is the second Painlevé  equation (\PIIeq)
   \beq\label{PIIstandard}
   q''=2q^3+tq+\tfrac12-\theta,
   \eeq
   where $'$ denotes the derivative with respect to $t$ and $\theta\in\mathbb C$ is an arbitrary parameter.
   All solutions of \PIIeq extend to meromorphic functions in the complex $t$-plane with only simple poles. The Riemann-Hilbert correspondence provides  a one-to-one  map between the set of all solutions and an  affine cubic surface $\mathcal S_\theta$ in~$\mathbb C^3$. The latter encodes the monodromy data of \PIIeq auxiliary linear problem and is defined~by\footnote{Strictly speaking, for the map to be bijective, the singular points of $\mathcal S_\theta$ should be resolved by blow-ups. These singularities however appear only when   ${s_1=s_2=s_3=\lb -1\rb^{\theta}}$ with $\theta\in\mathbb Z$. The corresponding \PIIeq transcendents are precisely the well-known special function solutions expressed in terms of Airy functions. The present paper is focused on the generic values of parameter $\theta$, therefore the discussion of this special case will be mostly omitted.} \cite{FN80}, \cite[Chapter~4]{FIKN06}
   \beq\label{monodromy_S}
   s_1s_2s_3-s_1-s_2-s_3+2\cos\pi\theta=0.
    \eeq
    Any solution of \PIIeq  becomes thereby completely specified by monodromy data $\mathbf{s}\in\mathcal S_\theta$ that can be viewed as a pair of integrals of motion for \eqref{PIIstandard}. Just as for classical special functions, the main task is then to describe the large time asymptotic behavior of $q=q\lb t,{\theta},\mathbf s\rb$   for given $\theta$, $\mathbf s$. 
    
    A complete description of such $t\to\infty$ asymptotics was accomplished in a series of works of Kapaev \cite{Kap88,Kap91,Kap92}. The resulting structure is remarkably rich. There are six special rays $\arg t=\frac{\pi p}{3}$,  ${p\in \mathbb Z/6\mathbb Z}$ along which the asymptotics is of trigonometric type. It involves two asymptotic parameters proper for each ray and its form further depends on whether $p$ is even or odd. Inside the sectors bounded by the rays the asymptotics is described by modulated elliptic functions \cite{Bou13,JK88}. We will recall the relevant results  in the main body of the paper, and develop them further in a form adapted to our purposes. At this point, let us simply note that expressing each pair of asymptotic parameters in terms of monodromy automatically gives the corresponding connection formulas between different rays and sectors.
  
    The main goal of the present paper is to derive connection formulas for the \PIIeq tau function defined by
      \beq\label{PIItauDef}
    \lb\ln \tau\rb'=\frac{q'^2-\lb q^2+\frac t2\rb^2}2+\theta q,
    \eeq
    between different rays of trigonometric degeneration. The motivation for studying the tau function is two-fold: 
    \begin{itemize}
    	\item First, it is the object that most frequently appears in applications. For example, the celebrated Tracy-Widom distribution from random matrix theory \cite{TWAiry} is the tau function associated to a special  solution of \PIIeq with $\theta=0$. The connection problem for this particular \PIIeq tau function was solved in \cite{DIK,BBDF}. Also note that every \PIIeq tau function yields a KdV tau function by
    		\beq
    	\tau_{\mathrm{KdV}}\lb x,t\rb=e^{\frac{x^3}{72t}}
    	\tau\lb \frac{x}{\sqrt[3]{3t}}\rb,
    	\eeq
        i.e. the function $u\lb x,t\rb=-2\lb\ln\tau_{\mathrm{KdV}}\rb_{xx}$ satisfies \eqref{KdV}.
        \item The tau function is in general a better behaved object with more structure than the transcendent $q$ itself. It is an entire function in the complex $t$-plane, and hence plays a role analogous to that of the Jacobi theta function in the theory of elliptic functions, whereas $q$ is a counterpart of the  Weierstrass $\wp$-function recovered by taking logarithmic derivatives. This  analogy can be made more precise, see e.g. \cite[Section 4.2]{BGMT24}, \cite[Remark~2.3]{IILZ25}. As we will see below, the description of \PIIeq asymptotics is also drastically simpler when expressed in terms of the tau function. Furthermore, Painlevé tau functions are closely related to Virasoro conformal blocks,  partition functions of $\mathcal N=2$ SUSY gauge theories, and topological recursion \cite{GIL12,GIL13,BS14,Nagoya15,BLMST,GG18,Nagoya18,NU19,Iwaki19,GMS20,PP23,BST25,IILZ25}. This links the corresponding connection problems  to fusion transformations in conformal field theory and Poisson geometry of the associated monodromy manifolds \cite{ILT13,LNR18,DMDG25}. 
    \end{itemize}
    
    Let us introduce some notation and make a number of genericity assumptions on monodromy. While not all of them are essential, they allow us to state the result without introducing additional technicalities discussed later. We require that
    \beq\label{assumption1}
    s_1,s_2,s_3,1-s_1s_2,1-s_2s_3,1-s_1s_3\notin\mathbb R_{\leq 0},
    \eeq
    and introduce the following quantities:
    	\begin{subequations}\label{nuomegas}
    \begin{gather}
    \nu_0=\frac{\ln s_2}{2\pi i},\qquad 
    \nu_1=\frac{\ln s_3}{2\pi i},\qquad
    \nu_2=\frac{\ln s_1}{2\pi i},  \\
    \omega_0=\frac{\ln \lb 1-s_1s_3\rb}{2\pi i},\qquad
    \omega_1=\frac{\ln \lb 1-s_1s_2\rb}{2\pi i},\qquad
    \omega_2=\frac{\ln \lb 1-s_2s_3\rb}{2\pi i}.
    \end{gather}
    \end{subequations}
    The logarithm branches are chosen so that $|\Re \nu_p|,|\Re \omega_p|<\frac12$ for $p=0,1,2$. It will be convenient to extend the values of $p$ to all integers by introducing $\nu_{p\pm3}=\nu_p$, $\omega_{p\pm3}=\omega_p$. Among six quantities \eqref{nuomegas}, independent are  at most two. Indeed, we have identities such as 
    \beq\label{nuomegaidentities}
    {e^{2\pi i \lb \nu_{p-1}+\nu_{p+1}\rb}=1-e^{2\pi i \omega_{p}}},\qquad e^{2\pi i\lb\omega_{p-1}+\omega_{p+1}\rb}=\lb e^{2\pi i\nu_p}-e^{\pi i \theta}\rb
    \lb e^{2\pi i\nu_{p}}-e^{-\pi i \theta}\rb,
    \eeq
    for every $p\in\mathbb Z/3\mathbb Z$.
    Any pair of these quantities can therefore be seen as local coordinates on the two-dimensional space of \PIIeq initial conditions.

    Under the above assumptions, it can be deduced from the results of \cite{Kap92} that the leading asymptotic behavior of the \PIIeq tau function \eqref{PIItauDef} as $t\to\infty$ along the special rays is given by
    \beq\label{astauspecial}
    \tau\lb t\rb=\begin{cases}
    	A_p\,\lb 4\sqrt2\, e^{\frac{i\pi}{2}}|t|^{\frac32}\rb^{\frac{\theta^2}{12}-\nu_p^2}e^{\frac{2i\sqrt2\,\nu_p}{3}\,|t|^{\frac32}}\left[1+o\lb 1\rb\right],\qquad & \arg t=\frac{2\pi p}{3},\\
    	B_p\,\lb 8e^{\frac{i\pi}{2}}|t|^{\frac32}\rb^{\frac{\theta^2}{3}-\frac1{12}-\frac{\omega_p^2}{2}}e^{\frac{|t|^3}{24}+\frac{2i \omega_p}{3}|t|^{\frac32}}\left[1+o\lb 1\rb\right],\qquad & \arg t=\pi+\frac{2\pi p}{3},
    	\end{cases}
    \eeq
    with $p\in\mathbb Z/3\mathbb Z$. The coefficients $A_p,B_p$ in \eqref{astauspecial} are individually undetermined since the tau function is defined by \eqref{PIItauDef} only up to a multiplicative normalization constant. However their ratios such as $A_p/B_q$ ($p,q\in\mathbb Z/3\mathbb Z$) are completely fixed by monodromy/initial conditions. Such ratios will be called \textit{connection coefficients}.
    
    Our main new result is given by
    \begin{customthm}{A} \label{THMA} Painlevé II connection coefficients between the special rays are given by the formulas
    	\begin{subequations}
        \label{conncoefsTHMA}
    	\begin{align}
    		\frac{B_{p+2}}{A_p}=&\,
    		2^{\frac{1-5\theta^2}{3}}
    		e^{\zeta'\lb-1\rb+\frac{\pi i\lb 1-\theta^2\rb}{12}+2\pi i \nu_p\omega_{p+2}-\pi i \nu_{p}^2}
    		\frac{\lb 2\pi\rb^{-\nu_p-\frac{\omega_{p+2}}{2}} G\lb 1+\omega_{p+2}\rb}{G\lb 1-\nu_p+\frac{\theta}{2}\rb G\lb 1-\nu_p-\frac{\theta}{2}\rb},\\
    			\frac{B_{p+1}}{A_p}=&\,
    		2^{\frac{1-5\theta^2}{3}}
    			e^{\zeta'\lb-1\rb-\frac{\pi i\theta^2}{6}-2\pi i \nu_p\omega_{p+1}+\frac{\pi i\omega_{p+1}^2}{2}}
    		\frac{\lb 2\pi\rb^{\nu_p+\frac{\omega_{p+1}}{2}}G\lb 1-\omega_{p+1}\rb}{G\lb 1+\nu_p+\frac{\theta}{2}\rb G\lb 1+\nu_p-\frac{\theta}{2}\rb},
    	\end{align}
    	\end{subequations}
    where $G\lb z\rb$ denotes Barnes $G$-function and $p\in\mathbb Z/3\mathbb Z$.
    \end{customthm}

    The first rigorous solution of the connection problem for Painlevé tau functions for  generic monodromy was obtained  for the Painlevé III equation of type $D_8$ by Its and Prokhorov \cite{IP15}, who proved a conjectural formula for the connection coefficients proposed in \cite{ILT14}. Their analysis was inspired by the pioneering work \cite{Ber09} that initiated the study of the dependence of isomonodromic tau functions on monodromy data. A general framework for solving the corresponding connection problem was subsequently developed in \cite{ILP16} and applied to Painlevé VI, Painlevé I, and the homogeneous ($\theta=\frac12$) \PIIeq equation.

    The main tool of \cite{ILP16} is an extension of the Jimbo-Miwa-Ueno differential  defining the tau function for fixed initial conditions \cite{JMU81} to a 1-form  that also incorporates the variation of monodromy. This extended 1-form can be expressed in terms of  solution of appropriate isomonodromic Riemann-Hilbert (RH) problem, and the connection coefficients can be computed in a straightforward way once a sufficiently accurate global asymptotic description of such a solution is achieved. In all cases mentioned above except for Painlevé~VI, the analysis is crucially simplified by the fact that the relevant differentials may be expressed directly in terms of the Painlevé transcendent $q$ and its derivatives with respect to time and monodromy  (in the terminology of \cite{ILP16}, extended 1-forms \textit{completely localize}). Solving the tau function connection problem therefore reduces to the study of asymptotics and connection properties of the Painlevé transcendent itself. For Painlevé VI, the lack of complete localization was the main source of technical difficulty.

    A similar phenomenon occurs for the \textit{inhomogeneous} \PIIeq equation considered in the present work. In this case, the extended differential  involves, in addition to  $q$ and its derivatives, an auxiliary function $\lambda$ that satisfies
   ${{\lambda'}/{\lambda}=-q}$ and thus represents, essentially, the primitive of $q$. This function will be referred to as the \textit{gauge factor}. The computation of \PIIeq tau function connection coefficients  requires solving the connection problem for $\lambda$ in addition to that of $q$. The nontrivial part of this problem is the calculation of differences of the \textit{finite terms} in the asymptotic expansions of the gauge factor along the special rays. This issue is closely related to the evaluation of integrals of \PIIeq transcendents, which have been studied for special families of solutions in \cite{BBDF,Mil18,Kok18,DXZ19}.

   Rather than relying on the asymptotic analysis of the RH problem, we develop a different approach, expected to extend at least partially to other Painlevé equations with parameters. We show that the structure of the extended differential, together with the known asymptotics of the \PIIeq transcendent, determines the tau function connection coefficients  between the special rays up to prefactors that may depend on $\theta$ but are independent of the Stokes data  $\mathbf{s}\in\mathcal S_\theta$. The $\theta$-dependence of the connection coefficients $A_p/A_q$, $B_p/B_q$ between the rays of the same type is fixed by analyzing the case of $\mathbb Z_3$-symmetric monodromy with $s_1=s_2=s_3$. For the rays of different types, the $\theta$-dependence of  $A_p/B_q$ is strongly constrained, in particular, by the Toda equations relating \PIIeq tau functions with B\"acklund transformed parameters, and can in fact be determined completely using the results of \cite{DXZ19} on the generalized Hastings-McLeod solutions.
   
   We also rederive the connection coefficients $A_p/A_q$ by a resurgent calculation. Motivated by the results of \cite{KRSSTlocal}, we match the tau function transseries obtained from the two-parameter solutions of \cite{BSSV,Vega,Vega23} to its Zak transform representation. We then recover the connection coefficients through Stokes continuation and the $\mathbb Z_3$-symmetry. The Zak transform expansions available throughout the Painlevé family \cite{BLMST} provide analogous local expressions against which resurgent constructions can be checked.

   The paper is organized as follows. In Section 2, we  provide the necessary background on \PIIeq, the associated linear system and its monodromy. We describe the asymptotic expansions of the \PIIeq transcendent $q\lb t\rb$ and of the tau function $\tau\lb t\rb$ along the special rays, and establish the correspondence between the relevant asymptotic parameters in order to be able to use Kapaev's formulas. The main result is Theorem~\ref{TheoConn} expressing the asymptotic parameters appearing in Zak transform representations of the \PIIeq tau function in terms of Stokes data. Section~\ref{sec_CC} is devoted to the proof of Theorem~\ref{THMA}. Extended differentials governing the variation of the tau function with respect to monodromy are introduced in Definition~\ref{defexttau}. Corollary~\ref{varForm1} relates the connection coefficients to the finite parts of the asymptotics of the gauge factor, while Propositions~\ref{propCC01} and \ref{propXipm} determine their dependence on monodromy --- completely in the case of $A_p/A_q$, $B_p/B_q$ (cf Corollary~\ref{corsametyperays}) and up to a $\theta$-dependent prefactor for $A_p/B_q$. The explicit form of this prefactor is obtained in Corollary~\ref{finalcorXi}. As a byproduct, Corollary~\ref{corU} yields the connection formula for the gauge factor between the special rays. Section~\ref{sec_numerics} presents numerical verification of our results. A resurgent derivation of the connection coefficients $A_p/A_q$ is given in Section~\ref{sec:resurgence}. Appendices collect the necessary properties of the Barnes $G$-function and discuss several families of solutions of \PIIeq from the perspective of Zak representations of the tau function and monodromy. \vspace{0.2cm}

   \noindent
  { \small \textbf{Acknowledgements}.  We are grateful to Pavlo Gavrylenko and Kohei Iwaki for useful discussions. N. Iorgov and Yu. Zhuravlov have been supported by the National Research Foundation of Ukraine through the project 
  2025.07/0437 ``Integrability, Geometry, and Quantum Field Theory: Modern Trends and Interplay''.   N. Tamarin thanks CAMGSD for its institutional support and acknowledges the support of the IRN:QFS (Quantum Fields and Strings)  for his visit to Institut Denis-Poisson in Tours in June 2026.
  }
    
    \section{Asymptotics and monodromy}\label{sec_AM}
    \subsection{Generalities}
    We start by recalling the relation between the standard and $\sigma$-forms of the Painlevé II equation. 
    \begin{lemma}
    If $q$ is any solution of \eqref{PIIstandard}, then the logarithmic derivative
    \beq\label{sigmafunction}
    \sigma=\lb\ln\tau\rb'
    \eeq
     of the  tau function defined by \eqref{PIItauDef} satisfies
      \beq\label{sigmaPII}
     \lb \sigma''\rb^2=2\sigma'\lb \sigma-t\sigma'\rb-4 \lb\sigma'\rb^3+\frac{\theta^2}{4}.
     \eeq
     Conversely, if $\sigma$ is any non-affine solution of \eqref{sigmaPII}, then
     	\beq\label{qintermsoftau}
     	q=-\frac{2\sigma''+\theta}{4\sigma'}
     	\eeq
     satisfies  \eqref{PIIstandard}.
     \end{lemma}
     \pf Differentiating \eqref{PIItauDef} with respect to $t$ and using \eqref{PIIstandard}, we obtain
     \beq\label{sigma1and2viaQ}
     2\sigma'=q'-q^2-\frac t2,\qquad 2\sigma''=-4\sigma' q-\theta.
     \eeq 
     Substituting these two relations (together with \eqref{PIItauDef}) into \eqref{sigmaPII}, it is straightforward to check that the latter equation is satisfied. 
     
     To prove the converse statement, it suffices to differentiate \eqref{sigmaPII}. When $\sigma''$ does not vanish identically, the resulting equation implies that
     \beq
     \sigma'''=\sigma-2t\sigma'-6\lb\sigma'\rb^2.
     \eeq
     Use this relation to express $q''$ in terms of $\sigma$, $\sigma'$ and $\sigma''$, and note that all even powers of $\sigma''$ can be written in terms of $\sigma$, $\sigma'$  using \eqref{sigmaPII}.
       \epf
     
     \begin{rmk} The above allows one to reconstruct   the conventional \PIIeq transcendent $q$ from the tau function. To describe potentially missing solutions corresponding to $\sigma=\alpha t+\beta$, one may again use the relations \eqref{sigma1and2viaQ}. The second of them implies that $4\alpha q+\theta=0$. The only constant solution of \eqref{PIIstandard} is $q=0$ which  exists only for $\theta=\frac12$. Therefore, we must have $\alpha=\theta=0$, which further implies the Riccati equation $q'=q^2+\frac t2$ (the first of relations \eqref{sigma1and2viaQ}) and $\beta=0$ (coming from the expression \eqref{PIItauDef} for $\sigma$). The corresponding Riccati solutions can be written in terms of the general solution of Airy equation
     	\beq\label{AiryBasic}
     	q\lb t\rb=-\frac{d}{dt}\left[\ln\lb C_1\operatorname{Ai}\lb -2^{-\frac13}t\rb+C_2\operatorname{Bi}\lb -2^{-\frac13}t\rb\rb\right],\qquad 
     	C_1,C_2\in\mathbb C^2\backslash\{0\}.
     	\eeq
    Therefore, to any solution $q\lb t\rb$ of \PIIeq in the standard form except for the one-parameter family of $\theta=0$ solutions we have just described, corresponds a non-affine solution of the $\sigma$-equation \eqref{sigmaPII}, and the former can be recovered from the latter using \eqref{qintermsoftau}.
     	\end{rmk}

     Next, recall the following discrete symmetries:
     \begin{itemize}
     	\item \textit{$\mathbb Z_3$-symmetry}: The \PIIeq equation remains invariant under transformation $t\mapsto \zeta t$, $q\mapsto \zeta^2 q$ with $\zeta^3=1$, which implies that if $q\lb t\rb$ solves \eqref{PIIstandard}, then $e^{\pm\frac{2\pi i}{3}}
     	q\bigl(e^{\pm\frac{2\pi i}{3}}t\bigr)$ are also solutions. Likewise, if $\tau\lb t\rb$ is a \PIIeq tau function, then so are $\tau\bigl( e^{\pm\frac{2\pi i}3}t\bigr)$ for the same parameter  $\theta$.
     	\item \textit{Bäcklund transformations}: Obviously, if $q$ is a solution of \PIIeq with parameter $\theta$, then $\tilde q=-q$ is a solution of \PIIeq with parameter $\tilde \theta=1-\theta$. On the other hand, the $\sigma$-equation \eqref{sigmaPII} is invariant under the sign flip $\theta\mapsto -\theta$. The interplay of these two $\mathbb Z_2$-symmetries generates affine Bäcklund transformations for \PIIeq. Indeed, the function
     	\beq\label{sigmatilde}
     	\tilde \sigma=\frac{\tilde q'^2-\lb \tilde q^2+\frac t2\rb^2}2+\tilde \theta \tilde q=\sigma-q
     	\eeq
     	is a solution of \eqref{sigmaPII} with $\theta\mapsto \theta-1$, and therefore, according to \eqref{qintermsoftau}, 
     	\beq
     	Q=-\frac{2\tilde\sigma''+\theta-1}{4\tilde\sigma'}=
     	\tilde q-\frac{\theta-1}{2\tilde\sigma'}=-q+\frac{\theta-1}{q'+q^2+\frac t2}
     	\eeq
     	satisfies \eqref{PIIstandard} with $\theta\mapsto\theta-1$, cf \cite[Eq. 32.7.2]{DLMF}. In particular, iteratively applying this transformation to the seed solution \eqref{AiryBasic}, one obtains the aforementioned special function solutions of \PIIeq.
     \end{itemize}

     \subsection{Asymptotics of $\tau$}
     As already mentioned in the Introduction, there are two families of special rays in the description of Painlevé II asymptotics as $t\to\infty$.
     Along the rays $\operatorname{arg}t=\frac{2\pi p}{3}$ ($p\in\mathbb Z/3\mathbb Z$), the tau function \eqref{PIItauDef} associated to generic solution has the following asymptotic expansions, cf \cite[Eqs. (3.32)]{BLMST}:
     \beq\label{expsI}
     \tau\lb t\rb= \mathcal N_{p,+}\sum_{n\in\mathbb Z}\mathcal D\lb 4\sqrt 2\,e^{\frac{i\pi}{2}}|t|^{\frac32}\,\bigl|\,\nu_p+n,\theta\rb e^{2\pi i n \rho_p },
     \eeq
     with
     \begin{gather}\label{expsID}
     	\mathcal D\lb s\,|\, \nu,\theta\rb\simeq\frac{G\lb 1+\nu+\tfrac{\theta}{2}\rb
     		G\lb 1+\nu-\tfrac{\theta}{2}\rb}{\lb2\pi\rb^{\nu}} s^{\frac{\theta^2}{12}-\nu^2}
     	\exp\left\{\frac{\nu s}{6}+\sum_{k=1}^{\infty}\mathcal D_k\lb \nu,\theta\rb s^{-k}\right\}.
     \end{gather}
     The first expansion coefficients of the building blocks $\mathcal D\lb s\,|\, \nu,\theta\rb$ are explicitly given by 
     \begin{subequations}\label{Dkcoefs}
     	\begin{align}
     		\mathcal D_1\lb \nu,\theta\rb=&\;\frac{\nu\lb 68\nu^2-9\theta^2+2\rb}{6} ,\\
     		\mathcal D_2\lb \nu,\theta\rb=&\;125 \nu ^4-\frac{43 }{2}\theta ^2 \nu ^2+
     		\frac{11 }{48}\theta ^4+15 \nu ^2-\frac{17}{12} \theta ^2,\\
     		\mathcal D_3\lb \nu,\theta\rb=&\;\nu\lb \frac{7126 }{3}\nu^4-495 \theta ^2 \nu ^2+\frac{101 }{8}\theta ^4+\frac{6202 }{9}\nu ^2-\frac{197 }{2}\theta ^2+\frac{35}{3}\rb,\\
     		\nonumber\mathcal D_4\lb \nu,\theta\rb=&\; 58366 \nu ^6-14210 \,\theta ^2 \nu ^4+\frac{4775 }{8}\theta ^4 \nu ^2-\frac{115 }{48}\theta ^6+32286 \nu ^4\\
     		&\; -\frac{11621 }{2}\theta ^2 \nu ^2 +\frac{673 }{12}\theta ^4+2308 \nu ^2-239 \theta ^2,\\
     		\nonumber\mathcal D_5\lb \nu,\theta\rb=&\; \nu\,\Biggl( \frac{8353064 }{5}\nu ^6-\frac{2321634 }{5}\theta ^2 \nu ^4+\frac{54927 }{2}\theta ^4 \nu ^2-\frac{11519 }{40}\theta ^6+\frac{7684772 }{5}\nu ^4-328122\, \theta ^2 \nu ^2\Biggr.\\
     		&\;\Biggl. +\frac{30949 }{4}\theta ^4+\frac{1341064 }{5}\nu ^2-\frac{203002}{5} \theta ^2+\frac{10010}{3} \Biggr).
     	\end{align}
     \end{subequations}
      Along the complementary rays $\operatorname{arg}t=\pi +\frac{2\pi p}{3}$ ($p\in\mathbb Z/3\mathbb Z$), the expansions take the form \cite[Eq. (3.28)]{BLMST}
     \beq\label{expsII}
     \tau\lb t\rb= \mathcal N_{p,-}\sum_{n\in\mathbb Z}\mathcal G\lb 8\,e^{\frac{i\pi}{2}}|t|^{\frac32}\,\bigl|\,\omega_p+n,\theta\rb e^{2\pi i n \xi_p },
     \eeq
     with 
     \begin{gather}
     	\label{expsIIG}
     	\mathcal G\lb s\,|\, \omega,\theta\rb= 
     	\frac{G\lb 1+\omega\rb}{\lb2\pi\rb^{\frac{\omega}{2}}} s^{\frac{\theta^2}{3}-\frac{1}{12}-\frac{\omega^2}{2}}\exp\left\{-\frac{s^2}{1536}+\frac{\omega s}{12}
     	+\sum_{k=1}^{\infty}\mathcal G_k\lb \omega,\theta\rb s^{-k}\right\},
     \end{gather} 
     where the first  coefficients of the building blocks $\mathcal G\lb s\,|\, \omega,\theta\rb$ are given by
     \begin{subequations}
     	\label{Gkcoefs}
     	\begin{align}
     		\mathcal G_1\lb \omega,\theta\rb=\,&\frac{\omega\lb34\,\omega^2-96\,\theta^2+31\rb}{6},\\
     		\mathcal G_2\lb \omega,\theta\rb=\,&\frac{375\,\omega^4-3\lb 832\,\theta^2-293\rb\omega^2+4\lb4\theta^2-1\rb \lb4\theta^2-9\rb}{6},\\
     		\mathcal G_3\lb \omega,\theta\rb=\,&\frac{\omega}{3}\lb 3563\, \omega ^4-43840\, \theta ^2 \omega ^2+5632\, \theta ^4+\frac{48557}{3} \omega ^2-16480\, \theta ^2+\frac{8831}{2}\rb,\\
     		\mathcal G_4\lb \omega,\theta\rb=\,&\,29183\, \omega ^6-\left(580160\, \theta ^2-220936\right) \omega ^4+\left(194560\, \theta ^4-599104\, \theta ^2+168659\right) \omega ^2\\
     		\nonumber &\,-\frac{64}{3}\left(4 \theta ^2-1\right) \left(4 \theta ^2-9\right) \left(4 \theta ^2-21\right),\\
     		\mathcal G_5\lb \omega,\theta\rb=\,&\, \frac{\omega}{5}   \Biggl(4 \left(1044133\, \omega ^6-30777264\, \theta ^2 \omega ^4+20472320\, \theta ^4 \omega ^2-729088\, \theta ^6\right)\Bigr.\\
     		\nonumber	&\,\quad+2 \left(23973623\, \omega ^4 -129457120\, \theta ^2 \omega ^2 +12910080\, \theta ^4\right)\\
     		\nonumber &\,\quad\Biggl.+74773352\, \omega ^2-52287968 \,\theta ^2+\frac{38950835}{3}\Biggr).
     	\end{align}
     \end{subequations}
     
     Let us make a few comments:
     \begin{itemize}
     	\item  We have 6 different pairs of asymptotic parameters $\lb\nu_p,\rho_p\rb$, $\lb\omega_p,\xi_p\rb$, $p\in\mathbb Z/3\mathbb Z$ corresponding to the rays $\arg t=\frac{2\pi p}{3},\pi+\frac{2\pi p}{3}$. Each of them can be viewed as encoding two \PIIeq initial conditions. Below we will relate these pairs between themselves via the Stokes data of the linear system using the results of \cite{Kap88,Kap92}.
     	\item Zak transform structure of the sums \eqref{expsI} and \eqref{expsII}  means that we are free to choose $-\frac12< \Re \nu_p,\Re \omega_p\leq \frac12$. If we further strengthen these assumptions to strict inequalities $|\Re \nu_p|,|\Re \omega_p|< \frac12$, then the leading contribution to the asymptotics of $\tau$ is determined by the $n=0$ term in \eqref{expsI} and \eqref{expsII}, with all corrections coming from $k\ge 1$ in \eqref{expsID}, \eqref{expsIIG} neglected. This corresponds to the asymptotic formulas \eqref{astauspecial} stated in the Introduction. 
     	
     	In the case when e.g. $ \nu_p=\frac12+i\alpha_p$ with $\alpha_p\in\mathbb R$, the leading contribution to the asymptotics of $\tau$ along the ray $\operatorname{arg}t=\frac{2\pi p}{3}$ is determined by the leading terms in $n=0$ and $n=-1$ building blocks:
     	\beq\label{compl_as}
     	\tau\lb t\rb\simeq \mathcal N_{p,+}\left[C_{p,+}s^{-i\alpha_p}e^{\frac{s}{12}}+C_{p,-}e^{-2\pi i\rho_p}s^{i\alpha_p}e^{-\frac{s}{12}}\right]s^{\frac{\theta^2}{12}-\frac14+\alpha_p^2}e^{\frac{i\alpha_p s}{6}},
     	\eeq
     	with 
     	\beq C_{p,\pm}=\frac{G\lb1+\frac{2i\alpha_p+\theta\pm1}{2}\rb G\lb1+\frac{2i\alpha_p-\theta\pm1}{2}\rb}{\lb2\pi\rb^{\pm\frac12+i\alpha_p}},\qquad 
     	s=4\sqrt 2 e^{\frac{i\pi}{2}}|t|^{\frac32}.
     	\eeq
     	\item We write $\mathcal D\lb s\,|\, \nu,\theta\rb$ and $\mathcal G\lb s\,|\, \omega,\theta\rb$ in the exponentiated form, since $\mathcal D_k\lb  \nu,\theta\rb$ and $\mathcal G_k\lb  \omega,\theta\rb$ turn out to be polynomials of degree $k+2$ in $\nu,\theta$ and $\omega,\theta$, respectively. Re-expanding, for instance, $\exp\sum\limits_{k=1}^{\infty}\mathcal D_k\lb \nu,\theta\rb s^{-k}$ as $\sum\limits_{k=0}^{\infty}\tilde{\mathcal D}_k\lb \nu,\theta\rb s^{-k}$, one finds that the resulting coefficients $\tilde{\mathcal D}_k\lb \nu,\theta\rb$ are polynomials of much larger degree $3k$. The cancellations taking place in the exponentiated form are expected to be a manifestation of the existence of a quasiclassical limit of a class of (as yet to be defined)  irregular Virasoro conformal blocks.
     	\item 
     	Note the symmetries
     	\beq\label{parityprops}
     	\begin{aligned}
     	\mathcal D_k\lb \nu,\theta\rb=\mathcal D_k\lb \nu,-\theta\rb=\lb -1\rb^k
     	\mathcal D_k\lb -\nu,\theta\rb ,\\
     	\mathcal G_k\lb \omega,\theta\rb=\mathcal G_k\lb \omega,-\theta\rb=
     	\lb -1\rb^k\mathcal G_k\lb -\omega,\theta\rb.
     	\end{aligned}
     	\eeq 
     	They imply that
     	\beq\label{DGsym}
     	\begin{aligned}
     		\mathcal D\lb e^{\pm i\pi}s\,|\,-\nu,\theta\rb=&\,
     		\frac{\lb2\pi\rb^{2\nu}
     			e^{\pm i\pi\lb \frac{\theta^2}{12}-\nu^2\rb}}{\hat G\lb\nu+\frac{\theta}{2}\rb \hat G\lb\nu-\frac{\theta}{2}\rb}\,\mathcal D\lb s\,|\,\nu,\theta\rb,\\
     		\mathcal G\lb e^{\pm i\pi}s\,|\,-\omega,\theta\rb=&\,\frac{\lb2\pi\rb^\omega
     		e^{\pm i\pi\lb \frac{\theta^2}{3}-\frac{1}{12}-\frac{\omega^2}{2}\rb}}{\hat G\lb\omega\rb}\,\mathcal G\lb s\,|\,\omega,\theta\rb,
     	\end{aligned}
     	\eeq
     	where $\hat G\lb z\rb=\ds\frac{G\lb 1+z\rb}{G\lb 1-z\rb}$.
     	\item The coefficients $\mathcal N_{p,\pm}$ coincide, up to suitable rescalings, with the coefficients $A_p$, $B_p$ in \eqref{astauspecial}. Thanks to $\mathbb Z_3$-symmetry, all possible connection coefficients $\mathcal N_{p,\epsilon}/\mathcal N_{q,\epsilon'}$ with $\epsilon,\epsilon'\in\{+,-\}$ and $p,q\in\mathbb Z/3\mathbb Z$ can be expressed in terms of two basic ones for the adjacent rays, for example, $\mathcal N_{2,-}/\mathcal N_{0,+}$ and $\mathcal N_{1,-}/\mathcal N_{0,+}$.
     \end{itemize}
     While the complete asymptotic expansions \eqref{expsI}--\eqref{expsID}, \eqref{expsII}--\eqref{expsIIG} are conjectural, their truncations at any finite order can be proved rigorously by the method of dominant balance applied to the $\sigma$-form \eqref{sigmaPII} of Painlevé~II. In what follows, we only need the first few terms of the asymptotics ($n=0,\pm 1$ and $k=1$), therefore the above formulas will be used without further justification.

  \subsection{Asymptotics of $q$}\label{subsec_Asq}
     
    We now construct the asymptotic expansions of the usual \PIIeq transcendent $q\lb t\rb$ as $t\to\infty$ along the special rays starting from the known leading behavior for generic initial conditions. The results will be subsequently compared with those obtained from the tau function expansions via \eqref{qintermsoftau} as well as with the asymptotic formulas of Kapaev \cite{Kap92}. Thanks to $\mathbb Z_3$-symmetry, it suffices to consider the rays with $\arg t=0,\pi$.
    \subsubsection{The ray $\arg t=0$}
    To describe  the asymptotics along the positive real axis, consider the ansatz   \beq\label{qexpreal}
    q\lb t\rb\simeq i\sqrt{2t}\sum_{k\in\mathbb Z}\sum_{l=0}^\infty q_{k,l}r^{-\frac {|k|}{2}-l-k\mu}e^{\frac{ikr}{6}} ,\qquad r=4\sqrt2\,|t|^{\frac32},\qquad |\Re \mu|<\frac12.
    \eeq
    Plugging it into \eqref{PIIstandard}, we find an overdetermined system of recurrence equations for the coefficients $\left\{q_{k,l}\right\}$. In particular, the leading asymptotic term corresponding to $\lb k,l\rb=\lb 0,0\rb$ satisfies $q_{0,0}^2=\frac14$. We accordingly set
    \begin{subequations}
    	\beq
    	q_{0,0}=\frac\epsilon2,\qquad \epsilon=\pm1.
    	\eeq
    	Different values of $\epsilon$ lead to different asymptotic behaviors of $q\lb t\rb$.
    	All other coefficients, as well as parameter $\mu$ in \eqref{qexpreal}, can be expressed recursively in terms of $q_{\pm1,0}$. These two coefficients therefore parameterize initial conditions for generic \PIIeq transcendent. One has, for example,
    	\begin{align}
    		&\mu=2iq_{1,0}q_{-1,0}+\epsilon\hat\theta,\\
    		&\epsilon q_{0,1}=i\lb 3\mu-\epsilon\hat\theta\rb,\qquad q_{\pm k,0}=\epsilon^{k-1}q_{\pm 1,0}^k,\quad k>0,\\
    		&q_{\pm 1,1}=\mp\frac{i}{12}\lb 102\mu^2\pm24\mu+17-18\hat\theta^2\pm12\epsilon\hat\theta\rb q_{\pm 1,0},\\
    		&\epsilon q_{0,2}=\frac{51\mu^2}{2}-9\epsilon\hat\theta\mu-\frac{9\hat\theta^2}{2}+2,\\ &\epsilon q_{\pm2,1}=\mp\frac{i}{6}\lb 102\mu^2\pm66\mu+41-18\hat\theta^2\pm18\epsilon\hat\theta\rb q_{\pm1,0}^2,\\
    		\nonumber &\qquad\ldots\qquad\ldots\qquad\ldots
    	\end{align}
    \end{subequations}
    where we introduce shifted parameter $\hat \theta=\theta-\frac12$.
    For $\hat\theta=0$, these expressions agree with homogeneous \PIIeq results from \cite[Subsection 4.3]{ILP16}.
    
    Next, we substitute \eqref{qexpreal} into \eqref{PIItauDef} to compute the corresponding expansions of the \PIIeq $\sigma$-function. Their leading terms are given by
    \beq
    \frac{\sigma\lb t\rb}{ 6\sqrt{2t}}=\begin{cases}
    	\ds	\frac{i\lb 2\mu+1\rb}{24}+\frac{iq_{-1,0}}{6}r^{\mu-\frac{1}{2}}e^{-\frac{ir}{6}}+ 
    	O\lb r^{\max\left\{2\Re\mu,0\right\}-1}\rb,\qquad &\epsilon=+1,\vspace{0.1cm}\\
    	\ds	\frac{i\lb 2\mu-1\rb}{24}+\frac{iq_{1,0}}{6}r^{-\mu-\frac{1}{2}}e^{\frac{ir}{6}}+
    	O\lb r^{\max\left\{-2\Re\mu,0\right\}-1}\rb ,\qquad &\epsilon=-1.
    \end{cases}
    \eeq
      It then becomes straightforward to match the asymptotic parameters $q_{\pm1,0}$ with the pair $\lb \nu_0,\rho_0\rb$ appearing in the Zak representation \eqref{expsI} of the \PIIeq tau function:
      
      \begin{itemize}
      	\item For $\epsilon=+1$, one has
      	\begin{subequations}
      		\label{nurhoplus}
      		\beq
      		\nu_0=\frac{2\mu+1}{4},\qquad q_{-1,0}=\frac{2\pi i\,e^{-2\pi i\lb \rho_0-\frac{\nu_0}{2}\rb}}{\Gamma\lb
      			\nu_0+\frac{\theta}{2}\rb \Gamma\lb
      			\nu_0-\frac{\theta}{2}\rb}.
      		\eeq
      		This case thus describes the situation where $0<\Re \nu_0<\frac12$.
      		\item  	Likewise, for $\epsilon=-1$ we find that
      		\beq
      		\nu_0=\frac{2\mu-1}{4},\qquad q_{1,0}=\frac{\Gamma\lb 1+
      			\nu_0+\frac{\theta}{2}\rb \Gamma\lb
      			1+\nu_0-\frac{\theta}{2}\rb}{2\pi i}e^{2\pi i\lb \rho_0-\frac{\nu_0}{2}\rb},
      		\eeq
      		which corresponds to the strip $-\frac12<\Re\nu_0<0$.
      		 \end{subequations}
      \end{itemize}
      
      \subsubsection{The ray $\arg t=\pi$}
        As $t\to-\infty$, $q\lb t\rb$ admits an expansion of  the form 
      \beq\label{qexprealbis}
      q\lb t\rb\simeq 2i\sqrt{|t|}\sum_{k\in\mathbb Z}\sum_{l=0}^\infty  q_{k,l}r^{-\frac{|k|}{2}-l-\frac{k\mu}{2}}e^{\frac{ikr}{12}} ,\qquad r=8\,|t|^{\frac32},\qquad |\Re \mu|<1.
      \eeq
      with 
      \beq
      q_{k,0}=0,\qquad k\in 2\mathbb Z.
      \eeq
      The latter constraint implies in particular that the leading asymptotics of $q\lb t\rb$ in \eqref{qexprealbis} is determined by the terms with $\lb k,l\rb=\lb\pm1,0\rb$. The coefficients $q_{k,l}$, as well as the parameter $\mu$, can be computed recursively in terms of $q_{\pm1,0}$. One has, e.g.
      \begin{subequations}
      	\begin{align}
      		&	\mu= -2iq_{1,0}q_{-1,0},\\
      		& q_{0,1}=4i\hat\theta,\qquad q_{\pm\lb 2k+1\rb,0}=q_{\pm1,0}^{2k+1},\qquad k\ge0,\\
      		& q_{\pm 1,1}=\mp \frac{i}{12}\lb 51\mu^2\pm36\mu+10-192\,\hat\theta^2\rb q_{\pm1,0},\\
      		& q_{0,2}=96\,\hat\theta\mu,\qquad q_{\pm2,1}=32i\,\hat\theta\,q_{\pm1,0}^2,\\
      		\nonumber &\qquad\ldots\qquad\ldots\qquad\ldots
      	\end{align}
      \end{subequations}
      In the homogeneous case $\hat\theta=0$, there are substantial simplifications; in this case $q_{k,l}=0$ for all $k\in2\mathbb Z$, $l\in\mathbb Z_{\ge 0}$. The expansion \eqref{qexprealbis} then takes the form given in \cite[Subsection 4.3]{ILP16}.
      
        The leading asymptotics of the \PIIeq $\sigma$-function inferred from \eqref{qexprealbis} is given by
      \beq
      -\frac{\sigma\lb t\rb}{12\sqrt{\left|t\right|}}= \frac{r}{768}+\frac{i\mu}{24}-\frac{i}{12\sqrt r}\lb q_{1,0} r^{-\frac{\mu}{2}}e^{\frac{ir}{12}}+q_{-1,0} r^{\frac{\mu}{2}}e^{-\frac{ir}{12}}\rb+O\lb r^{-1+2\left|\Re \mu\right|}\rb.
      \eeq
      Matching the asymptotic parameters with the pair $\lb\omega_0,\xi_0\rb$ appearing in the Zak representation \eqref{expsII}, we find that
      \beq
      \omega_0=\frac{\mu}{2},\qquad q_{1,0}=-\frac{\Gamma\lb1+\omega_0\rb}{\sqrt{2\pi}}e^{
      	-\frac{i\pi\lb 1+2\omega_0\rb}{4}+2\pi i\xi_0}.
      \eeq
    
    \subsection{Monodromy}
    Our next task is to relate various asymptotic parameters that appeared above to Stokes/monodromy data of the auxiliary linear problems associated with \PIIeq. There are two standard Lax representations of \PIIeq, due to Flaschka-Newell (FN) \cite{FN80} and Jimbo-Miwa (JM) \cite{JMU81b}. Below we will need both of them: we define the monodromy data in terms of the FN system to be able to use the results of \cite{Kap88,Kap92}, while the JM system will be used to construct the extension of the differential defining the \PIIeq
    tau function \eqref{PIItauDef} to the space of initial conditions.
    
    \subsubsection{Flaschka-Newell system}
     The FN system
    \beq\label{lsysKap}
    \partial_z\Phi=A\lb z\rb\Phi
    \eeq
    is defined by
    \beq\label{AKap}
    A\lb z\rb=z^2\lb\begin{array}{cc}
    	-4i & 0 \\ 0 & 4i
    \end{array}\rb +z\lb \begin{array}{cc} 0 & 4iq \\ -4iq & 0\end{array}\rb+
    \lb \begin{array}{cc} -i\lb t+2q^2\rb & -2p \\ -2p & i\lb t+2q^2\rb\end{array}\rb
    +z^{-1} \lb \begin{array}{cc} 0 & i\lb\theta-\frac12\rb \\ -i\lb\theta-\frac12\rb & 0 \end{array}\rb.
    \eeq
    It has a Fuchsian singularity at $z=0$ and an irregular singularity of Poincaré rank 3 at $z=\infty$. It is not a normal form of the general linear system on $\Cb\Pb^1$ with such singularity structure. Indeed, it has an extra symmetry
    $ A\lb z\rb=-\sigma_y A\lb -z\rb \sigma_y$,
    meaning that if $\Phi\lb z\rb$ solves \eqref{lsysKap}, then $\tilde\Phi\lb z\rb=\sigma_y\Phi\lb -z\rb$ solves it as well\footnote{Here and below, $\sigma_x=\lb \begin{array}{cc} 0 & 1 \\ 1 & 0 \end{array}\rb$,  $\sigma_y=\lb \begin{array}{cc} 0 & -i \\ i & 0 \end{array}\rb$, $\sigma_z=\lb \begin{array}{cc} 1 & 0 \\ 0 & -1 \end{array}\rb$ denote the Pauli matrices.}. While we do not need to recall the origin of the parameterization of $A\lb z\rb$, let us note that the local monodromy exponents around the Fuchsian singularity are given by $\left\{\theta-\frac12,-\theta+\frac12\right\}$ and thus encode the \PIIeq parameter.
    
    \begin{rmk} On the other hand, the system \eqref{lsysKap} can be shown to be equivalent to a normal form of $2\times 2$  linear system with one Fuchsian singularity and one ramified irregular singular point of Poincaré rank $\frac32$. Indeed, placing these singular points at $0$ and $\infty$ respectively by Möbius transformations, we can write
    \beq\label{lsysKap2}
    \partial_\xi\Psi=\tilde A\lb \xi\rb\Psi,\qquad A\lb \xi\rb=A_{1}\xi+A_0+A_{-1}\xi^{-1},
    \eeq
    where $2\times 2$ matrices $A_{\pm1,0}$ can be assumed traceless without any loss of generality. The ramified singularity at $\infty$ means that $A_1$ is non-diagonalizable.  Using constant gauge transformations, one can thus bring it to the form
    \beq
    A_1= \lb \begin{array}{cc} 0 & -2i \\ 0 & 0 \end{array}\rb.
    \eeq
    Further, using upper triangular gauge transformations and rescaling $\xi$, one can set
    \beq
    A_0=\lb \begin{array}{cc} 0 & \alpha \\ -2i & 0 \end{array}\rb,\qquad 
    A_{-1}=\lb \begin{array}{cc} \beta & \gamma \\ \delta & -\beta \end{array}\rb.
    \eeq
    Parameterize the remaining four parameters $\alpha,\beta,\gamma,\delta$ in terms of $p,q,t,\theta$ as
    \beq
    \begin{aligned}
    	\alpha=&\,i\lb p+q^2-\frac t2\rb,\\
    	\beta=&\, p q+q^3+\frac{q t}{2}-\frac{\theta-1}{2},\\
    	\gamma=&\, -i q\left(p q+q^3+\frac{q t}{2}- \theta+1\right),\\
    	\delta=&\, - i\lb p+q^2+\frac t2\rb.
    \end{aligned}
    \eeq
    It is then straightforward to check that if $\Phi\lb z\rb$ solves \eqref{lsysKap}, then
    \beq
    \Psi\lb \xi\rb=K\lb\xi\rb^{-1}\Phi\lb \xi^2\rb, \qquad K\lb\xi\rb=\frac1{\sqrt2}
    \lb\begin{array}{cc} 1 & -q \\ 0 & i\end{array}\rb
    \lb\begin{array}{cc} \xi^{\frac12} & 0 \\ 0 & \xi^{-\frac12} \end{array}\rb
    \lb\begin{array}{cc} 1 & -i \\ -i & 1 \end{array}\rb,
    \eeq
    solves \eqref{lsysKap2}, that is, we have
    \beq 
    \tilde A\lb\xi\rb=2\xi \,K\lb\xi\rb^{-1}A\lb\xi^2\rb K\lb \xi\rb-
    K\lb\xi\rb^{-1}\partial_\xi K\lb\xi\rb.
    \eeq
    \end{rmk}
    
    \subsubsection{Stokes data}
    
    The formal solution of the FN system \eqref{lsysKap} near the irregular singularity at $\infty$ has the form 
    \begin{gather}\label{GThFN1}
    	\Phi^{\lb \infty\rb}_{\text{form}}\lb z\rb=G^{\lb\infty\rb}\lb z\rb \exp\Theta^{\lb\infty\rb}\lb z\rb,\\
    	\label{GThFN2}
    	 G^{\lb\infty\rb}\lb z\rb =\mathbf 1+
    	\sum_{k=1}^\infty G_k^{\lb\infty\rb}z^{-k},\qquad \Theta^{\lb\infty\rb}\lb z\rb=-\lb\tfrac{4}{3}z^3+tz\rb i\sigma_z
    \end{gather}
    The formal monodromy contribution is absent because of the symmetry mentioned above. It also implies that $G^{\lb\infty\rb}\lb -z\rb=
    \sigma_y G^{\lb\infty\rb}\lb z\rb \sigma_y$, which means  that the expansion coefficients have the structure
    \beq\label{Gks}
    G_{2k+1}^{\lb\infty\rb}=\alpha_{2k+1}\sigma_z+\beta_{2k+1}\sigma_x,
    \qquad G_{2k}^{\lb\infty\rb}=\alpha_{2k}\mathbf 1+i\beta_{2k}\sigma_y.
    \eeq
    The first of these coefficients are explicitly given by
    \beq\label{alphabetak}
    	\begin{gathered}
    	\alpha_1=-\frac{i}{2}\lb p^2-q^4-tq^2+\lb 2\theta-1\rb q\rb,\qquad
    	\beta_1=\frac{q}{2},\qquad
    	\alpha_2=\frac{\alpha_1^2}{2}+\frac{q^2}{8},\qquad \beta_2=\frac{ip-2q\alpha_1}{4},\\
    	\alpha_3=\frac{4\alpha_1^3+\lb3q^2-2t\rb \alpha_1-i\lb pq+\lb\theta-\frac12\rb^2\rb}{24}.
    \end{gathered}
    \eeq
    Actual solutions of \eqref{lsysKap} can only be asymptotic to $\Phi^{\lb \infty\rb}_{\text{form}}$ inside appropriate Stokes sectors. The canonical solutions $ \Phi^{\lb \infty\rb}_l\lb z\rb$, $l\in\mathbb Z$ associated to  $z=\infty$ are uniquely defined by
    \beq
    \Phi^{\lb \infty\rb}_l\lb z\rb\simeq  \Phi^{\lb \infty\rb}_{\text{form}}\lb z\rb\qquad \text{as }z\to\infty,\quad \arg z\in\lb\frac{\pi \lb l-2\rb}{3},\frac{\pi l}{3}\rb.
    \eeq
    They are related by Stokes matrices with the usual triangular structure, $\Phi^{\lb \infty\rb}_{l+1}\lb z\rb=\Phi^{\lb \infty\rb}_{l}\lb z\rb S_l$.  The aforementioned symmetry implies that for any $l\in\mathbb Z$ we have \beq\label{cansolsymm}
    \Phi_{l+3}\lb ze^{i\pi}\rb=\sigma_y\Phi_l\lb z\rb\sigma_y,
    \eeq
    which in turn gives
    \beq
    S_{l+3}=\sigma_y S_l \sigma_y.
    \eeq
    Therefore, all Stokes matrices can be expressed in terms of $S_1,S_2,S_3$. Parameterize their nontrivial elements as\footnote{We flip the sign of $s_2$ compared to \cite{Kap88,Kap92} to make the resulting formulas more symmetric.}
    \beq
    S_1=\lb\begin{array}{cc} 1 & 0 \\ s_1 & 1\end{array}\rb,\qquad 
    S_2=\lb\begin{array}{cc} 1 & -s_2 \\ 0 & 1\end{array}\rb,\qquad 
    S_3=\lb\begin{array}{cc} 1 & 0 \\ s_3 & 1\end{array}\rb. 
    \eeq
    
    To derive further constraints on the Stokes data, fix the fundamental solution of \eqref{lsysKap} as $\Phi\lb z\rb =\Phi_1\lb z\rb$ and assume that $2\theta\notin\mathbb Z$ (the reader is referred to \cite{FN80} for the necessary modifications in the resonant case). The behavior of $\Phi\lb z\rb$ near the regular singularity at $z=0$ is then given by
    	\beq\label{connmatr0}
    \Phi\lb z\rb=G^{\lb 0\rb}\lb z\rb z^{\lb\theta-\frac12\rb\sigma_z} E,\qquad 
    G^{\lb 0\rb}\lb z\rb =G_{0}^{\lb 0\rb}\lb\mathbf 1+
    \sum_{k=1}^\infty G_k^{\lb0\rb}z^{k}\rb,
    \eeq
    where $E\in\mathrm{SL}\lb2,\mathbb C\rb$ is a constant connection matrix and $G_{0}^{\lb 0\rb}$ can be parameterized as
    \beq\label{gaugepar}
    G_{0}^{\lb 0\rb}=\frac{1}{\sqrt 2}\lb\begin{array}{cc}
    	e^{-\frac{i\pi}{4}} & e^{-\frac{i\pi}{4}} \\ -e^{\frac{i\pi}{4}} &
    	e^{\frac{i\pi}{4}}
    \end{array}\rb \lambda^{-\sigma_z},\qquad \lambda\in\mathbb C^*.
    \eeq
    The pair $\lb \lambda,E\rb$ is not defined uniquely, since the pair $\lb \Lambda^{-1}\lambda,\Lambda^{-\sigma_z}E\rb$ corresponds to the same fundamental solution $\Phi\lb z\rb$ for  any $\Lambda\in\mathbb C^*$. This ambiguity plays an important role in what follows. 
      Rewrite the relation \eqref{cansolsymm} as
    \beq
    \Phi\lb ze^{i\pi}\rb S_1S_2S_3=\sigma_y  \Phi\lb z\rb \sigma_y.
    \eeq
    Substituting the behavior \eqref{connmatr0} into this identity yields
    \beq
    S_1S_2S_3\sigma_y=-iE^{-1}e^{-i\pi\theta\sigma_z} E,
    \eeq
    and in particular $\operatorname{Tr}\lb S_1S_2S_3\sigma_y\rb=-2i\cos\pi\theta $, which is nothing else but our earlier cubic surface equation \eqref{monodromy_S}. Observe that for $\theta\notin\mathbb Z$ this equation determines $E$ in terms of $S_1,S_2,S_3$ up to the diagonal gauge freedom we have just mentioned.

    \subsubsection{Monodromy preserving deformation}
     In the setting of isomonodromic deformations, one varies the data $\lb t,q,p\rb$ of the linear system \eqref{lsysKap} simultaneously requiring that the monodromy data $\lb S_1,S_2,S_3,\theta\rb$ remain constant. This gives a second equation of the FN Lax pair,
     	\beq\label{2ndFN}
     \partial_t\Phi=B\lb z\rb\Phi,\qquad B\lb z\rb=
     \lb\begin{array}{cc}-iz & iq \\ -iq & iz\end{array}\rb.
     \eeq
     The compatibility of \eqref{lsysKap} and \eqref{2ndFN} leads to the equations
     \beq\label{defeqsFN1}
     q'=p,\qquad p'=2q^3+tq+\frac12-\theta,
     \eeq
      clearly equivalent to the standard form \eqref{PIIstandard} of \PIIeq. We need a further refinement of these equations. Namely, let us also require the connection matrix $E$ to be independent on the isomonodromic time $t$, a condition that can be ensured by adjusting the gauge parameter $\lambda$ in \eqref{gaugepar}. Substituting the local behavior \eqref{connmatr0} into \eqref{2ndFN}, one then finds that
      \beq\label{uform1}
      \lb \ln \lambda\rb'=-q.
      \eeq
      Comparing this equation with \eqref{sigmatilde}, it follows that
      \beq\label{uform2}
      \lambda\lb t\rb=\mathrm{const}\cdot\frac{\tilde \tau\lb t\rb}{\tau\lb t\rb},
      \eeq
      where $\tilde\tau$ denotes the tau function associated to the solution $\tilde q=-q$. Therefore, $\lambda$ is a meromorphic function of $t$ and its $t\to\infty$ asymptotics can be easily deduced from the asymptotics of the corresponding tau functions. The multiplicative ambiguity in the definition of $\lambda$ by \eqref{connmatr0}--\eqref{gaugepar} is also closely related to the tau function normalization ambiguity.

      \subsubsection{Asymptotic parameters in terms of monodromy}
      For generic monodromy data, the asymptotic behavior of $q\lb t\rb$ on the special rays is described by Theorems 3 and 4 of \cite{Kap92}, corresponding respectively to the rays $\arg t=\pi +\frac{2\pi p}{3}$ and $\arg t=\frac{2\pi p}{3}$, $p\in\mathbb Z/3\mathbb Z$. 
      
      Let us start with the ray $\arg t=0$. Part (i) of \cite[Theorem 4]{Kap92} states that if $s_2\ne 0$ and $\frac{\pi}{3}<|\arg s_2|<\frac{2\pi}{3}$, then
      \beq\label{asKapaevRay0}
      q\lb t\to+\infty\rb= i\epsilon \sqrt{\frac{t}{2}}+c \lb \frac23\rb^{i\gamma-\frac12}t^{\frac{3i\gamma}{2}-\frac14}e^{\frac{2i\sqrt 2}{3}t^{\frac32}}+d \lb \frac23\rb^{-i\gamma-\frac12}t^{-\frac{3i\gamma}{2}-\frac14}e^{-\frac{2i\sqrt 2}{3}t^{\frac32}}+O\lb t^{3|\Im \gamma|-1}\rb, 
      \eeq
      where
      \begin{subequations}\label{ident0}
      \begin{align}
      	&\epsilon=\operatorname{sgn} \Im s_2,\qquad \gamma=\frac{\ln s_2}{\pi}-\frac{i\epsilon}{2},\\
      	& c=\frac{\epsilon\pi 2^{\frac34}3^{-\frac12}\lb 6\sqrt2\rb^{i\gamma}e^{\frac{\pi\gamma}{2}-\frac{i\pi}{4}}}{\lb 1-s_2s_3\rb\Gamma\lb \frac{\epsilon\hat\theta+i\gamma}{2}\rb \Gamma\lb \frac{1-\epsilon\hat\theta+i\gamma}{2}\rb},\\
      	& d=-\frac{\epsilon\pi 2^{\frac34}3^{-\frac12}\lb 6\sqrt2\rb^{-i\gamma}e^{\frac{\pi\gamma}{2}+\frac{i\pi}{4}}}{\lb 1-s_1s_2\rb\Gamma\lb -\frac{\epsilon\hat\theta+i\gamma}{2}\rb \Gamma\lb \frac{1+\epsilon\hat\theta-i\gamma}{2}\rb}.
      \end{align}
      \end{subequations}
      Three contributions to the asymptotics in \eqref{asKapaevRay0} are consistent with the expansion \eqref{qexpreal} where they correspond to the terms with $k=0,\pm1$ and $l=0$. The parameter $\epsilon=\pm1$ appears in these expansions in the same way, and we further have
      \beq\label{ident1}
      \mu=-i\gamma,\qquad q_{1,0}=-2^{-\frac{3i\gamma}{2}+\frac14} 3^{-i\gamma+\frac12}\cdot ic,\qquad q_{-1,0}=-2^{\frac{3i\gamma}{2}+\frac14} 3^{i\gamma+\frac12}\cdot id.
      \eeq
      Next, comparing \eqref{ident0}--\eqref{ident1} with the expressions \eqref{nurhoplus} for asymptotic parameters of the tau function expansion, one finds for both $\epsilon=+1$ and $\epsilon=-1$ the \textit{same} remarkably simple result
      \beq\label{ident2}
      \nu_0=\frac{\ln s_2}{2\pi i},\qquad e^{2\pi i \rho_0}=1-s_1s_2.
      \eeq  
      Moreover, similar calculations give exactly the same identification \eqref{ident2} for Part (ii)  of \cite[Theorem 4]{Kap92} describing the ${t\to+\infty}$ asymptotics of $q\lb t\rb$ for $s_2\ne0$ with $\arg s_2\in\lb-\frac{\pi}{2},\frac{\pi}{2}\rb$ and $\arg s_2\in\lb \frac{\pi}{2},\frac{3\pi}{2}\rb$.
      
        Let us point out that, even though this coincidence results from rather nontrivial computations, it is not as surprising as it may seem at first sight. Indeed, it simply indicates that the tau function asymptotic expansion has the same form for generic Stokes data. That the corresponding expansions of $q\lb t\rb$ may have  different form in different regions of parameter space is merely an artefact of the denominators appearing in \eqref{qintermsoftau} upon differentiating $\ln\tau$, together with the dependence of the relative asymptotic ordering of different terms in \eqref{expsI}--\eqref{expsID} on the value of $\Re \nu_0$.
    
    \begin{rmk}\label{RmkTron}
    	An additional hypothesis $1-s_1s_2,1-s_2s_3\ne 0$, necessary for the formulas \eqref{asKapaevRay0}--\eqref{ident0} even to make sense, is left implicit in \cite{Kap92}. Instead, Remark 1 of loc. cit. asserts that these formulas admit the corresponding limit transitions. The same turns out to be true for the asymptotic expansion of the tau function. Note that e.g. $s_1s_2\to1$ means that $e^{2\pi i\rho_0}\to 0$. On the other hand, the monodromy condition \eqref{monodromy_S} implies that in this case either $\lb s_1,s_2\rb=\lb e^{i\pi\theta},e^{-i\pi\theta}\rb$ or $\lb s_1,s_2\rb=\lb e^{-i\pi\theta},e^{i\pi\theta}\rb$, and therefore the product of Barnes functions such as  $G\lb 1+\nu_0+n+\tfrac{\theta}{2}\rb
    	G\lb 1+\nu_0+n-\tfrac{\theta}{2}\rb$ vanishes for all but a finite number of values of $n\in\mathbb Z_{\leq 0}$. In particular, if we set $\nu_0=\frac{\theta}{2}$, $e^{2\pi i \rho_0}\to0$, only $n=0$ term appears in \eqref{expsI}, which yields the  asymptotics
    	\beq
    	\tau\lb t\rb\simeq\operatorname{const}\,\cdot\, \mathcal D\lb 4\sqrt 2\,e^{\frac{i\pi}{2}}|t|^{\frac32}\,\biggl|\,\frac{\theta}{2},\theta\rb
    	\eeq
    	describing the tau function corresponding to monodromy data $s_1^{-1}=s_2=s_3^{-1}=e^{i\pi\theta}$. In the case $\theta\in\mathbb Z$, there exists a more refined limit $s_1s_2\to 1$ of \eqref{expsI}--\eqref{expsID} leading to the asymptotic description of Airy function solutions of \PIIeq. Related limit transitions are also discussed in Subsection~\ref{subsec_HM}.
    \end{rmk}
    
\begin{figure}[h!]
    		\centering
   	\includegraphics[width=0.5\linewidth]{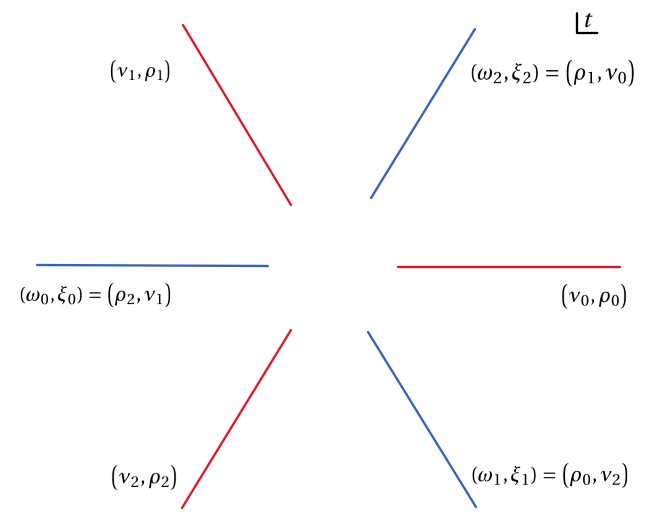}
    		\caption{Pairs of asymptotic parameters associated to different \PIIeq canonical rays.}
    		\label{figcanrays}
    	\end{figure}

    The expression of asymptotic parameters on the rays $\arg t=\pm\frac{2\pi}{3}$ in terms of monodromy can be now easily found using $\mathbb Z_3$-symmetry. Furthermore, the calculations analogous to the ones that have led to \eqref{ident2} can also be carried out for the complementary rays $\arg t=\pi +\frac{2\pi p}{3}$, $p\in\mathbb Z/3\mathbb Z$. Altogether, this leads to the following reformulation of Kapaev's results for generic monodromy:
    \begin{theo}\label{TheoConn} Under the assumptions
    	\beq\label{ess_genericity}
    	s_1,s_2,s_3,1-s_1s_2,1-s_2s_3,1-s_1s_3\ne 0,
    	\eeq
   the parameters of the asymptotic expansions \eqref{expsI}, \eqref{expsII} are given by 	
    	\begin{subequations}\label{monoparameters}
    		\begin{align}\label{asparsA}
    		& e^{2\pi i \nu_0}=e^{2\pi i\xi_2}=s_2,\qquad e^{2\pi i\rho_0}=e^{2\pi i \omega_1}=1-s_1s_2,\\
    		\label{asparsB}
    	& e^{2\pi i \nu_1}=e^{2\pi i\xi_0}=s_3,\qquad e^{2\pi i\rho_1}=e^{2\pi i \omega_2}=1-s_2s_3, \\
    	\label{asparsC}
    	& e^{2\pi i \nu_2}=e^{2\pi i\xi_1}=s_1,\qquad e^{2\pi i\rho_2}=e^{2\pi i \omega_0}=1-s_1s_3. 
    	\end{align}
    	\end{subequations}   	
    	\end{theo}
    	 Figure~\ref{figcanrays} schematically shows six rays of trigonometric degeneration in the complex $t$-plane along which the \PIIeq tau function has asymptotic expansions \eqref{expsI}--\eqref{expsID}, \eqref{expsII}--\eqref{expsIIG}, as well as the asymptotic parameters associated to each of them.

     \section{Connection coefficients}\label{sec_CC}
     \subsection{Jimbo-Miwa-Ueno tau function}
     The first point we want to emphasize is that  the isomonodromic tau function is determined by the linear problem undergoing the deformation (such as the FN system \eqref{lsysKap}) rather than by the isomonodromic system itself (such as \PIIeq). Different Lax representations of the same isomonodromic system may therefore give rise to different (though related) definitions of the tau function, which can sometimes be a source of confusion. Let us illustrate this phenomenon for \PIIeq.

     In terms of the notation \eqref{GThFN1}--\eqref{GThFN2}, the standard definition of the Jimbo-Miwa-Ueno (JMU) tau function for linear systems with regular and non-ramified irregular singularities \cite[Eqs. (1.23)--(1.24)]{JMU81} gives
     \beq\label{JMUdef}
     d_{\mathcal T}\ln\tau_{\mathrm{FN}}=-\operatorname{res}_{z=\infty}\operatorname{Tr}
     \lb {G^{\lb\infty\rb}\lb z\rb}^{-1}\partial_z G^{\lb\infty\rb}\lb z\rb d_{\mathcal T} \Theta^{\lb\infty\rb}\lb z\rb\rb.
     \eeq
     The subscript $\mathcal T$ in the above formula indicates that the differential is taken only with respect to the  time variable $t$; later we will deal with differentials extended to the space of monodromy data. Using the expression of $\Theta^{\lb\infty\rb}\lb z\rb$ and \eqref{Gks}--\eqref{alphabetak}, one obtains
     \beq
     d_{\mathcal T}\ln\tau_{\mathrm{FN}}=i\operatorname{Tr}\lb G_1^{(\infty)}\sigma_z\rb\;dt=2i\alpha_1dt=\lb p^2-q^4-tq^2+\lb 2\theta-1\rb q\rb dt.
     \eeq
     Since $p=q'$, the tau function associated to the FN system is  given by
     \beq
     \lb \ln \tau_{\mathrm{FN}}\rb'=2\lb \ln \tau\rb'-q+\frac{t^2}{4},
     \eeq
     where $\tau$ refers to the tau function introduced in our original definition \eqref{PIItauDef}. Further using \eqref{sigmatilde} (or \eqref{uform1}--\eqref{uform2}), we see that
     \beq
     \tau_{\mathrm{FN}}\lb t\rb=e^{\frac{t^3}{12}}\tau\lb t\rb\tilde\tau\lb t\rb.
     \eeq
     Recall that $\tilde\tau$ is the B\"acklund transformed tau function  associated to the solution $\tilde q=-q$ of \PIIeq with $\theta\mapsto 1-\theta$. Therefore, the JMU tau function of the FN system is a composite object, expressed via a product of two tau functions defined by \eqref{PIItauDef}.
     
     Next, let us consider the Jimbo-Miwa (JM) Lax pair for \PIIeq. Similarly to \eqref{lsysKap}, \eqref{2ndFN}, it reads
     \beq\label{JMLaxpair}
       \partial_z\Phi=A\lb z\rb\Phi,\qquad \partial_t\Phi=B\lb z\rb\Phi,
     \eeq
     where $A\lb z\rb$ and $B\lb z\rb$ are now defined by (cf \cite[Appendix C]{JMU81b})
     \begin{gather}
     A\lb z\rb=z^2\lb\begin{array}{cc} 1 & 0 \\ 0 & -1\end{array}\rb
     +z\lb\begin{array}{cc} 0 & \lambda \\ -2\lambda^{-1}P & 0\end{array}\rb+
     \lb\begin{array}{cc} P+\frac t2 & -\lambda q \\ -2 \lambda^{-1} \lb P q +\theta\rb & -P-\frac t2\end{array}\rb,\\
     B\lb z\rb=\frac{z}{2}\lb\begin{array}{cc} 1 & 0 \\ 0 & -1\end{array}\rb+\frac12
     \lb\begin{array}{cc} 0 & \lambda \\ -2\lambda^{-1}P & 0\end{array}\rb.
     \end{gather}
      The compatibility of \eqref{JMLaxpair} leads to the deformation equations
     \beq\label{deformeqs2}
     q'=P+q^2+\frac t2,\qquad P'=-2Pq-\theta,\qquad \lb\ln \lambda\rb'=-q,
     \eeq
     which coincide with \eqref{defeqsFN1}--\eqref{uform1} upon identification $p=P+q^2+\frac t2$. Note that, thanks to \eqref{sigma1and2viaQ}, we have a useful simple relation between $P$ and the tau function:
     \beq\label{PasSigmaprime}
     P=2\sigma'.
     \eeq
     The first equation in \eqref{JMLaxpair} is a normal form of $2\times2$ linear system with a single irregular singularity on $\mathbb C\mathbb P^1$ of Poincaré rank $3$, placed at $z=\infty$. The formal solution of the JM system near $\infty$ can be written in the form \eqref{GThFN1}, with
      \begin{gather}
  	\label{GThJM}
  	\Theta^{\lb\infty\rb}\lb z\rb=\lb \frac{z^3}{3}+\frac{tz}{2}-\theta\ln z\rb\sigma_z.
  \end{gather}
     The formal monodromy is now present and is encoded by the parameter $\theta$. The expressions of  matrix coefficients $G_k^{(\infty)}$ are  determined by the new system. In particular,
     \begin{subequations}\label{Ginfcoefs}
     \begin{align}
     \label{HamJM}& G_1^{(\infty)}=\lb\begin{array}{cc} -H & -\lambda/2 \\ -P/\lambda  & H \end{array}\rb,\qquad H=\frac{P^2}{2}+P\lb q^2+\frac t2\rb+\theta q,\\
     & G_2^{(\infty)}=\frac14\lb\begin{array}{cc}
     2H^2+P-\theta t & -2\lambda\lb H-q\rb \vspace{0.1cm}\\ 
     2\lambda^{-1}\lb PH-Pq-\theta\rb & 2H^2+P+\theta t\end{array}\rb,\\
     & G_3^{(\infty)}=\lb\begin{array}{cc}
     	-\frac{H^3-Pq-\theta\lb\theta+2\rb}{6}-\frac{H\lb 3P-\lb3\theta+2\rb t\rb}{12}
     	 & -\frac{\lambda\lb 2H^2-4qH-P+\lb\theta-2\rb t\rb}{8} \vspace{0.1cm} \\ 
     	 \frac{\lb\theta+1\rb \lb tP+2H\rb-2P\lb H-q\rb^2 +2\theta\lb H-q\rb}{4\lambda}
     	  & \frac{H^3-Pq-\theta\lb\theta-1\rb}{6}+\frac{H\lb 3P+\lb3\theta-2\rb t\rb}{12}\end{array}\rb.
     \end{align}
     \end{subequations}
     The first two coefficients were given in \cite[Eq. (C.14)]{JMU81b}; however, we will also need (the difference of diagonal elements of) $G_3^{(\infty)}$ in the proof of Theorem~\ref{Thm_Tau_Ext}  below.
     Computing the right side of \eqref{JMUdef} with these new $\Theta^{\lb\infty\rb}\lb z\rb$, $G^{\lb\infty\rb}\lb z\rb$, one then finds that
     \beq
     d_{\mathcal T}\ln\tau_{\mathrm{JM}}=-\frac12\operatorname{Tr}\lb G_1^{(\infty)}\sigma_z\rb\;dt=
     H dt.
     \eeq
     As expected, this reproduces Eq. (C.15) of \cite{JMU81b} and shows that $\tau_{\mathrm{JM}}=\tau$, i.e. the tau function defined by \eqref{PIItauDef} is nothing but the JMU tau function of the JM system.
     
     \subsection{Extended tau function}
     We now want to generalize the JMU differential so that it  controls the \PIIeq tau function dependence not only on the time variable $t$ but also on the formal monodromy exponent $\theta$ and Stokes coefficients $\mathbf{s}\in\mathcal S_\theta$. We will implement the general strategy of \cite{ILP16}. The starting point of the construction is a 1-form $\boldsymbol{\omega}$ (not to be confused with the asymptotic parameters $\omega_{0,1,2}$) which in the case at hand reads
     \beq\label{exttaudef0}
     \boldsymbol{\omega}=\operatorname{res}_{z=\infty}\operatorname{Tr}\lb A\lb z\rb dG^{\lb \infty\rb}\lb z\rb\, {G^{\lb \infty\rb}\lb z\rb}^{-1}\rb.
     \eeq
     The differential $d=d_{\mathcal T}+d_{\mathcal M}$ in the above formula is taken with respect to all possible parameters (here $\mathcal M$ stands for the space of monodromy data).
     \begin{theo}\label{Thm_Tau_Ext} We have the following:
     	\begin{enumerate}
     		\item The restriction of $\boldsymbol{\omega}$ to subspace with fixed monodromy coincides with $d_{\mathcal T}\ln\tau$.
     		\item The differential $d\boldsymbol{\omega}=\Omega$ is a closed 2-form on $\mathcal M$ independent of isomonodromic time.
     		\item The 1-form $\boldsymbol\omega$ is explicitly given by
     		\beq\label{omegaexpr1}
     		\boldsymbol{\omega}=\frac{2Pdq-qdP+2tdH-Hdt}{3}-\theta d\lb\ln \lambda\rb+\frac{d\lb \theta-\theta^2\rb}{3}.
     		\eeq
     	\end{enumerate}
     	\end{theo}
    \pf The first two assertions were proved in \cite[Section 4]{ILP16} in a much more general setting. Here we give a direct proof based on	\eqref{omegaexpr1}, which is itself obtained by direct computation. From \eqref{exttaudef0} it follows that
    \beq
    \begin{aligned}
    \boldsymbol{\omega}=&\,\operatorname{Tr} \sigma_z \lb dG_1^{\lb\infty\rb}\,\lb G_2^{\lb\infty\rb}-{G_1^{\lb\infty\rb}}^2\rb+dG_2^{\lb\infty\rb}\,G_1^{\lb\infty\rb}-dG_3^{\lb\infty\rb}\rb\\ &+\operatorname{Tr}\lb\begin{array}{cc} 0 & \lambda \\ -2\lambda^{-1}P & 0\end{array}\rb\lb dG_1^{\lb\infty\rb}\,G_1^{\lb\infty\rb}-dG_2^{\lb\infty\rb}\rb
    -\operatorname{Tr}\lb\begin{array}{cc} P+\frac t2 & -\lambda q \\ -2 \lambda^{-1} \lb P q +\theta\rb & -P-\frac t2\end{array}\rb dG_1^{\lb\infty\rb}.
    \end{aligned}
    \eeq
    Using \eqref{Ginfcoefs}, a somewhat tedious but straightforward calculation then  yields the stated result. An equivalent expression can be found in \cite[Eq. (33)]{IP18}.
    
    To prove the first assertion, it suffices to observe that the restriction of $\boldsymbol\omega$ to subspace with fixed monodromy is given by
    \beq
    \lb\frac{2Pq'-qP'+2tH'-H}{3}-\theta\lb \ln \lambda\rb'\rb dt=Hdt,
    \eeq 
	    where the equality is an easy corollary of the deformation equations \eqref{deformeqs2}. For the second assertion, we first note that
	    \beq
	    \Omega=dP\wedge dq-dH\wedge dt-d\theta\wedge d\ln \lambda.
	    \eeq
	 Next, differentiating with respect to time, we get
	 \beq
	 \begin{aligned}
	 \partial_t\Omega=&\,dP'\wedge dq+dP\wedge dq'-dH'\wedge dt+d\theta\wedge dq=\\
	 =&\,-d\lb\frac{\partial H}{\partial q}dq+\frac{\partial H}{\partial P}dP+\frac{\partial H}{\partial t}dt+\frac{\partial H}{\partial \theta}d\theta\rb=-d^2H=0.
	 \end{aligned} 
	 \eeq 
	 Here we have used that the deformation equations \eqref{deformeqs2} can be written as a non-autonomous Hamiltonian system with the Hamiltonian $H=H\lb P,q,t,\theta\rb$ defined by \eqref{HamJM}; that is, $q'=\frac{\partial H}{\partial P}$, $P'=-\frac{\partial H}{\partial q}$, $H'=\frac{\partial H}{\partial t}$.  
	 
	 These equations also immediately imply that $\Omega$ cannot contain the terms proportional to $dm\wedge d t$, where $m$ is an arbitrary parameter specifying the initial conditions. Indeed, the coefficient of such a term would be
	 \beq
	 P_m q'-P'q_m-H_m=0.
	 \eeq
	 The absence of a contribution proportional to $d\theta\wedge dt$ follows by a very similar argument. The two cases can also be treated uniformly by interpreting $H$ as a Hamiltonian of a 4-dimensional Hamiltonian system with coordinates $q,-\ln \lambda$ and respective conjugate momenta $P$, $\theta$.
    \epf
  
    Our next task is to find a 1-form $\boldsymbol{\omega}_0$ on $\mathcal M$ such that $d\boldsymbol{\omega}_0=\Omega$. Since the difference $\boldsymbol{\omega}-\boldsymbol{\omega}_0$ is then a closed 1-form on $\mathcal T\times\mathcal M$, we will be able to control the dependence of the tau function on monodromy by defining $d\ln\tau=\boldsymbol{\omega}-\boldsymbol{\omega}_0$. To carry out this program, we have to first compute $\Omega$. Note a useful rewrite:
    \beq\label{formOmega}
    \Omega=d_{\mathcal M}P\wedge d_{\mathcal M}q-d\theta\wedge d_{\mathcal M}\ln \lambda,
    \eeq
    where $d_{\mathcal M}$ denotes the differential with respect to monodromy data only. Let us now  make it more explicit using the previous asymptotic results. The idea is to compute the asymptotics of $\Omega$ up to $o\lb 1\rb$ terms as $t\to\infty$, which gives the exact result by Theorem~\ref{Thm_Tau_Ext}.

    To proceed, it is convenient to express all quantities appearing in \eqref{formOmega} in terms of tau functions. At first sight, this may look like a circular argument  since the extended tau function is precisely the object we seek to define. However, we will only be dealing with logarithmic derivatives which are independent of the relative normalization of the tau function asymptotics on different rays. In fact, these expressions are simply an auxiliary tool that simplifies the computation of relevant asymptotic behaviors. 
    
    For example, from  the relation \eqref{PasSigmaprime} we deduce
    \begin{lemma}\label{LemAsP}
    	Let the asymptotic parameters $\nu_p$, $\omega_p$ ($p\in\mathbb Z/3\mathbb Z$) in the expansions \eqref{expsI}--\eqref{expsID} and \eqref{expsII}--\eqref{expsIIG} satisfy the genericity condition $|\Re\nu_p|,|\Re\omega_p|<\frac12$. Then:
    	\begin{enumerate}
    	\item
    	As $t\to\infty$ with $\arg t=\frac{2\pi p}{3}$, we have
    	\beq\label{logaspplus}
    	P\lb t\rb= 2\frac{d^2}{dt^2}\left[\frac{\nu_p s}{6}+\ln\lb 1+I_+\lb\nu_p,\theta\rb e^{2\pi i \rho_p} s^{-1-2\nu_p}e^{\frac{s}{6}}+I_-\lb \nu_p,\theta\rb e^{-2\pi i\rho_p}s^{-1+2\nu_p}e^{-\frac{s}{6}}\rb\right]+o\lb|t|^{-\frac12}\rb,
    	\eeq
    	with $s=4\sqrt 2\,e^{\frac{i\pi}{2}}|t|^{\frac32}$ and
    	\beq
    	I_+\lb \nu_p,\theta\rb=\frac{\Gamma\lb 1+\nu_p+\frac{\theta}{2}\rb \Gamma\lb 1+\nu_p-\frac{\theta}{2}\rb}{2\pi},\qquad  I_-\lb \nu_p,\theta\rb=\frac{2\pi}{
    	\Gamma\lb \nu_p+\frac{\theta}{2}\rb \Gamma\lb \nu_p-\frac{\theta}{2}\rb}.
    	\eeq
    	\item As $t\to\infty$ with $\arg t=\pi +\frac{2\pi p}{3}$, we have
    	\beq\label{PasComp}
    	P\lb t\rb= -\frac{t}{2}+ 2\frac{d^2}{dt^2}\left[\frac{\omega_p s}{12}+\ln\lb 1+\frac{\Gamma\lb 1+\omega_p\rb}{\sqrt{2\pi}} e^{2\pi i \xi_p} s^{-\frac12-\omega_p}e^{\frac{s}{12}}+\frac{\sqrt{2\pi}}{\Gamma\lb \omega_p\rb} e^{-2\pi i\xi_p}s^{-\frac12+\omega_p}e^{-\frac{s}{12}}\rb\right]+o\lb|t|^{-\frac12}\rb,
    	\eeq
    	with $s=8\,e^{\frac{i\pi}{2}}|t|^{\frac32}$.
    	\end{enumerate}
    \end{lemma}
    
    \begin{rmk}
    	If we further assume that $\Re \nu_p\ne0$, then the expression under the logarithm in \eqref{logaspplus} can be simplified: the second term therein can be omitted when $0<\Re \nu_p<\frac12$, while the third term can be omitted when  $-\frac12<\Re \nu_p<0$.
    \end{rmk}
    
    Before we write a counterpart of Lemma~\ref{LemAsP} for $q$, let us note the following useful result:
          \begin{lemma}\label{LemBackMon}
    	If $q$ is a solution of \PIIeq corresponding to the monodromy data $\lb s_1,s_2,s_3,\theta\rb$, then $\tilde q=-q$ is a solution of \PIIeq that corresponds to monodromy 
    	\beq\lb \tilde s_1,\tilde s_2,\tilde s_3,\tilde\theta\rb=\lb -s_1,-s_2,-s_3,1-\theta\rb.
    	\eeq
    \end{lemma}
    \pf One way to prove this statement is to notice that the system \eqref{lsysKap} after replacements $q\to -q$, $p\to-p$, $\theta\to1-\theta$ is solved by $\tilde\Phi\lb z\rb=\sigma_z \Phi\lb z\rb \sigma_z$. The corresponding Stokes matrices are given by $\tilde S_l=\sigma_z S_l{}\sigma_z$, which immediately gives the result. 
    
    It is also instructive to give an alternative derivation using the asymptotics. For example,  it follows from \eqref{qexpreal} that $\tilde q$ corresponds to the same $\mu$ but reversed $\epsilon$ along the ray $\arg t=0$. Therefore, according to \eqref{nurhoplus}, the difference $\nu_0-\tilde\nu_0$ equals to either $\frac12$ or $-\frac12$, which in turn gives $\tilde s_2=-s_2$. Note that the asymptotic parameters of the transformed solution on the complementary rays are given by  $\lb\tilde\omega_p,\tilde\xi_p\rb=\lb\omega_p,\xi_p+\frac12\rb$.  \epf
    
    Lemma \ref{LemBackMon} thus allows to express the asymptotics of $q$ in a more convenient way using that $q=\sigma-\tilde\sigma$ (cf \eqref{sigmatilde}), instead of \eqref{qintermsoftau}. This yields
    \begin{lemma} \label{LemAsQ}
    Let the asymptotic parameters in \eqref{expsI}--\eqref{expsID} and \eqref{expsII}--\eqref{expsIIG} satisfy $|\Re\nu_p|,|\Re\omega_p|<\frac12$, $\Re\nu_p\ne0$. Then:
    \begin{enumerate}
    	\item As $t\to\infty$ with $\arg t=\frac{2\pi p}{3}$, we have
    	\begin{subequations}
    	\beq\label{logasqplusA}
    	\begin{aligned}
    		q\lb t\rb=&\, \frac{d}{dt}\,\biggl[\frac{ s}{12}+\frac{\theta+1-6\nu_p}{6}\ln s+\biggl.\ln\frac{ 1+I_-\lb \nu_p,\theta\rb e^{-2\pi i\rho_p}s^{-1+2\nu_p}e^{-\frac{s}{6}}}{
    			1+I_+\lb\nu_p-\frac12,1-\theta\rb e^{2\pi i \rho_p} s^{-2\nu_p}e^{\frac{s}{6}}}\biggr]+o\lb|t|^{-1}\rb,
    	\end{aligned}
    	\eeq
    	for $0<\Re\nu_p<\frac12$, and
    	\beq
    	\begin{aligned}\label{logasqplusB}
    		q\lb t\rb=&\, \frac{d}{dt}\,\biggl[\,-\frac{ s}{12}+\frac{\theta+1+6\nu_p}{6}\ln s+\biggl.\ln\frac{ 1+I_+\lb\nu_p,\theta\rb e^{2\pi i \rho_p} s^{-1-2\nu_p}e^{\frac{s}{6}}}{
    			1+I_-\lb \nu_p+\frac12,1-\theta\rb e^{-2\pi i\rho_p}s^{2\nu_p}e^{-\frac{s}{6}}}\biggr]+o\lb|t|^{-1}\rb,
    	\end{aligned}
    	\eeq
    	\end{subequations}
    	for $-\frac12<\Re\nu_p<0$,
    	where $s=4\sqrt 2\,e^{\frac{i\pi}{2}}|t|^{\frac32}$.
    	\item As $t\to\infty$ with $\arg t=\pi +\frac{2\pi p}{3}$, we have
    	\beq
    	q\lb t\rb= \frac{d}{dt}\left[\frac{2\theta-1}{3}\ln s+\ln\frac{ 1+\frac{\Gamma\lb 1+\omega_p\rb}{\sqrt{2\pi}} e^{2\pi i \xi_p} s^{-\frac12-\omega_p}e^{\frac{s}{12}}+\frac{\sqrt{2\pi}}{\Gamma\lb \omega_p\rb} e^{-2\pi i\xi_p}s^{-\frac12+\omega_p}e^{-\frac{s}{12}}}{ 1-\frac{\Gamma\lb 1+\omega_p\rb}{\sqrt{2\pi}} e^{2\pi i \xi_p} s^{-\frac12-\omega_p}e^{\frac{s}{12}}-\frac{\sqrt{2\pi}}{\Gamma\lb \omega_p\rb} e^{-2\pi i\xi_p}s^{-\frac12+\omega_p}e^{-\frac{s}{12}}}\right]+o\lb|t|^{-1}\rb,
    	\eeq
    	with $s=8\,e^{\frac{i\pi}{2}}|t|^{\frac32}$.
    \end{enumerate}
    	
    \end{lemma}
    \begin{rmk} This asymptotics is of course consistent with the results discussed earlier in Subsection~\ref{subsec_Asq}. The main motivation for us to write it in this form is to improve the error estimate compared to e.g. \eqref{asKapaevRay0} while keeping $\nu_p$, $\omega_p$ sufficiently generic. 
     \end{rmk}
    
    The last ingredient necessary for the computation of the 2-form  $\Omega$ is the asymptotics of the function $\lambda$ appearing in \eqref{formOmega}. Here again Lemma~\ref{LemBackMon} is extremely instrumental. Indeed, from \eqref{uform2} we almost immediately obtain
    \begin{lemma}\label{LemAsU}
    	Under assumptions of Lemma~\ref{LemAsQ}, we have:
    	\begin{enumerate}
    		\item As $t\to\infty$ with $\arg t=\frac{2\pi p}{3}$ ($p\in\mathbb Z/3\mathbb Z$),
    		\beq\label{lnuas1}
    		\ln \lambda\lb t\rb=
    		\begin{cases}
    		-\ds\frac{ s}{12}-\ds\frac{\theta+1-6\nu_p}{6}\ln s+\ln\ds\frac{\sqrt{2\pi}}{\Gamma\lb\nu_p+\frac{\theta}{2}\rb}-i\pi\rho_p+K_{p,+}+o\lb 1\rb,\qquad &\text{for } 0<\Re\nu_p<\frac12,\vspace{0.1cm}\\
    		\ds\frac{ s}{12}-\ds\frac{\theta+1+6\nu_p}{6}\ln s+\ln \ds\frac{\Gamma\lb 1+\nu_p-\frac{\theta}{2}\rb}{\sqrt{2\pi}}+i\pi  \rho_p+K_{p,+}+o\lb 1\rb,\qquad & \text{for }-\frac12<\Re\nu_p<0,
    		\end{cases}
    		\eeq
    		where  $s=4\sqrt 2\,e^{\frac{i\pi}{2}}|t|^{\frac32}$.
    		\item As $t\to\infty$ with $\arg t=\pi +\frac{2\pi p}{3}$ ($p\in\mathbb Z/3\mathbb Z$),
    		\beq\label{lnuas2}
    		\ln \lambda\lb t\rb= \frac{1-2\theta}{3}\ln s +K_{p,-}+o\lb 1\rb,
    		\eeq
    		where $s=8\,e^{\frac{i\pi}{2}}|t|^{\frac32}$.
    	\end{enumerate}
    \end{lemma}
    The constant contributions to this asymptotics, $K_{p,\pm}$, depend on the normalization of the function $\lambda$. However, their differences $K_{p,\epsilon}-K_{q,\epsilon'}$ are uniquely fixed by monodromy parameters $\theta$, $\mathbf s\in\mathcal S_\theta$. In fact, as we will see shortly, these differences are close relatives of the connection coefficients we want to compute.
    \begin{rmk}\label{RmqDiscont}
    	We parameterize the constant terms in the asymptotics of $\ln \lambda$ in \eqref{lnuas1} differently depending on the sign of $\Re \nu_p$ for the following reason. A possible way to normalise $\lambda$ is to simply set $\lambda=\tilde\tau/\tau$. In this case, $K_{p,+}$ introduced above are given (for both signs of $\Re\nu_p$!) by $K_{p,+}=i\pi \rho_p+\ln\;\tilde{\mathcal N}_{p,+}/\mathcal N_{p,+}$, where $\mathcal N_{p,+}$ is the tau function normalisation coefficient in \eqref{expsI} and $\tilde{\mathcal N}_{p,+}$ is its counterpart for the transformed tau function $\tilde\tau$ characterized by the asymptotic parameters $\lb\tilde\nu_p,\tilde\rho_p\rb=\lb\nu_p-\frac12,\rho_p\rb$. Therefore, even though the asymptotics of $\ln \lambda$ changes discontinuously when one crosses the surface $\Re\nu_p=0$ in the space of monodromy data, the quantities $K_{p,+}$ vary continuously.
    \end{rmk}

    \begin{theo}\label{thmOmega}
    	The 2-form $\Omega$  can be written as
    	\begin{subequations}
    	\begin{align}\label{OmegaPlus}
    	\Omega=&\,2\pi i d\nu_p\wedge d\rho_p-d\theta\wedge dK_{p,+}=\\
    	\label{OmegaMinus}
    	=&\,2\pi i d\omega_p\wedge d\xi_p-d\theta\wedge dK_{p,-},
    	\end{align}
    	\end{subequations}
    	for every $p\in\mathbb Z/3\mathbb Z$.
    \end{theo}
    \pf First consider the asymptotics of \eqref{formOmega}  on the rays with $\arg t=\frac{2\pi p}{3}$, $p\in\mathbb Z/3\mathbb Z$. Suppose e.g. that $0<\Re \nu_p<\frac 12$, then from \eqref{logaspplus}, \eqref{logasqplusA} it follows that
    \beq\label{auxOmega00}
    d_{\mathcal M}P\wedge d_{\mathcal M}q=2 d_{\mathcal M}\lb 
    \ln\lb 1+Q_-\rb
    \rb''
    \wedge d_{\mathcal M}\lb \ln\frac{ 1+Q_-}{
    	1+\tilde Q_+}\rb'+o\lb 1\rb,
    \eeq
    where 
    \beq
    Q_\pm=I_\pm\lb \nu_p,\theta\rb e^{\pm 2\pi i\rho_p}s^{-1\mp2\nu_p}e^{\pm\frac{s}{6}},\qquad 
    \tilde Q_\pm=I_\pm\lb\nu_p\mp\tfrac12,1-\theta\rb e^{\pm2\pi i \rho_p} s^{\mp2\nu_p}e^{\pm\frac{s}{6}}.
    \eeq
    In the limit we are interested in, we have  $Q_-,\tilde Q_+=o\lb 1\rb$, and also 
    \beq 
    Q_\pm'=\pm\ds\frac{Q_\pm s'}{6}+o\lb |t|^{-1}\rb,\qquad \tilde Q_\pm'=\pm\ds\frac{\tilde Q_\pm s'}{6}+o\lb |t|^{-1}\rb,\qquad Q_\pm''=\ds \frac{Q_\pm s'^2}{36}+o\lb |t|^{-\frac12}\rb,
    \eeq
    which allows to rewrite \eqref{auxOmega00} as
    \beq\label{Omega1stterm}
    \begin{aligned}
    d_{\mathcal M}P\wedge d_{\mathcal M}q=&\,-\frac{s'^3}{108}\,d_{\mathcal M} Q_-\wedge
    d_{\mathcal M} \tilde Q_++o\lb 1\rb=\\
    =&\,d_{\mathcal M}\lb I_-\lb \nu_p,\theta\rb e^{-2\pi i\rho_p}s^{2\nu_p}\rb\wedge d_{\mathcal M}\lb I_+\lb\nu_p-\tfrac12,1-\theta\rb e^{2\pi i \rho_p} s^{-2\nu_p} \rb+o\lb 1\rb=\\
    =&\,d\lb \nu_p-\tfrac{\theta}{2}\rb\wedge d_{\mathcal M}\ln\lb  I_+\lb\nu_p-\tfrac12,1-\theta\rb e^{2\pi i \rho_p} s^{-2\nu_p} \rb+o\lb 1\rb=\\
    =&\,d\theta\wedge d\nu_p\;\ln s+2\pi i d \nu_p\wedge d\rho_p-d\theta\wedge d\lb\pi i\rho_p+ \ln \Gamma\lb\nu_p+\tfrac{\theta}{2}\rb\rb+o\lb 1\rb.
    \end{aligned}
    \eeq
    To pass from the second to the third line in the above, we used that $I_-\lb \nu_p,\theta\rb I_+\lb\nu_p-\tfrac12,1-\theta\rb=\nu_p-\frac{\theta}{2}$. The asymptotics of the second term in \eqref{formOmega} is straightforward to compute using Lemma~\ref{LemAsU}:
    \beq\label{Omega2ndterm}
    d\theta\wedge d_{\mathcal M}\ln \lambda=d\theta\wedge d\nu_p\;\ln s-d\theta\wedge d\lb i\pi\rho_p-K_{p,+}+\ln\Gamma\lb\nu_p+\tfrac{\theta}{2}\rb\rb+o\lb 1\rb.
    \eeq
    In the difference of \eqref{Omega1stterm} and \eqref{Omega2ndterm}, the logarithmic term of the asymptotics cancels out as expected, while the $O\lb 1\rb$ term coincides with \eqref{OmegaPlus} and gives exact result by Theorem~\ref{Thm_Tau_Ext}. An analogous computation in the case $-\frac12<\Re \nu_p<0$ leads again to the same formula \eqref{OmegaPlus}, in accordance with Remark~\ref{RmqDiscont}.
    
    On the rays $\arg t=\pi+\frac{2\pi p}{3}$, $p\in\mathbb Z/3\mathbb Z$, one has 
    \beq
    d_{\mathcal M}P\wedge d_{\mathcal M}q=2 d_{\mathcal M}\lb 
    \ln\lb 1+R_++R_-\rb
    \rb''
    \wedge d_{\mathcal M}\lb \ln\frac{ 1+R_++R_-}{
    	1-R_+-R_-}\rb'+o\lb 1\rb,
    \eeq
    where 
    \beq
    R_+=\frac{\Gamma\lb 1+\omega_p\rb}{\sqrt{2\pi}} e^{2\pi i \xi_p} s^{-\frac12-\omega_p}e^{\frac{s}{12}},\qquad R_-= \frac{\sqrt{2\pi}}{\Gamma\lb \omega_p\rb} e^{-2\pi i\xi_p}s^{-\frac12+\omega_p}e^{-\frac{s}{12}}.
    \eeq
    Since (recall that here $|\Re\omega_p|<\frac12$)
    \beq
    R_{\pm}=o\lb 1\rb,\qquad R_{\pm}'=\pm\frac{R_\pm s'}{12}+o\lb |t|^{-1}\rb,\qquad
    R_{\pm}''=\frac{R_\pm s'^2}{144}+o\lb |t|^{-\frac12}\rb,
    \eeq
    we can write
    \beq
    \begin{aligned}
    d_{\mathcal M}P\wedge d_{\mathcal M}q=&\,4\lb\frac{s'}{12}\rb^3 d_{\mathcal M}\frac{R_++R_-}{\lb 1+R_++R_-\rb^2}\wedge d_{\mathcal M}\frac{R_+-R_-}{1-\lb R_++R_-\rb^2}+o\lb 1\rb=\\=&\,-8\lb\frac{s'}{12}\rb^3\frac{d_{\mathcal M} R_+\wedge d_{\mathcal M} R_-}{\lb 1+R_++R_-\rb^4}+o\lb 1\rb=2\pi i d\omega_p\wedge d\xi_p+o\lb 1\rb.
    \end{aligned}
    \eeq
   In the above calculation, one has to use several times that $R_+R_-=\omega_p/s$. From \eqref{lnuas2} it follows that $d\theta\wedge d_{\mathcal M}\ln \lambda=d\theta\wedge dK_{p,-}+o\lb 1\rb$, which then immediately gives \eqref{OmegaMinus}.
    \epf
    
    \begin{rmk}
     Temporarily viewing $s_1,s_2,s_3$ as independent coordinates on $\mathbb C^3$, one may equip  it with the Jacobian Poisson structure corresponding to the Casimir $W=s_1+s_2+s_3-s_1s_2s_3$:
     \beq
     \left\{f,g\right\}=\frac{df\wedge dg\wedge dW}{ds_1\wedge ds_2\wedge ds_3}.
     \eeq
     Affine cubic surfaces $\mathcal S_\theta$ defined by \eqref{monodromy_S}  are symplectic leaves for this structure. The symplectic form induced on each leaf   agrees (up to conventional factors of $2\pi i$) with the restriction of the 2-form $\Omega$ computed above to~$\mathcal S_\theta$. The pairs of asymptotic parameters $\lb\nu_p,\rho_p\rb$, $\lb \omega_p,\xi_p\rb$ give rise to different choices of Darboux coordinates on $\mathcal S_\theta$.
    \end{rmk}
    
    Let us now introduce six \PIIeq tau functions associated to generic monodromy data $\theta\in\mathbb C$, $\mathbf s\in\mathcal S_\theta$.
    \begin{defin}\label{defexttau}
    	We define the tau functions $\tau_{p,\epsilon}=\tau_{p,\epsilon}\lb t,\theta,\mathbf s\rb$ on $\mathcal T\times\mathcal M$ ($p\in\mathbb Z/3\mathbb Z$, $\epsilon=\pm$)  by
    	\begin{subequations}\label{TausExt}
    	\begin{align}\label{taupplus}
    		d\ln\tau_{p,+}=&\,\frac{2Pdq-qdP+2tdH-Hdt}{3}-\theta d\lb\ln \lambda-K_{p,+}\rb-2\pi i\nu_pd\rho_p+\frac{d\lb \theta-\theta^2\rb}{3},\\
    		\label{taupminus}
    		d\ln\tau_{p,-}=&\,\frac{2Pdq-qdP+2tdH-Hdt}{3}-\theta d\lb\ln \lambda-K_{p,-}\rb-2\pi i\omega_pd\xi_p+\frac{d\lb \theta-\theta^2\rb}{3}.
    	\end{align}
    	\end{subequations}
    \end{defin}
    Each of these tau functions is defined  by \eqref{TausExt} up to a multiplicative factor independent on both time and monodromy, and has a normalized asymptotic behavior along one of the six canonical rays in the $t$-plane that will be computed below. As functions of time, they are all proportional, and, as we are about to see, their ratios are essentially the connection coefficients that we are after.  
    
       \subsection{Connection coefficients}
    
    \subsubsection{Normalisation of $\tau_{p,\pm}$}
    
    \begin{prop}\label{propNorm}
    	Let $p\in\mathbb Z/3\mathbb Z$. For $|\Re\nu_p|,|\Re\omega_p|<\frac12$, $\Re\nu_p\ne0$, the tau function $\tau_{p,\pm}$ has the following asymptotic behavior:
    	\begin{enumerate}
    		\item As $t\to\infty$ with $\arg t=\frac{2\pi p}{3}$, one has
    		\beq
    		\tau_{p,+}\lb t\rb= N_{p,+}\frac{G\lb 1+\nu_p+\tfrac{\theta}{2}\rb
    			G\lb 1+\nu_p-\tfrac{\theta}{2}\rb}{\lb2\pi\rb^{\nu_p}} s^{\frac{\theta^2}{12}-\nu_p^2}e^{\frac{\nu_p s}{6}}\lb 1+o\lb 1\rb\rb,
    		\eeq
    		where $s= 4\sqrt 2\,e^{\frac{i\pi}{2}}|t|^{\frac32}$ and the prefactor $N_{p,+}$  is independent of monodromy.
    		\item As $t\to\infty$ with $\arg t=\pi+\frac{2\pi p}{3}$, one has
    		\beq
    		\tau_{p,-}\lb t\rb= N_{p,-}\frac{G\lb 1+\omega_p\rb
    			}{\lb2\pi\rb^{\frac{\omega_p}{2}}}  s^{\frac{\theta^2}{3}-\frac{1}{12}-\frac{\omega_p^2}{2}} e^{-\frac{s^2}{1536}+\frac{\omega_p s}{12}
    			}\lb 1+o\lb 1\rb\rb,
    		\eeq
    		where $s= 8\,e^{\frac{i\pi}{2}}|t|^{\frac32}$ and the prefactor $N_{p,-}$  is also independent of monodromy.
    	\end{enumerate}
    	\end{prop}
    \pf We already know from \eqref{expsI}--\eqref{expsID} that, as $t\to\infty$ with $\arg t=\frac{2\pi p}{3}$ and $|\Re\nu_p|<\frac12$, the asymptotics of the differential of $\ln\tau_{p,+}$ with respect to monodromy must have the form
    \beq\label{expectedAs1}
    d_{\mathcal M}\ln\tau_{p,+}=\frac{s}{6}d\nu_p+\ln s\; d\lb\frac{\theta^2}{12}-\nu_p^2\rb+d\ln\frac{G\lb 1+\nu_p+\tfrac{\theta}{2}\rb
    	G\lb 1+\nu_p-\tfrac{\theta}{2}\rb}{\lb2\pi\rb^{\nu_p}}+d\ln N_{p,+}+o\lb 1\rb,
    \eeq
    where $s= 4\sqrt 2\,e^{\frac{i\pi}{2}}|t|^{\frac32}$ and $N_{p,+}$ is the counterpart of $\mathcal N_{p,+}$ in \eqref{expsI} if we set $\tau=\tau_{p,+}$. Our task is precisely to compute this normalisation coefficient. To this end, it suffices to determine the asymptotics of $d_{\mathcal M}\ln\tau_{p,+}$, defined by \eqref{taupplus}, up to $o\lb 1\rb$ terms. We have
    \beq\label{derivtauppmon}
    \begin{aligned}
    	& d_{\mathcal M}\ln\tau_{p,+}=\frac{2Pd_{\mathcal  M}q-qd_{\mathcal M}P+2td_{\mathcal M}H}{3}-\theta d_{\mathcal M}\lb \ln \lambda-K_{p,+}\rb-2\pi i\nu_p d\rho_p +\frac{d\lb \theta-\theta^2\rb}{3}.
    	\end{aligned}
    \eeq
    
    The asymptotics of different terms in this formula can be obtained from Lemmas~\ref{LemAsP}, \ref{LemAsQ}, \ref{LemAsU} using arguments similar to those in the proof of Theorem~\ref{thmOmega} together with the notation introduced there. One has, in particular,
    \begin{gather}
    		Pd_{\mathcal M}q=\begin{cases} 
    		s\ds\frac{Q_-}{\lb 1+Q_-\rb^2}\lb\ds\frac{d_{\mathcal M}Q_-}{\lb 1+Q_-\rb^2}+\ds\frac{d_{\mathcal M}\tilde Q_+}{\lb 1+\tilde Q_+\rb^2}\rb+o\lb 1\rb,\qquad &\text{for }0<\Re\nu_p<\frac12,\vspace{0.1cm}\\  
    		-s\ds\frac{Q_+}{\lb 1+Q_+\rb^2}\lb\ds\frac{d_{\mathcal M}Q_+}{\lb 1+Q_+\rb^2}+\ds\frac{d_{\mathcal M}\tilde Q_-}{\lb 1+\tilde Q_-\rb^2}\rb+o\lb 1\rb,\qquad &\text{for }-\frac12<\Re\nu_p<0,\end{cases}\\
    	qd_{\mathcal M}P=\begin{cases}-d\nu_p+s\lb\ds\frac{Q_-}{1+Q_-}+\ds\frac{\tilde Q_+}{1+\tilde Q_+}-\ds\frac12\rb d_{\mathcal M}\ds\frac{Q_-}{\lb 1+Q_-\rb^2}+o\lb 1\rb,\qquad & \text{for }0<\Re\nu_p<\frac12,\vspace{0.1cm}\\
    		d\nu_p-s\lb\ds\frac{Q_+}{1+Q_+}+\ds\frac{\tilde Q_-}{1+\tilde Q_-}-\ds\frac12\rb d_{\mathcal M}\ds\frac{Q_+}{\lb 1+Q_+\rb^2}+o\lb 1\rb,\qquad &\text{for }-\frac12<\Re\nu_p<0,
    		\end{cases}\\
    	td_{\mathcal M}H=\begin{cases}\ds\frac{s}{4}d\nu_p+\ds\frac32d\lb \ds\frac{\theta^2}{12}-\nu_p^2\rb-\ds\frac{s}{4}\ds\frac{d_{\mathcal M}Q_-}{\lb 1+Q_-\rb^2}+o\lb 1\rb,\qquad\quad & \text{for }0<\Re\nu_p<\frac12,\vspace{0.1cm}\\
    	\ds\frac{s}{4}d\nu_p+\ds\frac32d\lb \ds\frac{\theta^2}{12}-\nu_p^2\rb+
    	\ds\frac{s}{4}\ds\frac{d_{\mathcal M}Q_+}{\lb 1+Q_+\rb^2}+o\lb 1\rb, \qquad\quad  & \text{for }-\frac12<\Re\nu_p<0,	\end{cases}\\
    	\theta d_{\mathcal M}\lb \ln \lambda-K_{p,+}\rb=\begin{cases}
    		-\ln s\;\ds\frac{\theta\, d\lb \theta-6\nu_p\rb}{6}+
    	\theta d\lb \ln\ds\frac{\sqrt{2\pi}}{\Gamma\lb\nu_p+\frac{\theta}{2}\rb}-i\pi\rho_p\rb +o\lb 1\rb,\quad & \text{for }0<\Re\nu_p<\frac12,\vspace{0.1cm}\\
    	-\ln s\;\ds\frac{\theta\, d\lb \theta+6\nu_p\rb}{6}+
    	\theta d\lb \ln\ds\frac{\Gamma\lb1+\nu_p-\frac{\theta}{2}\rb}{\sqrt{2\pi}}+i\pi\rho_p\rb +o\lb 1\rb,\quad & \text{for }-\frac12<\Re\nu_p<0.
    	\end{cases}
    \end{gather}
    Consider e.g. the case $0<\Re\nu_p<\frac12$. Summing up three first contributions to \eqref{derivtauppmon} and using that
     \beq
    Q_\pm\tilde Q_\mp=\frac{\nu_p\pm\frac{\theta}{2}}{s},
    \eeq
    we get
    \beq
    \begin{aligned}
    &\,\frac{2Pd_{\mathcal  M}q-qd_{\mathcal M}P+2td_{\mathcal M}H}{3}=\\=&\,\frac{s}{6}d\nu_p+ d\lb \ds\frac{\theta^2}{12}-\nu_p^2\rb+\frac{d\nu_p}{3}+\frac{2s Q_-d_{\mathcal M}\tilde Q_+}{3\lb 1+Q_-\rb^2\lb 1+\tilde Q_+\rb^2}-\frac{s \tilde Q_+d_{\mathcal M}Q_-}{3\lb 1+Q_-\rb^3\lb 1+\tilde Q_+\rb}+o\lb 1\rb=\\
    =&\,\frac{s}{6}d\nu_p+ d\lb \ds\frac{\theta^2}{12}-\nu_p^2\rb+\frac{d\nu_p}{3}+\frac{2\nu_p-\theta}{6} d_{\mathcal M} \lb 2\ln\tilde Q_+-\ln Q_-\rb+o\lb 1\rb=\\
    =&\,\frac s6\,d\nu_p-\ln s\;\lb 2\nu_p-\theta\rb  d\nu_p+ d\lb \ds\frac{\theta^2}{12}-\nu_p^2\rb+\frac{d\nu_p}{3}+\pi i\lb2\nu_p-\theta\rb d\rho_p+\\
    &\,+\frac{2\nu_p-\theta}{6} d_{\mathcal M}\lb 2\ln \lb\nu_p-\frac{\theta}{2}\rb+3\ln\Gamma \lb\nu_p-\frac{\theta}{2}\rb \Gamma \lb\nu_p+\frac{\theta}{2}\rb\rb+o\lb 1\rb
    \end{aligned}
    \eeq
    Combining this result with the remaining contributions to \eqref{derivtauppmon} and taking into account various cancellations, one finds that
    \beq\label{computedAs1}
    \begin{aligned} d_{\mathcal M}\ln\tau_{p,+}=&\,\frac s6\,d\nu_p+\ln s\; d\lb\frac{\theta^2}{12}-\nu_p^2\rb
     +\\+&\,\lb\nu_p+\frac{\theta}{2}\rb
     d\ln\Gamma \lb\nu_p+\frac{\theta}{2}\rb +\lb\nu_p-\frac{\theta}{2}\rb
     d\ln\Gamma \lb\nu_p-\frac{\theta}{2}\rb+d\lb\nu_p-\nu_p^2-\frac{\theta^2}{4}\rb+o\lb 1\rb.
     \end{aligned}
    \eeq
    It remains to compare this expression with \eqref{expectedAs1}. The  divergent terms in the first line of \eqref{computedAs1} have the expected form, while the $O\lb 1\rb$ term in the second line determines $d\ln N_{p,+}$ in \eqref{expectedAs1}.
    The logarithmic derivative of the Barnes $G$-function can be expressed in terms of the gamma function, cf Eq.~\eqref{diffBarnesG},
    which enables one to check  that $d\ln N_{p,+}=0$, as claimed. The calculation in the case $-\frac12<\Re\nu_p<0$ is completely analogous and gives the same result.
    
    On the ray with $\arg t=\pi+\frac{2\pi p}{3}$, the large $t$ asymptotics of $d\ln\tau_{p,-}$ is given by
    \beq
    d_{\mathcal M}\ln\tau_{p,-}=\frac{s}{12}d\omega_p+\ln s\;d\lb\frac{\theta^2}{3}-\frac{\omega_p^2}{2}\rb+d\ln \frac{G\lb 1+\omega_p\rb
    }{\lb2\pi\rb^{\frac{\omega_p}{2}}} +d\ln N_{p,-}+o\lb 1\rb
    \eeq
    where $N_{p,-}$ is a normalization coefficient that a priori depends on monodromy. This behavior has to be now compared with the expression following from \eqref{taupminus}. We have
    \begin{gather}
    Pd_{\mathcal M}q=-\frac{d\theta}{2}+\frac{s}{2}\lb\frac{R_++R_-}{\lb 1+R_++R_-\rb^2}-\frac14\rb d_{\mathcal M}\frac{R_+-R_-}{1-\lb R_++R_-\rb^2}+o\lb 1\rb,\\	
    qd_{\mathcal M}P=\frac{s}{2}\frac{R_+-R_-}{1-\lb R_++R_-\rb^2}d_{\mathcal M}\frac{R_++R_-}{\lb1+ R_++R_-\rb^2}+o\lb 1\rb,\\
    td_{\mathcal M}H=\frac{s}{8}d\omega_p+\frac32d\lb\frac{\theta^2}{3}-\frac{\omega_p^2}{2}\rb+\frac{s}{8}
    d_{\mathcal M}\frac{R_+-R_-}{1+R_++R_-}+o\lb 1\rb.
    \end{gather}
    This implies that
    \beq
    \begin{gathered}
    \frac{2Pd_{\mathcal  M}q-qd_{\mathcal M}P+2td_{\mathcal M}H}{3}=-\frac{d\theta}{3}
    +\frac{s}{12}d\omega_p+d\lb\frac{\theta^2}{3}-\frac{\omega_p^2}{2}\rb+\frac{s}{2}
    \lb R_-d_{\mathcal M} R_+-R_+d_{\mathcal M} R_-\rb+o\lb 1\rb=\\
    =\frac{s}{12}d\omega_p-\ln s\;d\lb\frac{\omega_p^2}{2}\rb+\omega_pd\ln\Gamma\lb\omega_p\rb+\frac{d\lb\omega_p-\omega_p^2\rb}{2}+2\pi i \omega_p d\xi_p-\frac{d\lb \theta-\theta^2\rb}{3}+o\lb 1\rb,
    \end{gathered}
    \eeq
    so that finally
    \beq
    d_{\mathcal M}\ln\tau_{p,-}=\frac{s}{12}d\omega_p+\ln s\;d\lb\frac{\theta^2}{3}-\frac{\omega_p^2}{2}\rb+d\ln \frac{G\lb 1+\omega_p\rb
    }{\lb2\pi\rb^{\frac{\omega_p}{2}}}+o\lb 1\rb,
    \eeq
    which completes the proof.
    \epf
    \begin{cor}\label{varForm1}
    	Let $\tau\lb t\rb$ be an arbitrarily normalized \PIIeq tau function associated to the monodromy data ${\theta\in\mathbb C}$, $\mathbf s\in\mathcal S_{\theta}$ satisfying the conditions $|\Re\nu_p|,|\Re\omega_p|<\frac12$, $\Re\nu_p\ne0$ for all $p\in\mathbb Z/3\mathbb Z$. The corresponding connection coefficients satisfy
    	\beq
    	d\ln\frac{\mathcal N_{q,-}}{\mathcal N_{p,+}}=2\pi i\lb \omega_q d\xi_q-\nu_p d\rho_p\rb-\theta d\lb K_{q,-}- K_{p,+}\rb.
    	\eeq
    \end{cor}
    \pf  For every $p,q\in\mathbb Z/3\mathbb Z$, we can express $\tau\lb t\rb$ in terms of the normalized tau functions as
    \beq
    \tau\lb t\rb=\frac{\mathcal N_{p,+}}{N_{p,+}}\tau_{p,+}\lb t\rb=\frac{\mathcal N_{q,-}}{N_{q,-}}\tau_{q,-}\lb t\rb.
    \eeq
    Proposition~\ref{propNorm} then implies that
    \beq
    d\ln\frac{\mathcal N_{q,-}}{\mathcal N_{p,+}}=d\ln\frac{\tau_{p,+}}{\tau_{q,-}},
    \eeq
    and the statement follows immediately from the definition \eqref{TausExt} of $\tau_{p,\pm}$. 
    \epf

     \subsubsection{Finite parts of the asymptotics of $\lambda$}
    
    We have shown that,  to compute the connection coefficients, we need to express in terms of monodromy the differences $K_{p,\epsilon}-K_{q,\epsilon'}$. These differences involve constant terms in the large $t$ asymptotics \eqref{lnuas1}--\eqref{lnuas2} of the gauge factor  $\lambda$ appearing in and \eqref{gaugepar} and \eqref{uform1}. Let us stress once again that $\lambda$ is defined only up to a multiplicative factor independent of $t$ whose modification shifts all $K_{p,\epsilon}$ by the same amount. The relative normalization of the asymptotics of $\lambda$ on different rays can be derived from a sufficiently accurate approximation to the solution of the FN linear system \eqref{lsysKap} or the associated Riemann-Hilbert problem. However, as we will now see, the structure of \eqref{TausExt} determines the differences $K_{p,\epsilon}-K_{q,\epsilon'}$ almost completely, allowing to bypass the construction of such an approximate solution.

    Let us introduce the following self-explanatory notation (cf Figure~\ref{figcanrays}) for the differences associated with adjacent rays:
    \beq
    \Delta_{p,\rcurvearrownw}=K_{p+2,-}-K_{p,+},\qquad \Delta_{p,\lcurvearrowsw}=K_{p+1,-}-K_{p,+},\qquad p\in\mathbb Z/3\mathbb Z.
    \eeq
    Any other difference can obviously be expressed in terms of these six functions on $\mathcal M$. Our goal is to express $\Delta_{p,\rcurvearrownw}$ and $\Delta_{p,\lcurvearrowsw}$ using the parameter triples $\lb \nu_p,\omega_{p+2},\theta\rb$ and  $\lb\nu_p,\omega_{p+1},\theta\rb$, respectively.

    \begin{lemma}\label{lemmaDeltas}
    	The functions $ \Delta_{p,\rcurvearrownw}\lb \nu_p,\omega_{p+2},\theta\rb$, $\Delta_{p,\lcurvearrowsw}\lb\nu_p,\omega_{p+1},\theta\rb$ have the form
    	\beq\label{Deltas}
    	\Delta_{p,\rcurvearrownw}\lb \nu_p,\omega_{p+2},\theta\rb=\frac12\ln\frac{e^{2\pi i\nu_p}-e^{-\pi i \theta}}{e^{2\pi i\nu_p}-e^{\pi i \theta}}+\delta_{p,\rcurvearrownw}\lb\theta\rb,\qquad
    	\Delta_{p,\lcurvearrowsw}\lb\nu_p,\omega_{p+1},\theta\rb=\delta_{p,\lcurvearrowsw}\lb\theta\rb,
    	\eeq
    	where the functions $\delta_{p,\rcurvearrownw}$, $\delta_{p,\lcurvearrowsw}$ depend only on the exponent $\theta$ of formal monodromy.
    \end{lemma}
    \pf It follows from Theorem~\ref{thmOmega} that 
    \begin{subequations}
    \begin{align}
    \boldsymbol\omega_{p,\rcurvearrownw}=&\,\lb K_{p+2,-}-K_{p,+}\rb d\theta-2\pi i\lb \xi_{p+2}d\omega_{p+2}-\rho_p d\nu_p\rb,\\
    \boldsymbol\omega_{p,\lcurvearrowsw}=&\lb K_{p+1,-}-K_{p,+}\rb d\theta-2\pi i\lb \xi_{p+1}d\omega_{p+1}-\rho_p d\nu_p\rb,
    \end{align}
    \end{subequations} 
    are closed 1-forms on $\mathcal M$. This implies
    \begin{subequations}
    \begin{align}\label{derivs1}
    	\frac{\partial \Delta_{p,\rcurvearrownw}}{\partial \nu_p}=2\pi i \frac{\partial\rho_p}{\partial\theta},\qquad 
    	\frac{\partial \Delta_{p,\rcurvearrownw}}{\partial \omega_{p+2}}=-2\pi i \frac{\partial\xi_{p+2}}{\partial\theta},\qquad
    	\frac{\partial\xi_{p+2}}{\partial\nu_p}=-\frac{\partial\rho_p}{\partial\omega_{p+2}},\\
    	\label{derivs2}
    		\frac{\partial \Delta_{p,\lcurvearrowsw}}{\partial \nu_p}=2\pi i \frac{\partial\rho_p}{\partial\theta},\qquad 
    	\frac{\partial \Delta_{p,\lcurvearrowsw}}{\partial \omega_{p+1}}=-2\pi i \frac{\partial\xi_{p+1}}{\partial\theta},\qquad
    	\frac{\partial\xi_{p+1}}{\partial\nu_p}=-\frac{\partial\rho_p}{\partial\omega_{p+1}},
    \end{align}
    \end{subequations}
    where $\rho_p,\xi_{p+2}$ in \eqref{derivs1} should be expressed in terms of $\lb \nu_p,\omega_{p+2},\theta\rb$, while $\rho_p,\xi_{p+1}$ in \eqref{derivs2} should be expressed in terms of $\lb\nu_p,\omega_{p+1},\theta\rb$. From \eqref{monoparameters} and \eqref{nuomegaidentities} we have
    \begin{subequations}\label{coordrels}
    \begin{gather}
    \rho_{p}\lb\nu_p,\omega_{p+2},\theta\rb=-\omega_{p+2}+\frac{\ln \lb e^{2\pi i\nu_p}-e^{\pi i \theta}\rb
    	\lb e^{2\pi i\nu_p}-e^{-\pi i \theta}\rb}{2\pi i},\qquad \xi_{p+2}\lb\nu_p,\omega_{p+2},\theta\rb=\nu_p,\\
    	\rho_{p}\lb\nu_p,\omega_{p+1},\theta\rb=\omega_{p+1},\qquad 
    	\xi_{p+1}\lb\nu_p,\omega_{p+1},\theta\rb=-\nu_{p}+\frac{\ln\lb 1-e^{2\pi i\omega_{p+1}}\rb}{2\pi i}.
    \end{gather} 
    \end{subequations}
    The last relations in both \eqref{derivs1} and \eqref{derivs2} are thus reassuringly satisfied, and we further obtain
    \begin{subequations}
    	\begin{gather}
    		\frac{\partial \Delta_{p,\rcurvearrownw}}{\partial \nu_p}=\pi i\lb\frac{1}{e^{\pi i\lb 2\nu_p+\theta\rb}-1}-\frac{1}{e^{\pi i\lb 2\nu_p-\theta\rb}-1}\rb,\qquad 
    		\frac{\partial \Delta_{p,\rcurvearrownw}}{\partial \omega_{p+2}}=0,\\
    		\frac{\partial \Delta_{p,\lcurvearrowsw}}{\partial \nu_p}=\frac{\partial \Delta_{p,\lcurvearrowsw}}{\partial \omega_{p+1}}=0.
    	\end{gather}
    \end{subequations}
    The result \eqref{Deltas} follows immediately.
    \epf

    \begin{prop}\label{propCC01}
    	 Under assumptions of Corollary~\ref{varForm1}, the connection coefficients can be expressed as
    	\begin{subequations}
    	\begin{align}
    		\label{conncoefratiosA}
    	\frac{\mathcal N_{p+2,-}}{\mathcal N_{p,+}}=&\,\Xi_{p,\rcurvearrownw}\lb\theta\rb
    	e^{2\pi i \nu_p\omega_{p+2}-\pi i \nu_p^2}\lb 2\pi\rb^{-2\nu_p}\frac{G\lb 1+\nu_p+\frac{\theta}{2}\rb G\lb 1+\nu_p-\frac{\theta}{2}\rb}{
    	G\lb 1-\nu_p+\frac{\theta}{2}\rb G\lb 1-\nu_p-\frac{\theta}{2}\rb},\\
    	\label{conncoefratiosB}
    	\frac{\mathcal N_{p+1,-}}{\mathcal N_{p,+}}=&\,\Xi_{p,\lcurvearrowsw}\lb\theta\rb 
    	e^{-2\pi i\nu_p\omega_{p+1}+\frac{\pi i\omega_{p+1}^2}{2}}\lb2\pi\rb^{\omega_{p+1}}\frac{G\lb 1-\omega_{p+1}\rb}{G\lb 1+\omega_{p+1}\rb},
    	\end{align}
    	\end{subequations}
    	where the prefactors $\Xi_{p,\rcurvearrownw}\lb\theta\rb$, $\Xi_{p,\lcurvearrowsw}\lb\theta\rb $ depend only on the formal monodromy exponent $\theta$.
     \end{prop} 
     \pf From Proposition~\ref{varForm1} and Lemma~\ref{lemmaDeltas} we now obtain
     \begin{subequations}
     \begin{align}
     	d\ln\frac{\mathcal N_{p+2,-}}{\mathcal N_{p,+}}=&\,2\pi i\lb \omega_{p+2} d\xi_{p+2}-\nu_p d\rho_p\rb-\theta d\lb\frac12\ln\frac{e^{2\pi i\nu_p}-e^{-\pi i \theta}}{e^{2\pi i\nu_p}-e^{\pi i \theta}}+\delta_{p,\rcurvearrownw}\lb\theta\rb\rb,\\
     	d\ln\frac{\mathcal N_{p+1,-}}{\mathcal N_{p,+}}=&\,
     	2\pi i\lb \omega_{p+1} d\xi_{p+1}-\nu_p d\rho_p\rb-\theta d\delta_{p,\lcurvearrowsw}\lb\theta\rb.
     \end{align}
     \end{subequations}
     Using once again equations \eqref{coordrels}, we find that
     \begin{subequations}
     	\begin{align}
     		&\begin{aligned}d\ln\frac{\mathcal N_{p+2,-}}{\mathcal N_{p,+}}=&\,2\pi i d\lb\nu_p\omega_{p+2}\rb
     		-\lb\nu_p+\frac{\theta}{2}\rb d\ln\lb 1-e^{2\pi i\lb \nu_p+\frac{\theta}{2}\rb}\rb-\lb\nu_p-\frac{\theta}{2}\rb d\ln\lb 1-e^{2\pi i\lb \nu_p-\frac{\theta}{2}\rb}\rb\\
     		&\,-\theta d\lb \delta_{p,\rcurvearrownw}\lb\theta\rb-i\pi\theta\rb,
     		\end{aligned}\\
     	&	d\ln\frac{\mathcal N_{p+1,-}}{\mathcal N_{p,+}}=-2\pi i d\lb\nu_p\omega_{p+1}\rb+\omega_{p+1}d\ln\lb 1-e^{2\pi i\omega_{p+1}}\rb-\theta d\delta_{p,\lcurvearrowsw}\lb\theta\rb.
     	\end{align}
     	\end{subequations}
        Both expressions can be integrated using the differentiation formula
        \eqref{diffGhat} satisfied by the Barnes $G$-functions.
        The unknown functions $\delta_{p,\rcurvearrownw}\lb\theta\rb$,  $\delta_{p,\lcurvearrowsw}\lb\theta\rb$ are absorbed into the prefactors $\Xi_{p,\rcurvearrownw}\lb\theta\rb$, $\Xi_{p,\lcurvearrowsw}\lb\theta\rb$.
     \epf	
    
    \subsubsection{Symmetric monodromy}
    To determine the form of the unknown functions $\Xi_{p,\rcurvearrownw}\lb\theta\rb$, $\Xi_{p,\lcurvearrowsw}\lb\theta\rb$, let us consider a special case of symmetric monodromy, where $s_1=s_2=s_3=\kappa$. In general, the cubic surface equation \eqref{monodromy_S} admits three possible solutions for $\kappa$:
    \beq \label{sym_mon}
    \kappa\in\left\{-2\cos\tfrac{\pi\theta}{3},-2\cos\tfrac{\pi\lb\theta+2\rb}{3},-2\cos\tfrac{\pi\lb\theta-2\rb}{3}\right\}.
    \eeq 
    These three options are, in a sense, equivalent: e.g. the solution of \PIIeq with parameter $\theta$ and $s_1=s_2=s_3=-2\cos\tfrac{\pi\lb\theta\pm2\rb}{3}$ can be obtained from the solution of \PIIeq with shifted parameter $\theta\pm 2$ and the same $s_{1,2,3}$ by B\"acklund transformations. 
    
    \begin{lemma}
    	For symmetric monodromy, we have
    	\beq
    	\tau\lb e^{\frac{2\pi i}{3}} t\rb=\begin{cases}
    		\tau\lb t\rb, \qquad & \text{for } \kappa=-2\cos\tfrac{\pi\lb\theta\pm2\rb}{3},\vspace{0.1cm}\\
    		e^{\frac{2\pi i }{3}}\tau\lb t\rb, \qquad & \text{for } \kappa=-2\cos\tfrac{\pi\theta}{3}.
    	\end{cases}
    	\eeq
     and, in particular,
    	\beq
    	\frac{\mathcal N_{p+1,+}}{\mathcal N_{p,+}}=\frac{\mathcal N_{p+1,-}}{\mathcal N_{p,-}}=\begin{cases}
    		1,\qquad & \text{for } \kappa=-2\cos\tfrac{\pi\lb\theta\pm2\rb}{3},\vspace{0.1cm}\\
    		e^{\frac{2\pi i }{3}}, \qquad & \text{for } \kappa=-2\cos\tfrac{\pi\theta}{3}.
    	\end{cases}
    	\eeq
    	for every  $p\in\mathbb Z/3\mathbb Z$.
    \end{lemma}
    \pf
    In the symmetric case, the $\mathbb Z_3$-symmetry implies that  $\tau\lb e^{\frac{2\pi i}{3}} t\rb$ is not merely a \PIIeq tau function with transformed Stokes data, but coincides with the original tau function up to a $t$-independent prefactor. Moreover, since $\tau$ is holomorphic, this prefactor (which could otherwise depend on $\theta$) must necessarily be a cubic root of unity:
    \beq
    \tau\lb e^{\frac{2\pi i}{3}} t\rb=\zeta \tau\lb t\rb,\qquad \zeta^3=1.
    \eeq
    Which of the three roots occurs may nevertheless depend on the choice of the solution of \eqref{sym_mon}.
    
    Let us first set $\kappa=-2\cos\tfrac{\pi\lb\theta-2\rb}{3}$. In this case, the correct root $\zeta=1$ can be easily  identified from the special monodromy $\theta=\frac12$, $\kappa=0$ corresponding to 
    \beq
    q\lb t\rb=0,\qquad \tau\lb t\rb=e^{-\frac{t^3}{24}}.
    \eeq
    This is the seed solution generating the sequence of the well-known (Yablonskii-Vorobiev) \PIIeq solutions, discussed in more detail in Appendix~\ref{App_YV}. The values of $\zeta$ corresponding to $\kappa=-2\cos\tfrac{\pi\theta}{3}$ and $\kappa=-2\cos\tfrac{\pi\lb\theta+2\rb}{3}$ can be identified similarly using the next two entries of this sequence.
    \epf
    
    \begin{prop}\label{propXipm}
    The functions $\Xi_{p,\rcurvearrownw}\lb\theta\rb$, $\Xi_{p,\lcurvearrowsw}\lb\theta\rb$ are independent of $p\in\mathbb Z/3\mathbb Z$, 
    \begin{align}\label{Xiprop}
    	\Xi_{p,\rcurvearrownw}\lb\theta\rb=\Xi_{\rcurvearrownw}\lb\theta\rb,\qquad 
    	\Xi_{p,\lcurvearrowsw}\lb\theta\rb=\Xi_{\lcurvearrowsw}\lb\theta\rb,
    \end{align}
    and, moreover,
    \beq
    \frac{\Xi_{\rcurvearrownw}\lb\theta\rb}{\Xi_{\lcurvearrowsw}\lb\theta\rb}=e^{\frac{\pi i\lb \theta^2+1\rb}{12}}.
    \eeq
    \end{prop}
    \pf Considering the ratios of \eqref{conncoefratiosA}, \eqref{conncoefratiosB} for symmetric monodromy with $\kappa=-2\cos\frac{\pi\lb \theta-2\rb}{3}$, we obtain
    \beq\label{normsymm}
    \frac{\mathcal N_{p+2,-}}{\mathcal N_{p+1,-}}=	\frac{\Xi_{p,\rcurvearrownw}\lb\theta\rb}{\Xi_{p,\lcurvearrowsw}\lb\theta\rb} F\lb \theta\rb=\frac{\mathcal N_{p,+}}{\mathcal N_{p-1,+}}=\frac{\Xi_{p-1,\rcurvearrownw}\lb\theta\rb}{\Xi_{p,\lcurvearrowsw}\lb\theta\rb} F\lb \theta\rb=1,
    \eeq
    where 
    \beq\label{1stEq_forF}
    F\lb \theta\rb=e^{4\pi i\nu\omega-\pi i\nu^2-\frac{\pi i\omega^2}{2}}\lb 2\pi\rb^{-2\nu-\omega}\hat{G}\lb \nu+\tfrac{\theta}{2}\rb \hat{G}\lb \nu-\tfrac{\theta}{2}\rb
    \hat{G}\lb\omega\rb,
    \eeq
    the function $\hat G\lb z\rb$ is defined by \eqref{defGhat}, and the functions $\nu=\nu\lb \theta\rb$, $\omega=\omega\lb \theta\rb$ are determined by
    \beq\label{nuomegatheta}
    e^{2\pi i\nu}=-2\cos\frac{\pi\lb\theta-2\rb}{3},\qquad e^{2\pi i\omega}=1-4\cos^2\frac{\pi\lb \theta-2\rb}{3}.
    \eeq
    The choice of solutions of \eqref{nuomegatheta} for $\nu,\omega$ is not important, since the right side of \eqref{1stEq_forF} remains invariant under integer shifts $\nu\mapsto\nu\pm1$, $\omega\mapsto\omega\pm1$, as can be proved using the recurrence relation \eqref{recGhat}.

    Equations \eqref{normsymm} already imply the first part of the statement. Our remaining task is to simplify the expression of $F\lb\theta\rb$. To this end, let us first note that $e^{2\pi i \omega}+e^{4\pi i \nu}=1$. The reflection formula \eqref{refl0} then allows to rewrite \eqref{1stEq_forF}  as
    \beq\label{aux0000}
    F\lb\theta\rb=e^{\pi i\nu^2-\frac{\pi i}{12}}\frac{\hat{G}\lb \nu+\tfrac{\theta}{2}\rb \hat{G}\lb \nu-\tfrac{\theta}{2}\rb}{\hat G\lb 2\nu\rb}.
    \eeq
    The differentiation formula \eqref{diffGhat} further implies that
    \beq
    \begin{gathered}
   \frac{d}{d\theta}\left[\ln \frac{\hat{G}\lb \nu+\tfrac{\theta}{2}\rb \hat{G}\lb \nu-\tfrac{\theta}{2}\rb}{\hat G\lb 2\nu\rb}+i\pi\lb\nu^2-\frac{\theta^2}{4}\rb\right]=\\
  = 
    -\lb\nu+\frac{\theta}{2}\rb \frac{d}{d\theta}\ln\lb 1-e^{2\pi i\lb \nu+\frac{\theta}{2}\rb}\rb -\lb\nu-\frac{\theta}{2}\rb \frac{d}{d\theta}\ln\lb 1-e^{2\pi i\lb \nu-\frac{\theta}{2}\rb}\rb+2\nu \frac{d}{d\theta}\ln\lb 1-e^{4\pi i\nu}\rb=-\frac{2\pi i\theta}{3}.
    \end{gathered}
    \eeq
    The simplifications in the above follow from the identities $ 1-e^{2\pi i\lb \nu\pm\frac{\theta}{2}\rb}=-e^{2\pi i \lb\omega\pm\frac{ \theta-2}{3}\rb}$ and $1-e^{4\pi i \nu}=e^{2\pi i \omega}$.
    One thereby obtains a surprisingly simple result 
    \beq
   \partial_\theta\ln F\lb \theta\rb=-\frac{i\pi\theta}{6},
    \eeq
    that is, $F\lb \theta\rb=\operatorname{const}\cdot e^{-\frac{\pi i\theta^2}{12}}$. The constant prefactor can be fixed using that for $\theta=0$ one has $\nu=0$, so that from \eqref{aux0000} we find that $F\lb 0\rb=e^{-\frac{\pi i}{12}}$.
    \epf
    
    Observe that at this point the connection problem for the \PIIeq tau function between rays of the same type is already solved! Indeed, Propositions~\ref{propCC01} and \ref{propXipm} explicitly express the ratios $\ds\frac{\mathcal N_{p,+}}{\mathcal N_{q,+}}$ and  $\ds\frac{\mathcal N_{p,-}}{\mathcal N_{q,-}}$ in terms of monodromy for any $p,q\in\mathbb Z/3\mathbb Z$. 
    \begin{cor}\label{corsametyperays}  Under assumptions of Corollary~\ref{varForm1}, we have
    		\begin{subequations}\label{conncoefppmm}
    \begin{align}\label{conncoefpp}
    	\frac{\mathcal N_{p+1,+}}{\mathcal N_{p,+}}=\frac{e^{2\pi i \omega_{p+2}\lb\nu_{p}+\nu_{p+1}\rb-\pi i \nu_p^2-\frac{\pi i\omega_{p+2}^2}{2}+\frac{\pi i\lb \theta^2+1\rb}{12}}}{\lb 2\pi\rb^{2\nu_p+\omega_{p+2}}}\hat G\lb \nu_p+\frac{\theta}{2}\rb \hat G\lb \nu_p-\frac{\theta}{2}\rb \hat G\lb \omega_{p+2}\rb,\\
    	\label{conncoefmm}
    \frac{\mathcal N_{p+2,-}}{\mathcal N_{p+1,-}}=\frac{e^{2\pi i \nu_p\lb\omega_{p+1}+\omega_{p+2}\rb-\pi i \nu_p^2-\frac{\pi i\omega_{p+1}^2}{2}+\frac{\pi i\lb \theta^2+1\rb}{12}}}{\lb 2\pi\rb^{2\nu_p+\omega_{p+1}}}\hat G\lb \nu_p+\frac{\theta}{2}\rb \hat G\lb \nu_p-\frac{\theta}{2}\rb \hat G\lb \omega_{p+1}\rb,  
    \end{align}
    	\end{subequations}
    	where $\omega_{p+2}$ in \eqref{conncoefpp} can be expressed in terms of asymptotic parameters $\nu_p,\nu_{p+1}$ on the corresponding rays and $\nu_p$ in \eqref{conncoefmm} can be expressed in terms of $\omega_{p+1},\omega_{p+2}$ using \eqref{nuomegaidentities}.
    \end{cor}
    
    As for the connection coefficients between rays of different type, the only missing ingredient is the function~$\Xi_{\lcurvearrowsw}\lb\theta\rb$  or, equivalently, $\Xi_{\rcurvearrownw}\lb\theta\rb$.
    
    \subsubsection{Recurrence relations for $\Xi_{\lcurvearrowsw}\lb\theta\rb$}
    It is clear already from \eqref{expsI} and \eqref{expsII} that $\Xi_{\lcurvearrowsw}\lb \theta\rb$ must be an even function of $\theta$. We now derive further constraints on its form. The first of them  comes from a ``bootstrap'' property of \eqref{omegaexpr1}, namely from the fact that the extended JMU differential involves the gauge factor $\lambda$, which is itself given by the ratio \eqref{uform2} of two tau functions with parameters $\theta$ and $\theta-1$.
    
    \begin{lemma}
    	We have
    	\beq\label{RecRelXi0}
    	\partial_\theta\ln\Xi_{\lcurvearrowsw}\lb \theta\rb=\theta \phi\lb\theta\rb, 
    	\eeq
    	where $\phi\lb\theta\pm1\rb=\phi\lb\theta\rb$.
    \end{lemma}
    \pf Recall that $\Xi_{\lcurvearrowsw}\lb \theta\rb$ is related to finite parts of the asymptotics of $\lambda$ by
     \beq\label{Xiaux02}
    d\ln  \Xi_{\lcurvearrowsw}\lb \theta\rb=-\theta  d\lb K_{p+1,-}-K_{p,+}\rb,
    \eeq
    cf Lemma~\ref{LemAsU}, Lemma~\ref{lemmaDeltas} and Proposition~\ref{propCC01}. Proposition~\ref{propXipm} shows that the right side of \eqref{Xiaux02} does not depend on $p\in\mathbb Z/3\mathbb Z$. On the other hand, the quantities $K_{p,\pm}$ are expressed in terms of the ratios of the normalization coefficients for the tau functions $\tau$ and $\tilde\tau$:
     \beq
    K_{p,+}=i\pi \rho_p+\ln\;\frac{\tilde{\mathcal N}_{p,+}}{\mathcal N_{p,+}},\qquad K_{p+1,-}=\ln\;\frac{\tilde{\mathcal N}_{p+1,-}}{\mathcal N_{p+1,-}},
    \eeq
    see Remark~\ref{RmqDiscont}. Recall once again that the transformed tau function is characterized by the asymptotic parameters $\tilde\nu_p=\nu_p-\frac12$ and $\tilde\omega_p=\omega_p$. Thus, using \eqref{coordrels} and \eqref{conncoefratiosB}, we find that
    \beq\label{Eq387}
     K_{p+1,-}-K_{p,+}=\ln\frac{\Xi_{\lcurvearrowsw}\lb\theta-1\rb 
     }{\Xi_{\lcurvearrowsw}\lb\theta\rb 
     }.
    \eeq
    Substituting this expression into \eqref{Xiaux02}, one can rewrite the resulting equation in the form
    \beq
    \frac{\partial_\theta\ln \Xi_{\lcurvearrowsw}\lb\theta\rb}{\theta}=\frac{\partial_\theta\ln \Xi_{\lcurvearrowsw}\lb\theta-1\rb}{\theta-1},
    \eeq
    which implies the statement of the lemma.
    \epf

      Another functional relation for $\Xi_{\lcurvearrowsw}\lb\theta\rb$ comes from the well-known Toda equation for \PIIeq tau functions.
    \begin{lemma}
    	The function $\Xi_{\lcurvearrowsw}\lb\theta\rb$ satisfies
    	\beq\label{RecRelXi}
    	\frac{\Xi_{\lcurvearrowsw}\lb\theta+1\rb\Xi_{\lcurvearrowsw}\lb\theta-1\rb}{\Xi_{\lcurvearrowsw}\lb\theta\rb^2}=2^{-\frac{10}{3}}e^{-\frac{i\pi}{3}}.
    	\eeq
    \end{lemma}
    \pf Let $\sigma$ be a solution of the $\sigma$-form \eqref{sigmaPII} of \PIIeq equation  with parameter $\theta$ and monodromy $\mathbf s=\lb s_1,s_2,s_3\rb\in\mathcal S_\theta$. Both $\theta$ and $\mathbf s$ are assumed here to be generic. Starting from $\sigma$, one can construct by B\"acklund transformations an infinite sequence of solutions of \PIIeq in the standard form, of which we need only three:
    \begin{itemize}
    	\item A solution $q$ of \eqref{PIIstandard} with parameter $\theta$, given by  $	q=-\ds\frac{2\sigma''+\theta}{4\sigma'}$, cf equation \eqref{qintermsoftau}. The corresponding $\sigma$-function is simply $\sigma$.
    	\item A solution $\tilde q=-q=\ds\frac{2\sigma''+\theta}{4\sigma'}$ of \eqref{PIIstandard} with parameter $1-\theta$. The corresponding $\sigma$-function is $\tilde \sigma=\sigma -q$, see~\eqref{sigmatilde}. It satisfies  \eqref{sigmaPII}  with parameter $\theta-1$ and  monodromy $-\mathbf s$.
    	\item A solution $\underaccent{\sim}{q}=\ds\frac{2\sigma''-\theta}{4\sigma'}$ of \eqref{PIIstandard} with parameter $1+\theta$. The corresponding $\sigma$-function is $\underaccent{\sim}{\sigma}=\sigma+\underaccent{\sim}{q}$, it solves \eqref{sigmaPII} with parameter $\theta+1$ and monodromy $-\mathbf s$.
    \end{itemize} 
    Therefore, one has
    \beq
    \underaccent{\sim}{\sigma}+\tilde\sigma-2\sigma=\underaccent{\sim}{q}-q=\lb \ln \sigma'\rb'.
    \eeq
    Rewriting the last relation in terms of the associated  tau functions (in addition to previously defined $\tau,\tilde\tau$, we similarly introduce $\underaccent{\sim}{\tau}$ by $\underaccent{\sim}{\sigma}=\lb\ln\underaccent{\sim}{\tau}\rb'$) and integrating it once, we obtain the Toda equation
    \beq\label{TodaEq}
    \lb\ln\tau\rb''=\operatorname{const}\,\cdot\, \frac{\tilde \tau \,\underaccent{\sim}{\tau}}{\tau^2}.
    \eeq
    The constant in the last equation depends on the normalization chosen for the tau functions $\tau$, $\tilde\tau$, $\underaccent{\sim}{\tau}$
    and is not relevant. What is important, however, is that the combination  $\ds \lb\ln\tau\rb''\frac{\tau^2}{\tilde \tau \,\underaccent{\sim}{\tau}}$ is independent on $t$. Thus, evaluating  this expression using the asymptotics on different canonical rays, we may derive a set of functional equations for the connection coefficients.
    
    Let us first consider the asymptotics as $t\to+\infty$. For example, for $0 <\Re \nu_0<\frac12$ one has 
    \beq\label{todaaux1}
    \begin{aligned}
   & \lb \ln\tau\rb''= -2^{-\frac23}e^{-\frac{i\pi}{3}}\frac{2\pi e^{-2\pi i \rho_0}s^{2\nu_0-\frac13}e^{-\frac s6}}{\Gamma\lb\nu_0+\frac{\theta}{2}\rb \Gamma\lb\nu_0-\frac{\theta}{2}\rb}\left[1+o\lb 1\rb\right],\\
   & \frac{\tilde\tau\underaccent{\sim}{\tau}}{\tau^2}= \frac{\mathcal N_{0,+}\lb\theta-1,-\mathbf s\rb \mathcal N_{0,+}\lb\theta+1,-\mathbf s\rb}{
   \mathcal N_{0,+}\lb\theta,\mathbf s\rb^2} \frac{2\pi s^{2\nu_0-\frac13}e^{-\frac s6}}{\Gamma\lb\nu_0+\frac{\theta}{2}\rb \Gamma\lb\nu_0-\frac{\theta}{2}\rb}\left[1+o\lb 1\rb\right],
    \end{aligned}
    \eeq
    with $s=4\sqrt 2\,e^{\frac{i\pi}{2}}t^{\frac32}$. The first estimate is obtained from \eqref{logaspplus} after recalling that $P=2\lb\ln\tau\rb''$. The second estimate follows from the fact that the asymptotic parameters of the B\"acklund transformed tau functions $\tilde \tau$, $\underaccent{\sim}{\tau}$ are given by $\tilde\nu_0=\underaccent{\sim}\nu_0=\nu_0-\frac12$.
    
    Likewise, the  $t\to\infty$ asymptotics along the ray with $\arg t=-\frac{\pi}{3}$ is given by
    \beq\label{todaaux2}
    \begin{aligned}
    & \lb \ln\tau\rb''= -\frac{t}{4}+o\lb |t|\rb,\\
    & \frac{\tilde\tau\underaccent{\sim}{\tau}}{\tau^2}= \frac{\mathcal N_{1,-}\lb\theta-1,-\mathbf s\rb \mathcal N_{1,-}\lb\theta+1,-\mathbf s\rb}{
    	\mathcal N_{1,-}\lb\theta,\mathbf s\rb^2}s^{\frac23}\lb 1+o\lb 1\rb\rb, 
    \end{aligned}
    \eeq
    with $s=8\,e^{\frac{i\pi}{2}}|t|^{\frac32}$. Here, the first estimate is determined by the leading term in \eqref{PasComp}, and the second follows from $\tilde\omega_1=\underaccent{\sim}{\omega}_1=\omega_1$.
    
    Combining \eqref{todaaux1}, \eqref{todaaux2} with \eqref{TodaEq}, we find that
    \beq
    \frac{\mathcal N_{1,-}\lb\theta-1,-\mathbf s\rb}{\mathcal N_{0,+}\lb\theta-1,-\mathbf s\rb}
    \frac{\mathcal N_{1,-}\lb\theta+1,-\mathbf s\rb}{\mathcal N_{0,+}\lb\theta+1,-\mathbf s\rb}
    \lb \frac{\mathcal N_{1,-}\lb\theta,\mathbf s\rb}{\mathcal N_{0,+}\lb\theta,\mathbf s\rb}\rb^{-2}=2^{-\frac{10}{3}}e^{-\frac{i\pi}{3}}e^{2\pi i \rho_0}.
    \eeq
    Using \eqref{conncoefratiosB}, the last result transforms into the recurrence relation \eqref{RecRelXi}.
    \epf
    
    \begin{lemma}~\label{LemXiper}
    	The function $\Xi_{\lcurvearrowsw}\lb\theta\rb$ has the form
    	\beq\label{Xiaux04}
    	\Xi_{\lcurvearrowsw}\lb\theta\rb=2^{-\frac{5\theta^2}{3}}e^{-\frac{i\pi\theta^2}{6}} e^{\phi_2\lb\theta\rb-\theta\partial_\theta\phi_2\lb\theta\rb},
    	\eeq
    	where $\phi_2\lb\theta\rb=\phi_2\lb -\theta\rb=\phi_2\lb\theta\pm1\rb$. 
    \end{lemma}
    \pf The general solution of the recurrence relation \eqref{RecRelXi} is
    \beq
    \Xi_{\lcurvearrowsw}\lb\theta\rb=2^{-\frac{5\theta^2}{3}}e^{-\frac{i\pi\theta^2}{6}} e^{\theta\phi_1\lb\theta\rb+\phi_2\lb\theta\rb},
    \eeq
    where $\phi_{1,2}\lb\theta\rb$ are arbitrary 1-periodic functions of $\theta$. This implies that
    \beq\label{Xiaux03}
    \partial_\theta\ln \Xi_{\lcurvearrowsw}\lb\theta\rb=\theta\lb-\frac{10}{3}\ln 2-\frac{i\pi}{3}+\partial_\theta\phi_1\lb\theta\rb\rb+\phi_1\lb\theta\rb+\partial_\theta\phi_2\lb\theta\rb.
    \eeq
    The relation \eqref{RecRelXi0} implies that $\partial_\theta\ln \ds\frac{\Xi_{\lcurvearrowsw}\lb\theta+1\rb}{ \Xi_{\lcurvearrowsw}\lb\theta\rb}=\phi\lb\theta\rb$. Computing the same quantity with \eqref{Xiaux03} and comparing the results, we see that
    \beq
    -\frac{10}{3}\ln 2-\frac{i\pi}{3}+\partial_\theta\phi_1\lb\theta\rb=\phi\lb\theta\rb,\qquad 
    \phi_1\lb\theta\rb+\partial_\theta\phi_2\lb\theta\rb=0,
    \eeq
    which proves \eqref{Xiaux04}.
    \epf
    
    Combining Propositions~\ref{propCC01} and \ref{propXipm} with Lemma~\ref{LemXiper}, we see that, to finish the proof of Theorem~\ref{THMA}, it remains only to show that the periodic function $\phi_2\lb\theta\rb$ in \eqref{Xiaux04} is in fact constant, and to determine its value.
    
    \subsubsection{$\Xi_{\lcurvearrowsw}\lb\theta\rb$ from generalized Hastings-McLeod solution\label{subsec_HM}}
    We define the generalized Hastings-McLeod (HM) solution of inhomogeneous \PIIeq as the one described by the following monodromy data:
    \beq\label{gHMmon}
    s_1=e^{-\pi i\theta},\qquad s_2=0,\qquad s_3=e^{\pi i\theta}.
    \eeq
    The properties of solutions of this type have been studied in \cite{IK01,CKV05}, as well as more recently, in a context more closely related to the present work, in \cite{Mil18,Kok18,DXZ19}.
    The monodromy \eqref{gHMmon} is singular from the point of view of generic asymptotics both as $t\to +\infty$ (note that $s_2=0$ corresponds to $\Im\nu_0\to+\infty$) and as $t\to-\infty$ (likewise, $s_1s_3\to1$ corresponds to $\Im\omega_0\to+\infty$). However, for the other four canonical rays it remains sufficiently generic, by which we mean that the corresponding asymptotics can be obtained from the generic formulas in appropriate limits. Moreover, it is known (see e.g. \cite{FIKN06}) that the asymptotics of such ``tronquée'' solution as $t\to\pm\infty$ extends from the rays $\arg t=0,\pi$ into the sectors of opening angle $\frac{2\pi}{3}$ centered along these rays. 
    
    We can therefore adopt a strategy inspired by (and related to) the one used in the homogeneous case in \cite[Subsection 4.5]{ILP16}:
    \begin{itemize}
    	\item The solution of the connection problem for the generalized HM tau function between the rays $\arg t=0$ and $\arg t=\pi$ can be extracted from \cite[Theorem 1.3]{DXZ19}.
    	\item Using the continuation of the asymptotics into wider sectors $\arg t\in[\frac{2 \pi}{3},\frac{4\pi }{3}]$ and $\arg t\in[-\frac{\pi}{3},\frac{\pi }{3}]$, one can obtain the solution of the connection problem for the same tau function e.g. between  the rays $\arg t=\frac{\pi}{3}$ and $\arg t=\frac{2\pi}{3}$. 
    	\item Since the latter solution should be covered by the formulas for generic monodromy, it allows us to determine the unknown function $\Xi_{\lcurvearrowsw}\lb\theta\rb$.
    \end{itemize}
    
   We start by explaining how to take the corresponding limits in the asymptotic expansions \eqref{expsI}, \eqref{expsII} on the rays $\arg t=\pm\frac{2\pi}{3}$, $\arg t=\pm\frac{\pi}{3}$. According to \eqref{monoparameters}, the pairs of asymptotic parameters on the rays $\arg t=\frac{2\pi}{3}$ and $\arg t=-\frac{\pi}{3}$ are 
   \beq
   \lb \nu_1,\rho_1\rb=\lb\tfrac{\theta}{2},0\rb,\qquad  \lb \omega_1,\xi_1\rb=\lb 0,-\tfrac{\theta}{2}\rb.
   \eeq
   The corresponding asymptotic expansions take the form
   \begin{align}\label{tronq01}
   	\tau_{\mathrm{gHM}}\lb t\rb=\begin{cases}\mathcal N_{1,+}\ds\sum_{n=0}^\infty \mathcal D\lb 4\sqrt 2\,e^{\frac{i\pi}{2}}|t|^{\frac32}\,\bigl|\,\frac{\theta}{2}+n,\theta\rb ,\qquad &\arg t=\frac{2\pi}{3},\\
   		\mathcal N_{1,-} \ds\sum_{n=0}^\infty\mathcal G\lb 8\,e^{\frac{i\pi}{2}}|t|^{\frac32}\,\bigl|\,n,\theta\rb e^{-\pi i n \theta },\qquad
   		&\arg t=-\frac{\pi}{3}.
   		\end{cases}
   	\end{align}
   	At first sight, the expansions along the rays $\arg t=\frac{4\pi}{3}$ and $\arg t=\frac{\pi}{3}$ appear more singular, since \eqref{monoparameters}  gives $e^{2\pi i\rho_2}=e^{2\pi i \xi_2}=0$ (while $\nu_2=-\frac{\theta}{2}$, $\omega_2=0$). The terms with $n>0$ in the Zak transforms \eqref{expsI}, \eqref{expsII} therefore vanish. The situation with $n<0$ terms is less clear. On the one hand, in this case we have ${e^{2\pi i n\rho_2},e^{2\pi i n\xi_2}\to\infty}$. On the other, constant prefactors such as $G\lb 1+\nu_2+\frac{\theta}{2}+n\rb$,  $G\lb 1+\omega_2+n\rb$ in  $\mathcal D\lb s\,|\,\nu_2,\theta\rb$, $\mathcal G\lb s\,|\,\omega_2,\theta\rb$, are equal to zero. Let us show that these two effects combine to give a finite result when the limit is examined more carefully. 
   	
   	Namely, set
   	\beq
   	s_1=\frac{e^{-\pi i \theta}}{1-e^{\pi i \theta}\varepsilon},\qquad s_2=\frac{2\varepsilon \cos\pi \theta}{2i\sin\pi\theta+\varepsilon},\qquad s_3=\frac{e^{\pi i \theta}}{1+e^{-\pi i \theta}\varepsilon}.
   	\eeq
   	It can be easily checked that this triple of Stokes parameters satisfies the monodromy constraint \eqref{monodromy_S}. Next, consider the limit $\varepsilon\to 0$, which yields the monodromy \eqref{gHMmon}. In this limit,  one has
   	\begin{subequations}
   	\begin{gather}
   		2\cos2\pi\nu_2-2\cos\pi\theta=\lb 1-e^{2\pi i \theta}\rb\varepsilon+O\lb\varepsilon^2\rb,\qquad e^{2\pi i \rho_2}=-2i\sin\pi\theta\;\varepsilon+O\lb\varepsilon^2\rb,\\
   		e^{2\pi i\omega_2}-1=i\varepsilon e^{\pi i \theta}\cot\pi\theta+O\lb\varepsilon^2\rb,\qquad e^{2\pi i \xi_2}=-i\varepsilon\cot\pi\theta+O\lb\varepsilon^2\rb.
   	\end{gather}
   	\end{subequations}
   	Note that we have, e.g.
   	\beq
   	\begin{gathered}
   	\sum_{n\leq 0}\mathcal G\lb s\,|\,\omega_2+n,\theta\rb e^{2\pi i n\xi_2}=
    e^{\mp i\pi\lb \frac{\theta^2}{3}-\frac{1}{12}\rb} \sum_{n\leq 0}	\frac{\hat G\lb\omega_2+n\rb 	e^{\pm  \frac{i\pi\lb\omega_2+n\rb^2}{2}}}{\lb2\pi\rb^{\omega_2+n}}\mathcal G\lb e^{\pm i\pi}s\,|\,-\omega_2-n,\theta\rb e^{2\pi i n\xi_2}=\\
    = \frac{e^{\mp i\pi\lb \frac{\theta^2}{3}-\frac{1}{12}-\frac{\omega_2^2}{2}\rb}\hat G\lb\omega_2\rb }{\lb 2\pi\rb^{\omega_2}}\sum_{n=0}^\infty \mathcal G\lb e^{\pm i\pi}s\,|\,-\omega_2+n,\theta\rb \lb\pm\frac{ e^{2\pi i\xi_2\pm\pi i \omega_2}}{2i\sin\pi\omega_2}\rb^{-n},
    \end{gathered}
   	\eeq
   	where the first equality follows from the symmetry \eqref{DGsym} and the second from the recurrence relation \eqref{recGhat}. The limit $\varepsilon\to0$ of the last expression is straightforward and leads to an expansion similar to the second expression in \eqref{tronq01}:
   	\beq\label{tronq02}
   	\tau_{\mathrm{gHM}}\lb t\rb=\mathcal N_{2,-}e^{ i\pi\lb \frac{\theta^2}{3}-\frac{1}{12}\rb}
   	\sum_{n=0}^\infty \mathcal G\lb 8\,e^{-\frac{i\pi}{2}}|t|^{\frac32}\,\bigl|\,n,\theta\rb e^{\pi i n\theta},\qquad \arg t=\frac{\pi}{3}.
   	\eeq
   	This highlights the fact that the asymptotics of the generalized HM solution continues from the positive real axis to the sector $\arg t\in[-\frac{\pi}3,\frac\pi 3]$. In fact, for $|\arg t|<\frac{\pi}{3}$ we have
   	\beq \label{gHMasright}
   	\begin{gathered}
   	\tau_{\mathrm{gHM}}\lb t\rb\simeq\mathcal N_R\cdot \mathcal G\lb 8e^{i\pi}t^{\frac32}\,\bigl|\,0,\theta\rb
   	=\\
   	=\mathcal N_R e^{ i\pi\lb \frac{\theta^2}{3}-\frac{1}{12}\rb}\cdot \lb 4t\rb^{\frac{\theta^2}{2}-\frac{1}{8}}\exp\left\{-\frac{t^3}{24}+\frac{\lb4\theta^2-1\rb \lb4\theta^2-9\rb}{96t^3}
    -\frac{\left(4 \theta ^2-1\right) \left(4 \theta ^2-9\right) \left(4 \theta ^2-21\right)}{192t^6}+O\lb t^{-9}\rb	\right\},
    \end{gathered}
   	\eeq
   	up to exponentially small corrections. Moreover, the leading asymptotic behavior extends to the rays $\arg t=\pm\frac{\pi}{3}$, which means that
   	\beq\label{ccmmaux}
   	\mathcal N_R=\mathcal N_{1,-}=\mathcal N_{2,-} e^{ -i\pi\lb \frac{\theta^2}{3}-\frac{1}{12}\rb}.
   	\eeq
   	
   	 An analogous computation gives
   	\beq\label{tronq03}
   	\tau_{\mathrm{gHM}}\lb t\rb=\mathcal N_{2,+} \frac{\lb 2\pi\rb^{\theta}e^{-\frac{i\pi\theta^2}{6}}}{\hat G\lb \theta\rb}
   	\sum_{n=0}^\infty \mathcal D\lb 4\sqrt 2\,e^{-\frac{i\pi}{2}}|
   	t|^{\frac32}\,\bigl|\,\frac{\theta}{2}+n,\theta\rb ,\qquad \arg t=\frac{4\pi}{3}.
   	\eeq
   	The leading terms in both \eqref{tronq03} and the first expression in \eqref{tronq01} agree with the asymptotics 
   	\beq\label{gHMasleft}
   	\begin{gathered}
   		\tau_{\mathrm{gHM}}\lb t\rb\simeq\mathcal N_L\cdot \mathcal D\lb 4\sqrt 2\,e^{-\frac{i\pi}{2}}t^{\frac32}\,\bigl|\,\frac{\theta}{2},\theta\rb
   		=\\
   		=\mathcal N_L \frac{G\lb 1+\theta\rb}{\lb 2\pi\rb^{\frac\theta 2}}\lb2^{-\frac52}e^{\frac{i\pi}{2}}\rb^{\frac{\theta^2}{6}}\cdot t^{-\frac{\theta^2}{4}}\exp\left\{-\frac{i\theta\sqrt 2}{3}t^{\frac32}+\frac{i\theta \lb4\theta^2+1\rb}{24\sqrt 2\,t^{\frac32}}
   		-\frac{\theta^2\lb 8\theta^2+7\rb}{96t^3}+O\lb |t|^{-\frac92}\rb	\right\},
   	\end{gathered}
   	\eeq 
   	valid inside the sector $\arg t\in\lb\frac{2\pi}{3},\frac{4\pi}{3}\rb$ up to exponentially small corrections. Consequently,
   	\beq\label{ccppaux}
   	\mathcal N_L = \mathcal N_{1,+}=\mathcal N_{2,+}\frac{\lb 2\pi\rb^{\theta}e^{\frac{i\pi\theta^2}{6}}}{\hat G\lb \theta\rb}.
   	\eeq
   	
   	\begin{rmk}
   		Notice that the connection coefficients $\ds\frac{\mathcal N_{2,-}}{\mathcal N_{1,-}}$ and $\ds\frac{\mathcal N_{2,+}}{\mathcal N_{1,+}}$ we have just obtained from the continuation of asymptotics agree with the general formulas found earlier. Indeed, from \eqref{conncoefmm} it follows that
   		\beq
   		\frac{\mathcal N_{2,-}}{\mathcal N_{1,-}}=e^{\frac{\pi i\lb 1+\theta^2\rb}{12}}\lim_{\varepsilon\to0} \frac{e^{2\pi i\nu_0\lb\omega_1+\omega_2\rb-\pi i\nu_0^2}}{\lb2\pi\rb^{2\nu_0}} \hat G\lb \nu_0+\frac{\theta}{2}\rb \hat G\lb \nu_0-\frac{\theta}{2}\rb= e^{ i\pi\lb \frac{\theta^2}{3}-\frac{1}{12}\rb},
   		\eeq
   		in agreement with \eqref{ccmmaux}.
   		Here we used  that $\omega_{1,2}=O\lb \varepsilon\rb$, $\nu_0=O\lb |\ln\varepsilon|\rb$ as well as the asymptotics \eqref{asGhat}. From \eqref{conncoefpp} we similarly find that
   		\beq
   		\frac{\mathcal N_{2,+}}{\mathcal N_{1,+}}=\frac{e^{\frac{\pi i\lb 1-2\theta^2\rb}{12}}\hat G\lb\theta\rb}{\lb 2\pi\rb^\theta}\lim_{\varepsilon\to 0} \frac{
   			e^{2\pi i \omega_{0}\lb\nu_{1}+\nu_{2}\rb-\frac{\pi i\omega_{0}^2}{2}}\hat G\lb \omega_{0}\rb}{\lb 2\pi\rb^{\omega_0}}=\frac{e^{-\frac{\pi i \theta^2}{6}}\hat G\lb\theta\rb}{\lb 2\pi\rb^\theta},
   		\eeq
   		which agrees with \eqref{ccppaux}.
   	\end{rmk}
   	\begin{lemma}
   		The function $\Xi_{\lcurvearrowsw}\lb\theta\rb$ can be expressed in terms of the connection coefficient of the generalized HM tau function between $-\infty$ and $+\infty$:
   		\beq\label{XiNRNL}
   		\Xi_{\lcurvearrowsw}\lb\theta\rb=e^{i\pi\lb \frac{\theta^2}{3}-\frac{1}{12}\rb}\frac{\mathcal N_R}{\mathcal N_L}.
   		\eeq
   	\end{lemma}
   	\pf From \eqref{conncoefratiosB} with $p=1$ and \eqref{Xiprop}, we find that
   	\beq
   	\Xi_{\lcurvearrowsw}\lb\theta\rb=\frac{e^{2\pi i \nu_1\omega_2-\frac{\pi i \omega_2^2}{2}}\hat G\lb\omega_2\rb}{\lb 2\pi\rb^{\omega_2}}\frac{\mathcal N_{2,-}}{\mathcal N_{1,+}}=\frac{\mathcal N_{2,-}}{\mathcal N_{1,+}}.
   	\eeq  
   	where the second equality follows from $\nu_1=\frac{\theta}{2}$, $\omega_2=0$. The statement of the lemma then follows immediately  from \eqref{ccmmaux} and \eqref{ccppaux}.
   	\epf
   	
   	Let us now explain how to extract the ratio $\ds \frac{\mathcal N_R}{\mathcal N_L}$ from the results of \cite{DXZ19}. The starting point is the function $H\lb t;\theta,0\rb$ from \cite[Eq. (1.24)]{DXZ19}, which in the notation of the present manuscript is directly related\footnote{The matching is not entirely trivial due to various sign differences in the notation for $q$, $s_{1,2,3}$, etc., but is otherwise uninstructive. The condition $\omega=0$ (in the notation of \cite{DXZ19}) is however crucial for our purposes. Although a similar comparison can be carried out for tronquées solutions with $\omega\ne0$, this would only yield a special case of Corollary~\ref{corsametyperays}.} to the logarithmic derivative of the tau function associated to the generalized Hastings-McLeod solution of \PIIeq:
   	\beq\label{HtaugHM}
   	H\lb t;\theta,0\rb=-2^{\frac13}\sigma_{\mathrm{gHM}}\lb -2^{\frac13}t\rb=\frac{d}{dt}\ln \tau_{\mathrm{gHM}}\lb -2^{\frac13}t\rb.
   	\eeq
   	The authors define a regularized integral \cite[Eq. (1.33)]{DXZ19}
   	\beq
   	I_2\lb t;\frac\theta2,0\rb=\int_c^{+\infty}\lb H\lb x;\theta,0\rb+\theta\sqrt x+\frac{\theta^2}{4x}\rb dx
   	+\int_t^c H\lb x;\theta,0\rb dx+\frac{2\theta}{3}c^{\frac32}+\frac{\theta^2}{4}\ln c,\qquad c>0.
   	\eeq
    Using \eqref{HtaugHM}, it is straightforward to see that
    \beq
    I_2\lb t;\frac\theta2,0\rb=\lim_{T\to+\infty}\ln\lb T^{\frac{\theta^2}{4}}e^{\frac{2\theta}{3}T^{\frac32}}\tau_{\mathrm{gHM}}\lb -2^{\frac13}T\rb\rb-\ln \tau_{\mathrm{gHM}}\lb -2^{\frac13}t\rb,
    \eeq
    so that, in particular, $I_2\lb t;\frac\theta2,0\rb$ is independent of $c$. This already automatically gives a part of the statement of Theorem~1.3 in \cite{DXZ19}. The part we actually need concerns the asymptotics of $I_2\lb t;\frac\theta2,0\rb$ as $t\to-\infty$:
    \begin{prop}
    	As $t\to-\infty$, one has\footnote{We note a minor correction to \cite[Eq. (1.34)]{DXZ19}: the term $\alpha$ therein should be absent. The discrepancy appears to arise from algebraic manipulations in deriving Eq. (4.19) of loc.~cit. from Eqs. (4.15) and (4.18). In fact, the term $-2\alpha\lb \alpha+1\rb$ in (4.19) should simply be $-2\alpha^2$.}
    	\beq\label{DXZas}
    	I_2\lb t;\frac\theta2,0\rb=-\frac{t^3}{12}-\frac{4\theta^2-1}{8}\ln\,|t|-\frac{\ln 2}{24}-\zeta'\lb -1\rb+\ln\frac{G\lb 1+\theta\rb}{\lb 2\pi\rb^{\frac{\theta}{2}}}+O\lb|t|^{-\frac32}\rb.
    	\eeq
    \end{prop}
    \begin{cor}\label{finalcorXi} The function $\Xi_{\lcurvearrowsw}\lb\theta\rb$ is equal to 
    	\beq
    	\Xi_{\lcurvearrowsw}\lb\theta\rb=2^{\frac{1-5\theta^2}{3}}e^{-\frac{i\pi\theta^2}{6}+\zeta'\lb -1\rb}.
    	\eeq
    \end{cor}
    \pf From \eqref{gHMasleft} it follows that
    \beq\lim_{T\to+\infty}\ln\lb T^{\frac{\theta^2}{4}}e^{\frac{2\theta}{3}T^{\frac32}}\tau_{\mathrm{gHM}}\lb -2^{\frac13}T\rb\rb=
    \ln \,\mathcal N_L +\ln\frac{G\lb 1+\theta\rb}{\lb 2\pi\rb^{\frac\theta 2}} -\frac{i\pi\theta^2}{6}-\frac{\theta^2}{2}\ln 2.
    \eeq
    On the other hand, \eqref{gHMasright} implies that, as $t\to-\infty$,
    \beq
    \ln \tau_{\mathrm{gHM}}\lb -2^{\frac13}t\rb=\frac{t^3}{12}+\frac{4\theta^2-1}{8}\ln\,|t|+\frac{7\lb 4\theta^2-1\rb}{ 24}\ln\,2+\ln \,\mathcal N_R+i\pi\lb\frac{\theta^2}{3}-\frac1{12}\rb+O\lb |t|^{-3}\rb.
    \eeq
    Comparing the difference of these two expressions with \eqref{DXZas} and using \eqref{XiNRNL}, we obtain the result, whose form clearly agrees with  Lemma~\ref{LemXiper}.
     \epf

     Combining Propositions~\ref{propCC01} and \ref{propXipm} with the previous corollary gives 
     \begin{theo}\label{TheoCC}
     Let the monodromy parameters satisfy the genericity conditions $0<|\Re\nu_p|<\frac12$, $|\Re\omega_p|<\frac12$ for all $p\in\mathbb Z/3\mathbb Z$. Then the connection coefficients are given by	\begin{subequations}\label{conncoefratiosV2}
    	\begin{align}
    		\label{conncoefratiosAV2}
    	\frac{\mathcal N_{p+2,-}}{\mathcal N_{p,+}}=&\,2^{\frac{1-5\theta^2}{3}}
    	e^{\zeta'\lb -1\rb+\frac{i\pi\lb1-\theta^2\rb}{12}+2\pi i \nu_p\omega_{p+2}-\pi i \nu_p^2}\frac{\hat G\lb \nu_p+\frac{\theta}{2}\rb \hat G\lb \nu_p-\frac{\theta}{2}\rb}{
    	\lb 2\pi\rb^{2\nu_p}},\\
    	\label{conncoefratiosBV2}
    	\frac{\mathcal N_{p+1,-}}{\mathcal N_{p,+}}=&\,2^{\frac{1-5\theta^2}{3}}
    	e^{\zeta'\lb -1\rb-\frac{i\pi\theta^2}{6}-2\pi i\nu_p\omega_{p+1}+\frac{\pi i\omega_{p+1}^2}{2}}\frac{\lb2\pi\rb^{\omega_{p+1}}}{\hat G\lb \omega_{p+1}\rb},
    	\end{align}
    	\end{subequations}
        where $\hat G\lb z\rb  =\ds\frac{G\lb 1+z\rb}{G\lb 1-z\rb}$.
     \end{theo}

     Theorem~\ref{THMA} from the Introduction is nothing but a more symmetric rewrite of Theorem~\ref{TheoCC} obtained by taking into account the normalization prefactors in \eqref{expsID} and  \eqref{expsIIG}. The genericity conditions on $\nu_p$, $\omega_p$ are purely technical. Analogous arguments should also apply to monodromy data corresponding to the subsets ${\Re\nu_p=0}$, $\Re\nu_p=\frac12$ and $\Re\omega_p=\frac12$, starting from a more complicated asymptotics (see e.g. the discussion around \eqref{compl_as}), and are expected to give the same final result \eqref{conncoefratiosV2}. Moreover, the quasiperiodicity properties of \eqref{conncoefratiosV2} with respect to integer shifts $\nu_p\mapsto\nu_p+1$, $\omega_p\mapsto\omega_p+1$ then extend these formulas to all finite values of $\nu_p,\omega_p$, in agreement with the Zak transform structure of \eqref{expsI}, \eqref{expsII}. The only essential condition is \eqref{ess_genericity}, which rules out divergent $\nu_p,\omega_p$.
 
   	 \begin{cor}\label{corU} Under assumptions $0<|\Re\nu_p|<\frac12$, $|\Re\omega_p|<\frac12$ for every $p\in\mathbb Z/3\mathbb Z$, the following holds:
    \begin{itemize} 
    \begin{subequations}
     \item For $\Re\nu_p>0$, we have
     \begin{align}
     \frac{\lim\limits_{T\to+\infty} T^{\theta-\frac12}\lambda\lb T e^{\frac{\pi i\lb 2p+1\rb }{3}}\rb}{\lim\limits_{T\to+\infty} T^{\frac{\theta+1}{4}-\frac{3\nu_p}{2}}e^{\frac{i\sqrt2}{3} T^{\frac32}}\lambda\lb T e^{\frac{2\pi ip }{3}}\rb}=&\,\frac{\sqrt \pi\; 2^{\frac{1+7\theta-10\nu_p}{4}}e^{-\frac{i\pi\lb \theta -6 \nu_p +12 \omega_{p+2} +1\rb}{12}}}{\Gamma\lb 1-\nu_p-\frac{\theta}{2}\rb},\\
     \frac{\lim\limits_{T\to+\infty} T^{\theta-\frac12}\lambda\lb T e^{\frac{\pi i\lb 2p-1\rb }{3}}\rb}{\lim\limits_{T\to+\infty} T^{\frac{\theta+1}{4}-\frac{3\nu_p}{2}}e^{\frac{i\sqrt2}{3} T^{\frac32}}\lambda\lb T e^{\frac{2\pi ip }{3}}\rb}=&\,\frac{\Gamma\lb \nu_p+\frac{\theta}{2}\rb\; 2^{\frac{-3+7\theta-10\nu_p}{4}}e^{\frac{i\pi\lb \theta -6 \nu_p +12 \omega_{p+1} +1\rb}{12}}}{\sqrt\pi}.
     \end{align}
     \item For $\Re\nu_p<0$, we have
     \begin{align}
     \frac{\lim\limits_{T\to+\infty} T^{\theta-\frac12}\lambda\lb T e^{\frac{\pi i\lb 2p+1\rb }{3}}\rb}{\lim\limits_{T\to+\infty} T^{\frac{\theta+1}{4}+\frac{3\nu_p}{2}}e^{-\frac{i\sqrt2}{3} T^{\frac32}}\lambda\lb T e^{\frac{2\pi ip }{3}}\rb}=&\,\frac{\Gamma\lb -\nu_p+\frac{\theta}{2}\rb\; 2^{\frac{-3+7\theta+10\nu_p}{4}}e^{-\frac{i\pi\lb \theta +6 \nu_p -12 \omega_{p+2} +1\rb}{12}}}{\sqrt\pi},\\
     \frac{\lim\limits_{T\to+\infty} T^{\theta-\frac12}\lambda\lb T e^{\frac{\pi i\lb 2p-1\rb }{3}}\rb}{\lim\limits_{T\to+\infty} T^{\frac{\theta+1}{4}+\frac{3\nu_p}{2}}e^{-\frac{i\sqrt2}{3} T^{\frac32}}\lambda\lb T e^{\frac{2\pi ip }{3}}\rb}=&\,\frac{\sqrt \pi\; 2^{\frac{1+7\theta+10\nu_p}{4}}e^{\frac{i\pi\lb \theta +6 \nu_p -12 \omega_{p+1} +1\rb}{12}}}{\Gamma\lb 1+\nu_p-\frac{\theta}{2}\rb}.
     \end{align}
     \end{subequations}
     \end{itemize}
     \end{cor}
     \pf Direct computation using Eq. \eqref{uform2}, Lemma~\ref{LemBackMon} and Theorem~\ref{TheoCC}. \epf
     
   Corollary~\ref{corU} may be viewed as a generic monodromy counterpart of results on total integrals of \PIIeq transcendents such as Theorem 1.3 in \cite{Mil18}, Theorems 1.1-1.2 in \cite{Kok18} and Theorem~1.1 in \cite{DXZ19}. 
    
 \section{Numerical implementation}\label{sec_numerics}
 We now briefly discuss numerical verification of the above results. It provides a useful check against errors/misprints that are not uncommon in the literature on Painlevé asymptotics. The interpretation of numerics, however, can itself be rather subtle, as demonstrate e.g. examples provided in \cite{KV23} for Painlevé~III equation of type $D_7$. We therefore explain the main issues and precautions needed to obtain reliable numerical results.

  The procedure decomposes into two steps. First, given Cauchy data $q\lb t_0\rb$, $q'\lb t_0\rb$ for \PIIeq at a finite point (e.g. $t_0=0$) together with parameter $\theta$, one should compute the corresponding Stokes coefficients $\mathbf s={\lb s_1,s_2,s_3\rb\in \mathcal S_\theta}$. By Theorem~\ref{TheoConn}, this gives a prediction for asymptotic parameters $\lb \nu_p,\rho_p\rb$, $\lb\omega_p,\xi_p\rb$ of Zak transforms \eqref{expsI}, \eqref{expsII} on the special rays. Next, the \PIIeq equation \eqref{PIIstandard} or its $\sigma$-form \eqref{sigmaPII} with these initial data should be integrated numerically so as to reach sufficiently large values of $|t|$ where the numerical solution can be compared with the expected asymptotic behavior, thereby allowing to check the results such as Theorem~\ref{THMA} and Corollary~\ref{corU}. Since the relevant large time expansions of $q$, $\sigma$ and $\tau$ are asymptotic rather than convergent, their accuracy depends on the truncation point, and including more terms does not necessarily improve the approximation. One must therefore determine appropriate (ideally optimal) truncation levels.

  There exist techniques for direct computation of the Stokes data from the linear problem, given in our case by the Flaschka-Newell system \eqref{lsysKap}--\eqref{AKap}. These techniques use the formal solution such as \eqref{GThFN1}--\eqref{GThFN2} and may  involve optimal truncation, Borel-Laplace summation and numerical integration of the linear system to relate its canonical solutions in the relevant sectors. In the present work, however, we adopt a more self-contained approach, where the numerical integration of \PIIeq  is used for both steps of the procedure.

 It is convenient to start from preliminary estimates of monodromy parameters appearing in the leading asymptotic behavior of the \PIIeq $\sigma$-function on a pair of rays, say, as $t\to\pm\infty$. In this case, whenever $\Re\nu_0,\Re\omega_0\ne \frac12\;\operatorname{mod}\; \mathbb Z$, we have
   \beq\label{numlimitsnuomega}
   \nu_0=\lim_{t\to +\infty}\frac{\sigma\lb t\rb}{i\sqrt{2t}},\qquad \omega_0=-\lim_{t\to-\infty}\frac{\sigma\lb t\rb+\frac{t^2}{8}}{i\sqrt{-t}}.
   \eeq
   These limiting values thus fix $s_2=e^{2\pi i \nu_0}$ and the product $s_1s_3=1-e^{2\pi i\omega_0}$. One can then compute $s_1$ and $s_3$ using the monodromy constraint \eqref{monodromy_S}. This yields two pairs $\lb s_1,s_3\rb$ related by the exchange $s_1\leftrightarrow s_3$. The correct choice can be easily identified numerically. Using the same procedure for the other  rays, we can similarly find approximate values of $\nu_1$, $\omega_1$ and  $\nu_2$, $\omega_2$.
 The exact monodromy parameters are related by identities such as \eqref{monodromy_S}, \eqref{nuomegaidentities} and \eqref{monoparameters}, which are therefore expected to hold approximately for the estimates.

\begin{eg}
Choose ``random'' parameter $\theta$ and initial conditions  as
\beq
  \theta = \frac 2 3 + \frac {2i} 7 ,\qquad 
  q\lb 0\rb=\frac{17}{59}-\frac{i}{3},\qquad 
  q'\lb 0\rb=-\frac{19}{53}+\frac{3i}{10}.
\eeq
The rational numbers are used instead of decimal approximations to make subsequent numerical computations with tools such as Mathematica $\mathtt{NDSolve}$ more efficient and to facilitate the reproduction of our results by the interested reader. 
The estimates for $\nu_0$ and $\omega_0$ are obtained from  \eqref{numlimitsnuomega} using the numerical solution for $|t|\in [ 90,100]$ and averaging over fast oscillations. Together with a similar computation for the other rays, this gives
\begin{subequations}
\begin{alignat}{2}
& \nu_0^{\mathrm{est}} = -0.436(1) -0.075(1) i, \qquad && 
\omega_0^{\mathrm{est}} = 0.143(2) -0.023(2) i,\\
& \nu_1^{\mathrm{est}} = 0.210(1) + 0.050(1) i, \qquad
&&\omega_1^{\mathrm{est}} = -0.271(2) - 0.086(2) i,\\
&\nu_2^{\mathrm{est}} = -0.412(2)-0.042(2) i,\qquad
&&\omega_2^{\mathrm{est}} = 0.151(2) -0.056(1) i.
\end{alignat}
\end{subequations}
The errors in the above are estimated from analogous calculations for intervals of different lengths and positions with $t\sim 10^2\text{--}10^3$.
 \end{eg}
 
 At the next step, the preliminary estimates are refined. This requires a high-precision numerical solution for \PIIeq function $\sigma\lb t\rb$ and its asymptotic expansions following from \eqref{expsI}--\eqref{expsID}, \eqref{expsII}--\eqref{expsIIG}. The asymptotic series should be truncated to finite sums, with the truncation error assessed using the previous estimates. Matching the numerical and asymptotic evaluation at two distinct values of $t$ yields a system of two equations for the corresponding pair of asymptotic parameters, whose solution then gives their improved values. 
 \begin{eg}\label{numerics_eg2}
 Consider the ray $\arg t=0$. For the truncation error analysis, we use explicit formulas for $\mathcal{D}_k\lb \nu,\theta\rb$ for $k\le 10$. One may characterize the ratio of consecutive terms in the series \eqref{expsID} by
 \beq
 \delta_k\lb t\rb=\left|\frac{\mathcal D_{k}\lb \nu_0^{\textrm{est}},\theta\rb}{4\sqrt 2 \,t^{\frac32}\mathcal D_{k-1}\lb \nu_0^{\textrm{est}},\theta\rb}\right|.
 \eeq
 A few values of $\delta_k\lb t\rb$ are shown in the table below. The optimal truncation at $k=10$ corresponds to the value $t\approx 5$. For the actual computation of Stokes data, we will conservatively evaluate the expansion of $\mathcal D\lb 4\sqrt 2\,e^{\frac{i\pi}{2}}|t|^{\frac32}\,\bigl|\,\nu,\theta\rb$ truncated at 10th order at $t\sim t_{\mathrm{as}}=10^2$.
 \begin{center}
 \begin{tabular}{|c|c|c|c|}
 \hline 
 $k$ & $\delta_k\lb 10.\rb$ & $\delta_k\lb 25.\rb$ & $\delta_k\lb 100.\rb$ \\
 \hline
    $2$     & 0.0354 & 0.0090 & 0.0011 \\
    $5$     & 0.1432 & 0.0362 & 0.0045 \\
    $10$    & 0.3147 & 0.0796 & 0.0099 \\
  \hline
 \end{tabular}
 \end{center}
 One also has to understand how to restrict the values of $n$ in the sum \eqref{expsI}. Introduce the ratios of leading terms of each block,
 \beq
 r_n=\left|e^{2\pi i n\rho_0^{\mathrm{est}}} \frac{G\lb 1+\nu_0^{\mathrm{est}}+n+\tfrac{\theta}{2}\rb
     		G\lb 1+\nu_0^{\mathrm{est}}+n-\tfrac{\theta}{2}\rb}{\lb2\pi\rb^{n}G\lb 1+\nu_0^{\mathrm{est}}+\tfrac{\theta}{2}\rb
     		G\lb 1+\nu_0^{\mathrm{est}}-\tfrac{\theta}{2}\rb} \lb 4\sqrt 2\,e^{\frac{i\pi}{2}} t^{\frac32}\rb^{-n^2-2n\nu_0^{\mathrm{est}}}\right|,
 \eeq
 where $\rho_0^{\mathrm{est}}$ can be determined e.g. from $\rho_0=\omega_1$, cf \eqref{asparsA}.
Several approximate values of $r_n$ at  $t= t_{\mathrm{as}}$ are collected in the following table:
\begin{center}
\begin{tabular}{|r|c||r|c|}
\hline $n$ & $r_n$ & $n$ & $r_n$ \\
\hline
$-1$ & $1.4\times 10^{-8}$ & $1$ & $0.213$\\
$-2$ & $1.3\times 10^{-23}$ & $2$ & $4.0\times 10^{-10}$ \\
$-3$ & $2.1\times 10^{-45}$ & $3$ & $5.7\times 10^{-26}$ \\
\hline
\end{tabular}
\end{center}
We may therefore safely restrict the range of summation in \eqref{expsI} to $n=-3,\ldots,3$. 
\end{eg}

\begin{eg} Finally, using high-precision numerical solution of \PIIeq up to $t=t_{\mathrm{as}}$, compare it with $\sigma\lb t\rb$ computed from the restricted asymptotic expansion of $\tau\lb t\rb$ at 
two values of $t$ close to  $t_{\mathrm{as}}$. Solving these two equations numerically for $\nu_0$ and $\rho_0$ using $\nu_0^{\mathrm{est}}$, $\rho_0^{\mathrm{est}}$ as starting values, we find
\begin{subequations}\label{numvalsmonpars}
\beq
\nu_0 =-0.4363144624- 0.0754041540 i, \qquad 
\rho_0 = -0.2708581616 - 0.0863637857 i.
\eeq
Similar analysis for the other rays gives the other pairs of asymptotic parameters: 
\begin{alignat}{2}
&\nu_1=0.2103704177 + 0.0503168236 i,\qquad &&\rho_1= 0.1515561850 - 0.0558738550 i,\\
&\nu_2= -0.4127973633 - 0.0426779499 i,\qquad &&\rho_2= 0.1436157519 - 0.0237111845 i,\\
&\omega_0= 0.1436157519  - 0.02371 11845 i,\qquad  &&\xi_0= 0.2103704177 + 0.05031 68236 i,\\
&\omega_1= -0.2708581619  - 0.0863637858 i,\qquad &&\xi_1= -0.4127973632 - 0.0426779499 i,\\
&\omega_2= 0.1515561850  - 0.0558738550 i,\qquad  &&\xi_2= -0.4363144624 - 0.0754041541 i.
\end{alignat}
\end{subequations}
The numerical errors in all of the above values are estimated to be of order $10^{-10}$. Additional check with the relations \eqref{monodromy_S}, \eqref{nuomegaidentities} and \eqref{monoparameters}  shows 
that they are indeed verified with this accuracy (in particular, it is manifest that $\rho_{p}=\omega_{p+1}$, $\xi_p=\nu_{p+1}$).
 \end{eg}

\begin{figure}[h!]
     \centering
     \includegraphics[width=0.9\linewidth]{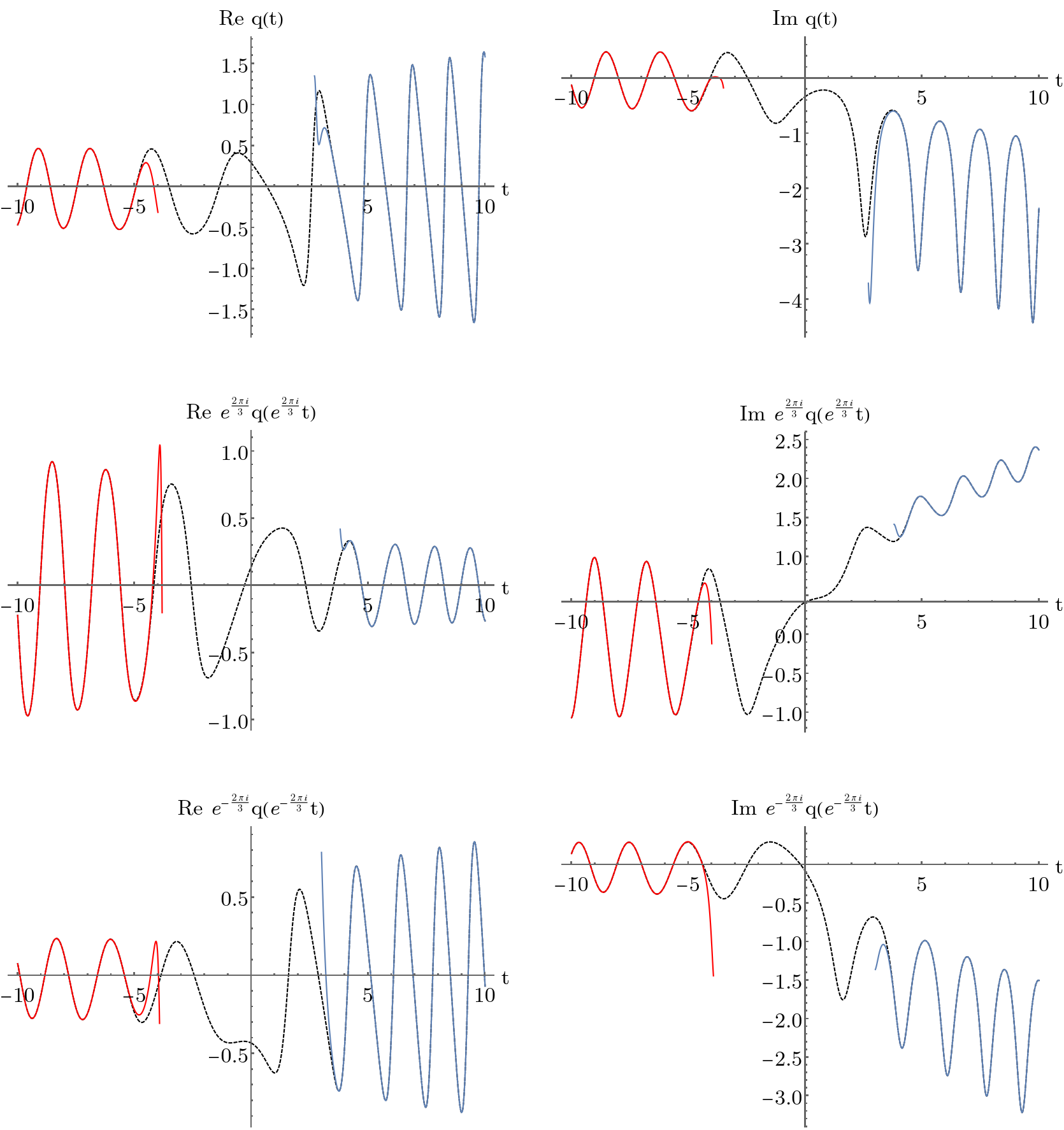}
     \caption{Numerical solution vs truncated asymptotics for $q\lb t\rb$ along the special rays.}
     \label{FiguresForQ}
 \end{figure}

 Figure~\ref{FiguresForQ} shows the numerical solution for $q\lb t\rb$ (black dashed line) along with its expression computed from the asymptotic expansions \eqref{expsI}--\eqref{expsID} and \eqref{expsII}--\eqref{expsIIG}  (shown respectively in blue and red) calculated with the above monodromy. The asymptotic expression thus provides a surprisingly accurate approximation of $q\lb t\rb$ even for intermediate values of $|t|$. 

 It is worth mentioning that the approximation of $q$ is obtained by substituting the truncated expansion of $\tau$ and its derivatives into \eqref{qintermsoftau} without subsequent reexpansion  of the result. In the terminology of \cite{KV23}, we avoid taking the ``asymptotics of asymptotics''. The reason is that, depending on monodromy, a single subleading term in the asymptotics of $\tau$ can generate under reexpansion a large number of relevant contributions to the asymptotics of $q$. This may be seen in the above example where the Cauchy data were  deliberately chosen ill-conditioned for the ray $\arg t=0$, with $\Re\nu_0$ being close to one of the endpoint values $\pm\frac12$. The relative magnitude of the ${n=1}$ contribution to \eqref{expsI}, as compared to $n=0$ contribution,  is of order $O\lb t^{-3\lb 2\Re\nu_0+1\rb/2}\rb\approx O\lb t^{-0.19}\rb$ as $t\to+\infty$, explaining why $r_1$ in the second table of Example~\ref{numerics_eg2} is not particularly small. If this slowly decaying term is neglected, achieving a relative accuracy of $\tau\lb t\rb$ of just $10^{-2}$ would require  $t$ to be of order $10^{10}$, well beyond the reach of direct numerical integration of the differential equation. This term enters the denominator of $q\lb t\rb$ as a correction, cf e.g. 
 \eqref{logasqplusA}--\eqref{logasqplusB}. A similar phenomenon explains the surprising behavior of \cite[Example 6]{KV23} illustrated by Figures~1 and~2 therein.

 At this point, the Stokes data are determined and  Theorem~\ref{TheoConn} is  checked. The corresponding tau function may now be computed by numerically evaluating the integral
 \beq\label{taunormzero}
 \tau_{\mathrm{num}}\lb t\rb=\exp\int_0^t\sigma\lb \eta\rb d\eta
 \eeq
 along different rays, i.e. we choose the normalization $\tau_{\mathrm{num}}\lb 0\rb=1$. To check Theorem~\ref{THMA}, one has to compare this numerical evaluation with the asymptotics \eqref{astauspecial}. In order to achieve a reasonable accuracy for moderately large~$t$, the leading-order asymptotics has to be replaced by the truncated asymptotic expansion as explained above.

 \begin{eg}
 Figure~\ref{tau_rat} shows the ratio of the tau function $\tau_{\mathrm{num}}\lb t\rb$ computed with \eqref{taunormzero} and  the asymptotic expansion $\tau_{\mathrm{tr}}\lb t\rb$ along the ray $\arg t=0$, given by
 \eqref{expsI}. The latter is truncated at $k\leq 10$, $n=-3,\ldots,3$  and normalized by $\mathcal N_{0,+}=1$. It indeed approaches a constant $c_{0,+\infty}$ with 8-digit accuracy already for $t\sim 10$. For larger values $t\sim 10^2$, the accuracy is primarily limited by the precision of numerical integration in \eqref{taunormzero}. The connection coefficients $\mathcal N_{p,\epsilon}/\mathcal N_{q,\epsilon'}$ with $\epsilon,\epsilon'\in\left\{+,-\right\}$ and $p,q\in\mathbb{Z}/3\mathbb{Z}$ are given by the ratios of such limiting constants computed on different rays. The results are in full agreement with equations \eqref{conncoefratiosV2}.
 \begin{figure}[h!]
     \centering
     \includegraphics[width=0.95\linewidth]{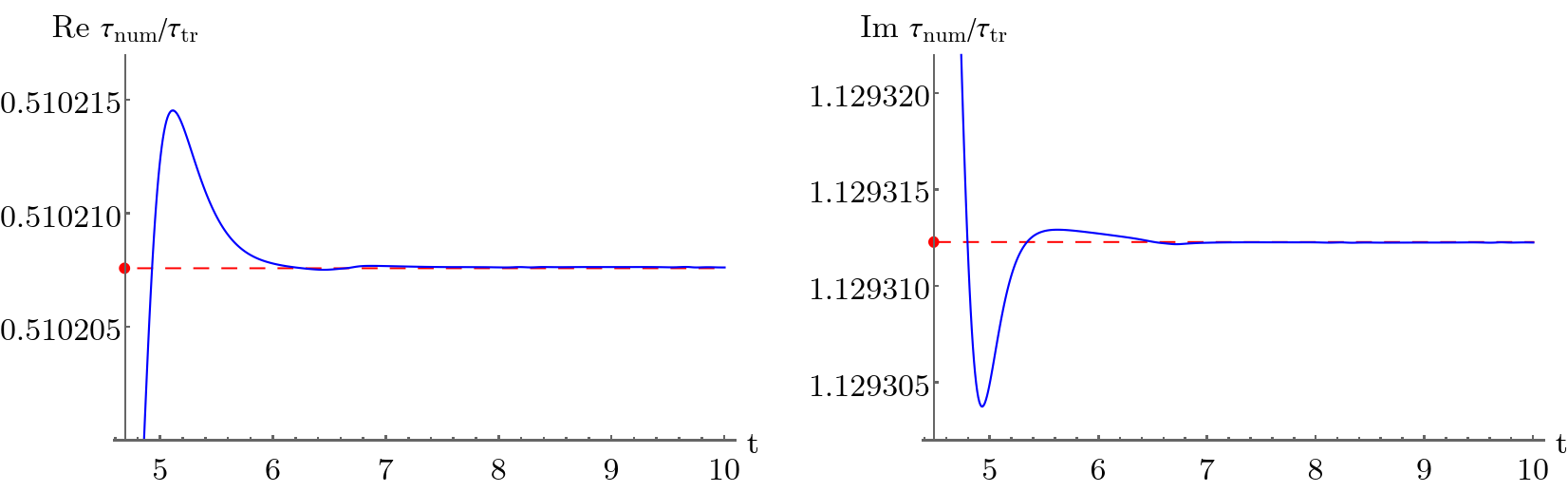}
     \caption{Ratio $\ds\frac{\tau_{\mathrm{num}}}{\tau_{\mathrm{tr}}}$ along the ray $\arg t=0$. The limiting value  $c_{0,+\infty}\approx 0.51020765 + 1.1293123 i$. }
     \label{tau_rat}
 \end{figure}
 \end{eg}

   \section{Resurgent derivation}

\label{sec:resurgence}

   In this section, we show that the connection coefficients of Corollary~\ref{corsametyperays}, namely the ratios of the normalization factors $\mathcal N_{p,+}$ on the rays $\arg t=\frac{2\pi p}{3}$ ($p=0,1,2$), can also be derived from the resurgent structure of the two-parameter transseries solutions of Painlev\'e II constructed in \cite{BSSV,Vega,Vega23}. The connection coefficients  $\mathcal N_{p,-}$ between the complementary rays involve the second transseries background of \cite{Vega,Vega23} and will not be considered here, though the corresponding computation is analogous. The derivation is independent of the isomonodromic arguments of Section~\ref{sec_CC}. Its main inputs are the nonlinear transseries and their Stokes data \cite{BSSV,Vega,Vega23}, together with the Zak transform (ZT) representation of the tau function. One also needs the dilogarithmic normalization accompanying a Stokes jump of the Zak representations \cite{IM,KRSST}. The results of this section are therefore conditional.

The central step is to connect two descriptions of the same formal tau function. The resurgent construction naturally gives a rectangular expansion in the two transseries parameters $\lb\sigma_1,\sigma_2\rb$. Using the Hamiltonian expression of the time derivative of $\ln\tau$, it can be converted into a tau function transseries and then reorganized into the form \eqref{expsI}--\eqref{expsID}\footnote{The explicit relation between rectangular resurgent transseries
and the corresponding Zak transforms is highly nontrivial and was developed in
\cite[Sections~3.1--3.3]{KRSSTlocal}
for Painlevé I and Hermitian matrix models, including their
double-scaling limits. The rectangular transseries construction for Painlevé I itself goes back to \cite{gikm10}.}. This identifies the ZT variables $\lb\nu,\rho\rb$ in terms of the transseries parameters and places the Stokes data of the nonlinear solution in the same variables that appear in the asymptotics on the special rays.

We begin by reviewing the two-parameter transseries, including its resonant logarithmic sectors and Stokes geometry. We then construct the corresponding ZT and check the reorganization explicitly, sector by sector, in Appendix~\ref{app:sectors}. Once the correspondence is established, the Stokes transition can be written directly in the ZT variables, leading to an alternative derivation of the relative normaliza\-tion \eqref{conncoefpp}. 

\subsection{The two-parameter Painlev\'e II transseries}
\label{subsec:transseries}

We first translate to the normalization used in the resurgent analysis of \cite{BSSV,Vega,Vega23}. Setting
\begin{equation}\label{eq:zu}
z=2^{-\frac13}e^{i\pi}t,\qquad q(t)=2^{-\frac13}u\lb z\rb,\qquad \alpha=\frac12-\theta,
\end{equation}
the \PIIeq equation \eqref{PIIstandard} becomes
\begin{equation}\label{eq:PIIu}
\partial_{zz}u=2u^3-2zu+2\alpha .
\end{equation}
The function $\sigma\lb t\rb=\lb \ln\tau\rb'$ of \eqref{PIItauDef} becomes $\sigma_z:=\partial_z\ln\tau=-2^{\frac13}\sigma(t)$, and substituting \eqref{eq:zu} into \eqref{PIItauDef} gives
\begin{equation}\label{eq:Hz}
\sigma_z=\frac{\lb  u^2-z\rb^2-\lb\partial_z u\rb^2}{4}-\theta u.
\end{equation}
Thus $\arg z=\arg t+\pi$ along the continuation considered below.

Starting from \eqref{eq:PIIu}, the analysis of \cite{BSSV,Vega,Vega23} proceeds by a two-parameter transseries ansatz around the formal background $u\simeq\sqrt z$ (the other formal backgrounds are $u\simeq-\sqrt z$ and $u\simeq\alpha/z$). In the resurgent variables $w=z^{-\frac34}$ and $x=w^2=z^{-\frac32}$, the resulting transseries solution takes the form
\begin{equation}\label{eq:Vega-ts}
\begin{aligned}
u\lb z;\sigma_1,\sigma_2\rb &=\sqrt{z}\sum_{n,m\ge0}\sigma_1^{\,n}\sigma_2^{\,m}\,e^{-\frac{(n-m)A}{w^2}}\,\Phi_{(n,m)}(w),\\
\Phi_{(n,m)}\lb w\rb&=\sum_{k=0}^{k_{nm}}\ln^k w\sum_{g\ge0}u_{2g}^{(n,m)[k]}\,w^{\,2g+\beta^{[k]}_{(n,m)}}.
\end{aligned}
\end{equation}
Substitution in the differential equation fixes the instanton action $A=\frac43$ and recursively determines the sector coefficients, exponents and logarithmic powers. For example, $k_{nm}=\min\lb n,m\rb-m\delta_{nm}$ and $\beta^{[k]}_{(n,m)}=n+m-2\lfloor\frac{k_{nm}+k}{2}\rfloor-2\lb n-m\rb\alpha$. The parameters $\sigma_1$ and $\sigma_2$ multiply the elementary exponentials $e^{-Az^{\frac32}}$ and $e^{+Az^{\frac32}}$, respectively, and are normalized by $u^{(1,0)[0]}_0=u^{(0,1)[0]}_0=1$.

Let us examine the structure of this transseries solution in a bit more detail. Note that the exponential weight of a sector depends only on $n-m$. Thus, for example, $\lb 1,0\rb$ and $\lb 0,1\rb$ carry the two elementary actions $+A$ and $-A$, while adding the same integer to both indices leaves the exponential factor unchanged. Since the two elementary actions are opposite, $A+(-A)=0$, the transseries is said to be resonant \cite{mss22}. Sectors such as $(1,0)$ and $(2,1)$ therefore lie on the same exponential scale, which in fact is the origin of the logarithms appearing explicitly in \eqref{eq:Vega-ts} \cite{ASV}.

These logarithmic sectors can be packaged in a more convenient form. Introducing
\begin{equation}\label{eq:mu}
\mu:=\sigma_1\sigma_2,
\end{equation}
the logarithms can be resummed into a power of $z$, and after redefining $\Phi_{(n,m)}\lb w\rb$ the same transseries may be written as\footnote{Note that while $\beta^{[k]}_{(n,m)}$ contained the $\alpha$ dependence explicitly, after log resumming the transseries it is more convenient to take that factor out, which is why $\beta_{nm}$ is instead used in the equations below.}
\begin{equation}\label{eq:transseries}
\begin{aligned}
u\lb z;\sigma_1,\sigma_2\rb&=\sqrt z\sum_{n,m\ge0}\sigma_1^{\,n}\sigma_2^{\,m}\,
e^{-(n-m)A z^{\frac32}}\,z^{\frac32\lb n-m\rb\lb\alpha+4\mu\rb}\,\Phi_{(n,m)}\lb w\rb,\\
\Phi_{(n,m)}\lb w\rb &=\sum_{g\ge0}u^{(n,m)}_{2g}\,w^{\,\beta_{nm}+2g},
\end{aligned}
\end{equation}
with $\beta_{nm}=n+m-2\lfloor k_{nm}/2\rfloor$. In particular, expanding in $\mu=\sigma_1\sigma_2$ the second factor in $z^{\frac32\lb n-m\rb\lb \alpha+4\mu\rb}
=z^{\frac32\lb n-m\rb\alpha}e^{6\lb n-m\rb\mu\ln z}$
 reproduces precisely the powers of $\ln z$ generated by resonance. We use the resummed form \eqref{eq:transseries} from now on. The first sectors needed below are collected in Appendix~\ref{app:sectors}.

Extending the same kind of resurgent analysis as above to the Borel plane also determines the Stokes geometry and the associated Stokes transformations. The coefficients of the formal sectors have factorial large-order growth, and the large-order relations between different sectors identify the Borel singularities together with the residues that control their discontinuities \cite{BSSV,Vega,Vega23}. In the Borel plane of $x=z^{-\frac32}$ these singularities lie at the integer action lattice $A\mathbb Z\backslash\{0\}$. When the Laplace direction reaches one of these singular directions, the two lateral Borel resummations differ; the corresponding jump is compensated by a transformation of the transseries parameters. The induced Stokes transformations obtained in this way will be used in Subsection~\ref{subsec:stokes-symmetry}.

Since $A=\frac43$ is real, the singular Borel directions are $\arg x=0$ and $\arg x=\pi$ modulo $2\pi$, corresponding to
$\arg z\in\left\{0,\pm\frac{2\pi}{3}\right\}$.
The complementary rays
$\arg z\in\left\{\pi,\pm\frac\pi3\right\}$
are anti-Stokes rays, where the two elementary exponentials have the same magnitude. Under \eqref{eq:zu}, these anti-Stokes rays are those with $\arg t=\frac{2\pi p}{3}$ on which the ZT asymptotics \eqref{expsI}--\eqref{expsID} are written, while the Stokes rays correspond to the complementary directions of \eqref{expsII}--\eqref{expsIIG}.

\subsection{From the rectangular tau transseries to Zak transform}
\label{subsec:fourier}

The transseries \eqref{eq:transseries} is a solution of the nonlinear Painlev\'e equation, naturally organized as a double series in the rectangular parameters $\lb \sigma_1,\sigma_2\rb$. Equation \eqref{eq:Hz} turns it directly into a tau function transseries,
\begin{equation}\label{eq:taurect}
\tau_{\rm rect}\lb z;\sigma_1,\sigma_2\rb=C\lb \sigma_1,\sigma_2\rb\exp\left\{\int^z\sigma_z\lb z';\sigma_1,\sigma_2\rb\,d z'\right\},
\end{equation}
where the integration constant $C$ may depend on the transseries parameters but is independent of $z$. At this stage the result is still naturally organized as a rectangular double series. The identification of rectangular resurgent transseries with Zak representations was made explicit in \cite{KRSSTlocal} for Painlevé I and matrix-model partition functions. Here we make this relation precise for the inhomogeneous Painlevé II transseries, determine the parameter dictionary, and use it to compute the connection coefficients between rays of the same type. This is the step that relates the resurgent parameters $\lb \sigma_1,\sigma_2\rb$ to the variables $\lb \nu,\rho\rb$ used in the asymptotic description of Section~\ref{sec_AM}.

For the continuation used below, $\arg t:0\to-\frac{2\pi}{3}$ corresponds under \eqref{eq:zu} to $\arg z:\pi\rightarrow\frac{\pi}{3}$.
Throughout this sector we write
\begin{equation}\label{eq:sdef}
s:=8e^{-i\pi}z^{\frac32},\qquad
\ln s=\ln\lb 8|z|^{\frac32}\rb+\frac{3i}{2}\arg z-i\pi,
\qquad \frac{\pi}{3}\leq\arg z\leq\pi .
\end{equation}
On the initial ray $\arg t=0$, where $\arg z=\pi$, this gives
$s=4\sqrt2\,e^{\frac{i\pi}{2}}|t|^{\frac32}$,
which is exactly the variable used in \eqref{expsI}. In terms of $z$, the complete building block \eqref{expsID} is
\begin{equation}\label{eq:blockz}
\begin{aligned}
\mathcal D\lb s\,|\,\nu,\theta\rb\simeq{}
8^{\frac{\theta^2}{12}-\nu^2}e^{i\pi\nu^2-\frac{i\pi\theta^2}{12}}
\frac{G\left(1+\nu+\frac\theta2\right)G\left(1+\nu-\frac\theta2\right)}{(2\pi)^\nu}
z^{-\frac{3\nu^2}2+\frac{\theta^2}{8}}\exp\!\left\{-\frac43\nu z^{\frac32}
+\sum_{k\ge1}\mathcal D_k\lb \nu,\theta\rb s^{-k}\right\}.
\end{aligned}
\end{equation}
For an appropriate choice of $C$, we express the tau function
transseries in the Zak coordinates
$\nu=\nu(\boldsymbol{\sigma})$ and
$\rho=\rho(\boldsymbol{\sigma})$, where
$\boldsymbol{\sigma}=(\sigma_1,\sigma_2)$.
Using the same notation in these coordinates, we write
\begin{equation}\label{eq:DFTtau}
\tau_{\rm rect}\lb z;\nu,\rho\rb=\sum_{n\in\mathbb Z}e^{2\pi i n\rho}\,\mathcal D\lb s\,|\,\nu+n,\theta\rb.
\end{equation}
As we will soon see, the induced Stokes transformations along the direction $\arg x=-\pi$ will be conveniently packaged in the coordinate \cite[Eqs.~(3.3.109), (3.3.110)]{Vega}\footnote{There is a typo in \cite{Vega,Vega23}, where the relations are denoted with respect to $\arg x=\pi$ instead of $\arg x=-\pi$.}
\begin{equation}\label{eq:rhopi}
\rho_{-\pi}:=\sigma_1N^{(-1)}\lb \mu\rb,\qquad
N^{(-1)}\lb \mu\rb=
\frac{i\,2^{3\theta-2-12\mu}e^{-i\pi(\theta-4\mu)}}
{\Gamma\lb 1-2\mu\rb\Gamma\lb \theta-2\mu\rb}.
\end{equation}
As is explained in Appendix~\ref{app:sectors}, direct comparison of the Fourier sum with the rectangular tau transseries then fixes
\begin{equation}\label{eq:DFT}
\nu=\frac\theta2-2\mu,\qquad
 e^{2\pi i\rho}
=-\frac{\pi\,8^{2\nu}e^{-2\pi i\nu}}
{\Gamma\lb 1+\nu-\frac\theta2\rb\Gamma\lb \nu+\frac\theta2\rb}\,\sigma_1
=4\pi i\rho_{-\pi}.
\end{equation}

\subsection{Stokes continuation}
\label{subsec:stokes-symmetry}

We now continue the ZT from the $p=0$ ray to the $p=2$ ray. Under \eqref{eq:zu}, the clockwise continuation
${\arg t:0\rightarrow-\frac{2\pi}{3}}$ corresponds to
$\arg z:\pi\rightarrow\frac{\pi}{3}$.
Along this path the resurgent variable $x=z^{-\frac32}$ has
$\arg x:-\frac{3\pi}{2}\rightarrow-\frac{\pi}{2}$,
and crosses the Stokes line at $\arg x=-\pi$ (that is, $\arg z=\frac{2\pi}{3}$).
\begin{figure}[ht]
\centering
\begin{tikzpicture}[scale=1.15,>=stealth]
\draw[->,thick] (0,0) -- (180:2.6) node[left,align=center] {$p=0$\\$\arg z=\pi$};
\draw[->,thick] (0,0) -- (-60:2.6) node[below right,align=center] {$p=1$\\$\arg z=-\frac{\pi}{3}$};
\draw[->,thick] (0,0) -- (60:2.6) node[above right,align=center] {$p=2$\\$\arg z=\frac{\pi}{3}$};
\draw[->,dashed] (0,0) -- (0:2.6) node[right] {$(\omega_0,\xi_0)$};
\draw[->,dashed] (0,0) -- (120:2.6) node[above left] {$(\omega_1,\xi_1)$};
\draw[->,dashed] (0,0) -- (-120:2.6) node[below left] {$(\omega_2,\xi_2)$};
\draw[->,red!70!black,thick] (180:1.35) arc (180:60:1.35);
\node[red!70!black] at (120:1.67) {\small $\arg x=-\pi$};
\node at (0.62,-0.18) {\small $z$-plane};
\end{tikzpicture}
\caption{Clockwise continuation from the $p=0$ ray to the $p=2$ ray. Along the path $\arg z:\pi\to\frac{\pi}{3}$, the continuously tracked resurgent (Borel) variable $x=z^{-\frac32}$ crosses the Stokes line at $\arg x=-\pi$ ($\arg z=\frac{2\pi}{3}$), corresponding to the complementary ray $\lb \omega_1,\xi_1\rb$.}
\label{fig:rays}
\end{figure}
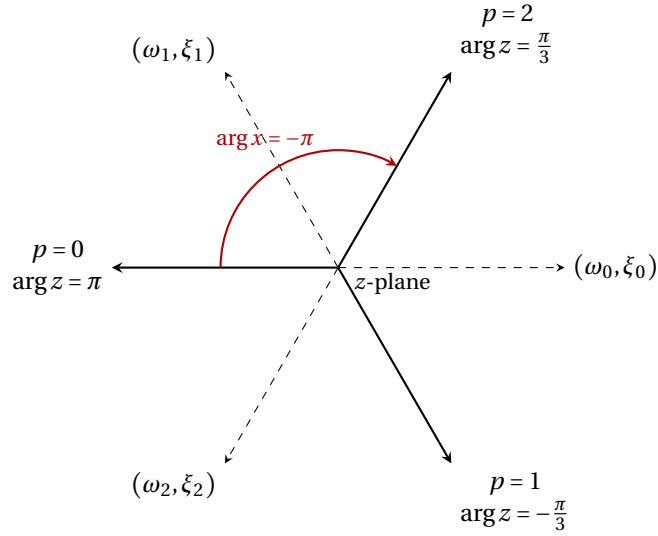

As mentioned above, to continue our solution across a singular Borel direction we must consider the lateral Laplace resummations. In other words, after crossing the Stokes line along $\arg x=-\pi$ our transseries parameters $\lb \sigma_1,\sigma_2\rb$ should be modified according to the induced Stokes transformation. Translating the transformations obtained in \cite{BSSV,Vega,Vega23} to the coordinate $\rho_{-\pi}$ of \eqref{eq:rhopi} and $\mu$, and using the inverse of their formal Stokes-automorphism convention \cite[Eq.~(3.3.118)]{Vega}, continuation across this ray acts by\footnote{There appears to be a typo in that equation, which should be denoted $S_{-\pi}$ rather than $S_{\pi}$, and similarly $\rho_{\pi}$ should be $\rho_{-\pi}$.}
\begin{equation}\label{eq:backwardStokes}
\lb\mu,\rho_{-\pi}\rb\longmapsto
\left(\mu+\frac{\ln\lb 1-4\pi i\rho_{-\pi}\rb}{4\pi i},\ \rho_{-\pi}\right).
\end{equation}
Thus we note that the coordinate $\rho_{-\pi}$ is invariant across this Stokes jump. Using \eqref{eq:DFT}, this becomes
\begin{equation}\label{eq:nushift}
\lb \nu,\rho\rb\longmapsto
\left(\nu-\frac{\ln\lb 1-e^{2\pi i\rho}\rb}{2\pi i},\ \rho\right).
\end{equation}
Therefore note that the coordinate $\rho$ itself is unchanged by the Stokes crossing, as was also found in \cite{IM}. At the level of the tau function, the same transformation carries a dilogarithmic normalization, and overall after the Stokes jump one has \cite{KRSST,IM}\footnote{This dilogarithm prefactor can also be obtained, as shown in \cite{IM}, by using the definition of the extended tau function itself \cite{LR16}. Nonetheless, it also has an independent resurgence derivation, which is what we choose to emphasise, since it highlights that the  connection coefficient computation done in this section is genuinely independent from other considerations in this work. It is still worth noting that the approaches are consistent with one another: differentiating the prefactor gives
$d\Li(e^{2\pi i\rho})
=-2\pi i \ln\lb 1-e^{2\pi i\rho}\rb\,d\rho$,
which exactly compensates for the change of the $-2\pi i\nu\,d\rho$ term in the extended differential \eqref{taupplus}.
}
\begin{equation}\label{eq:StokesDFT}
\tau_{\rm rect}\lb z;\nu,\rho\rb \longmapsto
\exp\left\{\frac{\Li\lb e^{2\pi i\rho}\rb}{2\pi i}\right\}
\tau_{\rm rect}\left(z;\nu-\frac{\ln\lb 1-e^{2\pi i\rho}\rb}{2\pi i},\rho\right).
\end{equation}

Finally, continuing to rotate until we reach the endpoint $\arg z=\frac{\pi}{3}$, the continuously followed variable \eqref{eq:sdef} is
$s=e^{-i\pi}\tilde s$, 
where $\tilde s:=4\sqrt2\,e^{\frac{i\pi}{2}}|t|^{\frac32}$ is the standard radial variable of \eqref{expsI} on the $p=2$ ray. The reflection identity \eqref{DGsym} therefore rewrites the continued Fourier blocks directly in the endpoint ZT.

\subsection{Connection coefficients}
\label{subsec:derivation}

\begin{prop}\label{prop:main}
Let $\tau_{\rm rect}$ be normalized as in \eqref{eq:DFTtau}, and let $\tau=\mathcal N_{0,+}\tau_{\rm rect}(z;\nu_0,\rho_0)$ have the asymptotics \eqref{expsI} on the ray $\arg t=0$ ($\arg z=\pi$), with parameters $\lb\nu_0,\rho_0\rb$ and normalization $\mathcal N_{0,+}$. Then, by continuing the transseries solution from $\arg t=0$ to $\arg t=-\frac{2\pi}{3}$\footnote{This means both appropriately transforming across the Stokes line at $\arg t=-\frac{\pi}{3}$ as well as making use of the discrete symmetry  \eqref{DGsym}.}, its asymptotics on the ray $\arg t=-\frac{2\pi}{3}$  ($\arg z =\frac{\pi}{3}$) is given by \eqref{expsI} with $p=2$, where the induced parameters $\lb\nu_2,\rho_2\rb$ satisfy
\begin{subequations}\label{eq:kinematics}
\begin{align}
\nu_2&=-\nu_0+\frac{\ln\lb 1-e^{2\pi i\rho_0}\rb}{2\pi i},\label{eq:kin1}\\
e^{2\pi i(\rho_0+\rho_2)}&=\lb 1-e^{2\pi i\lb \nu_2-\frac\theta2\rb}\rb\lb 1-e^{2\pi i\lb \nu_2+\frac\theta2\rb}\rb,\label{eq:kin2}
\end{align}
\end{subequations}
which are the same relations as can be obtained from \eqref{monoparameters}. Moreover, the prefactor accumulated along this continuation is
\begin{equation}\label{eq:N20}
\frac{\mathcal N_{2,+}}{\mathcal N_{0,+}}=
\exp\left\{\frac{\Li\lb e^{2\pi i\rho_0}\rb}{2\pi i}\right\}
\frac{e^{i\pi\lb \nu_2^2-\frac{\theta^2}{12}\rb}\lb 2\pi\rb^{2\nu_2}}
{\hG\lb\nu_2+\frac\theta2\rb\hG\lb \nu_2-\frac\theta2\rb}.
\end{equation}
\end{prop}
\pf
On the initial ray $\arg t=0$, equations \eqref{eq:sdef} and \eqref{eq:DFTtau} give
\[
\tau=\mathcal N_{0,+}\tau_{\rm rect}(z;\nu_0,\rho_0),
\qquad
s=4\sqrt2\,e^{\frac{i\pi}{2}}|t|^{\frac32}.
\]
After crossing $\arg x=-\pi$, \eqref{eq:StokesDFT} gives
\begin{equation}\label{eq:continuedDFT}
\tau=\mathcal N_{0,+}
\exp\left\{\frac{\Li\lb e^{2\pi i\rho_0}\rb}{2\pi i}\right\}
\tau_{\rm rect}(z;\nu'_0,\rho_0),
\qquad
\nu_0':=\nu_0-\frac{\ln\lb 1-e^{2\pi i\rho_0}\rb}{2\pi i}.
\end{equation}
At the endpoint, $s=e^{-i\pi}\tilde s$. A key step at this point is to note that the $\mathbb Z_3$-symmetry of \PIIeq allows us to lift the reflection identity \eqref{DGsym} to the ZT itself. Indeed, applying this identity mode by mode and reindexing $n\mapsto-n$ gives a relation similar to \eqref{DGsym} but at the level of $\tau_{\rm rect}$. In particular, besides producing the identical normalisation prefactor, the shift $\nu\mapsto\nu+n$ alongside the reindexing $n\mapsto-n$ modifies the Fourier parameter in a nontrivial way. After simple algebraic manipulations using the recurrence relation \eqref{recGhat}, one obtains
\begin{equation}\label{eq:reflectedDFT}
\tau_{\rm rect}(z;\nu'_0,\rho_0)
=
\frac{e^{i\pi\lb \nu_2^2-\frac{\theta^2}{12}\rb}\lb 2\pi\rb^{2\nu_2}}
{\hG\lb \nu_2+\frac\theta2\rb\hG\lb \nu_2-\frac\theta2\rb}
\tau_{\rm rect}(e^{\frac{2\pi i}{3}}z;\nu_2,\rho_2),
\end{equation}
where
\beq
\nu_2=-\nu_0',
\qquad
e^{2\pi i\rho_2}
=-4\sin\pi \lb \nu_2-\frac\theta2\rb
\sin\pi \lb \nu_2+\frac\theta2\rb e^{2\pi i\lb \nu_2-\rho_0\rb},
\eeq
which is equivalent to the relations  \eqref{eq:kinematics}.

As can be directly verified by using the relations of Theorem~\ref{TheoConn} as well as \eqref{monodromy_S}, equations \eqref{eq:kinematics} are indeed the corresponding monodromy relations satisfied by the asymptotic parameters along the ray with $p=2$. Hence the series on the right of \eqref{eq:reflectedDFT} is the unit-normalized ZT \eqref{expsI} on the $p=2$ ray. Comparing \eqref{eq:continuedDFT}--\eqref{eq:reflectedDFT} with
$\tau=\mathcal N_{2,+}\tau_{\rm rect}(e^{2\pi i/3}z;\nu_2,\rho_2)$
gives \eqref{eq:N20}, thus confirming that the accumulated prefactor along the continuation (including the subsequent rewriting of the variable with rotated phase) precisely equals the corresponding connection constant between the two rays.
\epf

\begin{cor}\label{cor:main}
For every $p\in\mathbb Z/3\mathbb Z$, we have
\beq
\frac{\mathcal N_{p+1,+}}{\mathcal N_{p,+}}=
\exp\left\{-\frac{\Li\lb e^{2\pi i\omega_{p+2}}\rb}{2\pi i}\right\}
 \frac{e^{i\pi\lb \frac{\theta^2}{12}-\nu_p^2\rb}}{\lb 2\pi\rb^{2\nu_p}}\hG\lb \nu_p+\frac\theta2\rb\hG\lb \nu_p-\frac\theta2\rb,\label{eq:Np2b}
\eeq
which coincides with \eqref{conncoefpp}.
\end{cor}
\pf
Equation \eqref{eq:Np2b} is obtained from Proposition~\ref{prop:main} after a cyclic rotation and redefinition of parameters \cite{KRST,Tamarin} followed by inversion.
To rewrite it in the form \eqref{conncoefpp}, it suffices to rewrite the dilogarithmic prefactor using \eqref{BarnesLi2} together with  $e^{2\pi i\omega_{p+2}}=1-e^{2\pi i(\nu_p+\nu_{p+1})}$.
\epf

Let us finally briefly reiterate on the approach taken in this section to calculate the connection coefficients, since it reveals a striking yet simple underlying relation between Stokes-continued formal transseries solutions on the one hand, and asymptotics of solutions to nonlinear equations on the other. Indeed, a similar calculation has also been demonstrated for the case of Painlevé I \cite{Tamarin,KRST}, and it is expected that this holds more generally for other Painlevé-type equations. This calculation followed the analytic continuation of the transseries solution across a clockwise rotation through $\frac{2\pi}{3}$. Along this continuation, the Stokes crossing transformed $\nu$ while leaving the conjugate coordinate $\rho$ invariant and contributed a universal dilogarithmic-type normalization. Finally, at the endpoint the overall rotation was then absorbed by the underlying discrete symmetry of solutions, realised here by the reflection identity of the Fourier blocks, which further transformed the parameters, this time both $\nu$ as well as $\rho$. These transformed parameters then turned out to be exactly those of the adjacent ray, leaving the accumulated scalar as the required connection coefficient.

   \section{Concluding remarks}
   The connection problem for generic \PIIeq transcendent $q$  along the special rays was solved in \cite{Kap88} using WKB analysis. Riemann–Hilbert methods, on the other hand, have so far been applied mainly to  distinguished classes of solutions of inhomogeneous \PIIeq,  such as tronquées and generalized Hastings–McLeod solutions \cite{Mil18,Kok18,DXZ19}. In the present work, we have been able to compute the connection coefficients for the \PIIeq tau function by exploiting existing results. It would nevertheless be interesting  to provide a Riemann-Hilbert rederivation of Kapaev's asymptotic formulas for generic monodromy  (in the form of Theorem~\ref{TheoConn}) as well as of our solution of the connection problem for the gauge factor $\lambda$ (Corollary~\ref{corU}). This would give an independent streamlined proof of the \PIIeq tau function connection coefficients  obtained here.

   The calculation in Section~\ref{sec:resurgence}, together with its Painlevé I counterpart \cite{Tamarin,KRST}, suggests extending the resurgent approach to the other Painlevé equations. The Zak transform expansions  \cite{BLMST} provide both a consistency check and a guide for this extension. Once the transseries parameters are identified with the Zak coordinates, as outlined in \cite{KRSSTlocal}, the resurgent construction must reproduce these expansions. Their Fourier structure and explicit block coefficients therefore constrain the resulting tau transseries and guide its reorganization. A subsequent calculation similar to the one in this work could then further be used to obtain an independent resurgence computation for the corresponding connection constants.

   We conclude by summarizing the current status of the connection problem for generic tau functions of the other Painlevé equations between the relevant special rays and regular-type singular points. For P$_\text{III}$ of type $D_8$, P$_\text{I}$  and P$_\text{VI}$, the corresponding connection formulas are known and have been rigorously proved in \cite{IP15}, \cite{LR16}  and \cite{ILP16}, respectively. Our work adds \PIIeq to this list of solved cases. For P$_\text{V}$, the corresponding formulas are known in conjectural form \cite[Conjecture D]{LNR18}. For the remaining P$_\text{IV}$ and P$_\text{III}$ equations of type $D_6$ and $D_7$, even conjectural results are not  available yet. We plan to address these questions in future work.

    \appendix
    \section{Barnes $G$-function}\label{App_Barnes}
    The Barnes $G$-function $G\lb z\rb$ is an entire function defined by the infinite product representation
    \beq
    \frac{G\lb 1+z\rb}{\lb 2\pi\rb^{\frac z2}}=e^{-\frac{\lb 1+\gamma\rb z^2+z}{2}}
    \prod_{n=1}^{\infty}\lb 1+\frac{z}{n}\rb^n e^{\frac{z^2}{2n}-z},
    \eeq
    where $\gamma=-\Gamma'\lb 1\rb$ denotes the Euler's constant.
    The main properties of $G\lb z\rb$ relevant for the present work are the recurrence relation
    \beq
    G\lb 1+z\rb=\Gamma\lb z\rb G\lb z\rb,
    \eeq
    and the differentiation formula
    \beq\label{diffBarnesG}
    \frac{d}{dz}\left[\ln \frac{G\lb 1+z\rb}{\lb 2\pi\rb^{\frac z2}}-\frac{z\lb 1-z\rb}{2}\right]=z  \frac{d}{dz}\ln\Gamma\lb z\rb.
    \eeq
    The only zeros of $G\lb z\rb$ in $\mathbb C$ are non-positive integers, the zero at $z=-n$ ($n\in\mathbb Z_{\ge 0}$) having multiplicity $n+1$.

    We also use the combination
    \beq\label{defGhat}
    \hat G\lb z\rb=\frac{G\lb 1+z\rb}{G\lb 1-z\rb}.
    \eeq
    It satisfies the recurrence relation
    \beq\label{recGhat}
    \hat G\lb z+n\rb=\lb -1\rb^{\frac{n\lb n+1\rb}{2}}\lb\frac{\pi}{\sin \pi z}\rb^n \,\hat G\lb z\rb,\qquad n\in\mathbb Z,
    \eeq
    and the differentiation formula
    \beq\label{diffGhat}
    \frac{d}{dz}\left[ \ln\hat G\lb z\rb-\frac{i\pi z^2}{2}-z\ln2\pi\right]=- z \frac{d}{dz}\ln \lb 1-e^{2\pi i z}\rb.
    \eeq
    It has the asymptotics
    \beq\label{asGhat}
    \hat G\lb z\rb\simeq \lb 2\pi \rb^z e^{\frac{\pi i}{2}\lb z^2-\frac16\rb}\left[1+o\lb |z|^{-N}\rb\right]
    \eeq
    valid for every $N>0$ in any sector with $0<\arg z<\pi$ as $z\to\infty$.
    
    The function $\hat G\lb z\rb$ is a close relative of the classical dilogarithm function. We have, e.g.
    \beq\label{BarnesLi2}
    \hat G\lb z\rb=\exp\left\{-\frac{\operatorname{Li}_2\lb e^{2\pi i z}\rb}{2\pi i }
    -z\ln\frac{1-e^{2\pi i z}}{2\pi}+\frac{\pi i}{2}\lb z^2-\frac1{6}\rb\right\},
    \qquad z\in i\mathbb R_{>0}.
    \eeq
    The analytic structure of $\hat G\lb z\rb$ is however much simpler: it is meromorphic in the entire complex plane, with a zero $z=-n$ of order $n$ and pole $z=n$ of order $n$ for every $n\in\mathbb Z_{>0}$. The choice of the branch on the right of \eqref{BarnesLi2} is therefore irrelevant.
    One can establish an analog of the well-known dilogarithm reflection formula, 
    \beq\label{dilogrefl}
    \operatorname{Li}_2\lb z\rb+\operatorname{Li}_2\lb 1-z\rb={\frac{\pi^2}{6}-\ln z\ln\lb 1-z\rb}.
    \eeq
    \begin{lemma} 
    	Let $z_1,z_2\in\mathbb C$ satisfy
    	\beq\label{condi0}
    	e^{2\pi i z_1}+e^{2\pi i z_2}=1,
    	\eeq
    	then
    	\beq\label{refl0}
    	\hat G\lb z_1\rb \hat G\lb z_2\rb=\lb2\pi\rb^{z_1+z_2}e^{\frac{\pi i}{2}\lb z_1^2+z_2^2\rb-2\pi i z_1z_2-\frac{i\pi}{12}}.
    	\eeq
    \end{lemma}
    \pf According to \eqref{diffGhat}, we have
    \beq
    \begin{gathered}
    d\left[	\ln\lb \hat G\lb z_1\rb \hat G\lb z_2\rb\rb-\frac{\pi i\lb z_1^2+z_2^2\rb}{2}-\lb z_1+z_2\rb\ln 2\pi\right]=\\
    =-z_1d\ln\lb 1-e^{2\pi i z_1}\rb-z_2d\ln\lb 1-e^{2\pi i z_2}\rb=-2\pi id\lb z_1 z_2\rb,
    \end{gathered}
    \eeq
    where the second equality easily follows from \eqref{condi0}. Therefore, \beq\lb2\pi\rb^{-z_1-z_2}e^{-\frac{\pi i}{2}\lb z_1^2+z_2^2\rb+2\pi i z_1z_2} \hat G\lb z_1\rb \hat G\lb z_2\rb \eeq
    can only be a constant independent on the choice of $z_1,z_2$. The value of this constant is readily determined by setting ${z_1=-z_2=\frac16}$.
    \epf
    
     Using \eqref{dilogrefl}, one can also rewrite \eqref{BarnesLi2} as
    \beq\label{BarnesLi2V2}
    \hat G\lb z\rb=\lb2\pi\rb^z\exp\left\{\frac{\operatorname{Li}_2\lb 1-e^{2\pi i z}\rb}{2\pi i }
    +\frac{\pi i z^2}{2}\right\}.
    \eeq
    
    \section{Yablonskii-Vorobiev polynomials}\label{App_YV}
      There exists a family of rational solutions of \PIIeq \eqref{PIIstandard} with $\theta\in\mathbb Z+\frac12$ that can be generated from the solution $q=0$ of the equation with $\theta=\frac12$ by B\"acklund transformations.  These correspond to a sequence of tau functions  given by\footnote{Recall that  the equation \eqref{sigmaPII} for $\sigma=\lb\ln \tau\rb'$ is invariant under the sign change $\theta\mapsto -\theta$.}
    \beq\label{tauYV}
    \tau^{\text{YV}}_l\lb t\rb=e^{-\frac{t^3}{24}}Q_l\lb t\rb,\qquad \theta=\pm\lb l+\tfrac12\rb,\quad l\in\mathbb N,
    \eeq
    where $Q_{l}\lb t\rb$ are polynomials of degree $\frac{l\lb l+1\rb}{2}$ in $t$ usually referred to as Yablonskii-Vorobiev (YV) polynomials. In particular\footnote{The expression of $Q_5\lb t\rb$ appears in \cite[Eqs. 32.8.7]{DLMF} with a typo in the coefficient of $t^3$.},
    \beq
    \begin{gathered}
    	Q_0\lb t\rb=1,\qquad Q_1\lb t\rb=t,\qquad Q_2\lb t\rb=t^3+4,\\ Q_3\lb t\rb
    	=t^6+20t^3-80,\qquad
    	Q_4\lb t\rb=t^{10}+60 t^7+11200t,\\ Q_5\lb t\rb= t^{15} + 140 t^{12} + 2800 t^9 + 78400 t^6 - 3136000 t^3 - 6272000,\\
    	\ldots\qquad\ldots\qquad \ldots
    \end{gathered}
    \eeq
   These solutions are characterized by monodromy data
    \beq\label{YVmonodromy}
    \lb s_1,s_2,s_3\rb_{\mathrm{YV}}=\lb 0,0,0\rb.
    \eeq
    
     From the perspective of asymptotic expansions \eqref{expsI}--\eqref{expsID} on the rays $\arg t=\frac{2\pi p}{3}$, $p\in\mathbb Z/3\mathbb Z$, the monodromy \eqref{YVmonodromy} is clearly singular, cf \eqref{monoparameters}. However, YV solutions can be matched with  generic asymptotic expansions \eqref{expsII}--\eqref{expsIIG} on the rays $\arg t=\pi+\frac{2\pi p}{3}$, $p\in\mathbb Z/3\mathbb Z$ through a limiting procedure which amounts to setting $\omega_{0}=\omega_1=\omega_2=0$, $e^{2\pi i \xi_{0,1,2}}\to 0$. In this case, only the $n=0$ term survives in the Zak transform \eqref{expsII}. The asymptotic series $\mathcal G\lb s\,|\,0,l+\frac12\rb$ terminates and is described by the polynomials $Q_l\lb t\rb$. Indeed, 
     defining the polynomials $P_l\lb z\rb$ of degree $\Bigl\lfloor\frac{l\lb l+1\rb}{6}\Bigr\rfloor$ by 
     \beq t^{-\frac{l\lb l+1\rb}{2}}Q_l\lb t\rb=P_l\lb t^{-3}\rb,
     \eeq
     one has
     \beq
     \mathcal G\lb s\,|\,0,l+\tfrac12\rb= s^{\frac{l\lb l+1\rb}{3}}e^{-\frac{s^2}{1536}}P_l\lb 64 s^{-2}\rb,\qquad l\in\mathbb N.
     \eeq
     \section{Riccati solutions}
     Another well-known family of \PIIeq solutions, now with $\theta\in\mathbb Z$, is generated by the seed solution \eqref{AiryBasic}. The corresponding tau functions can be expressed as Hankel determinants
    \beq\label{tauRic}
    \tau^{\mathrm{Riccati}}_l\lb t\rb=\operatorname{det}\left \|\psi^{(j+k)}\lb t\rb\right\|_{j,k=0,\ldots l-1},\qquad \theta=l,\quad l\in\mathbb N^*,
    \eeq
     where 
    \beq\label{psiAiry}\psi\lb t\rb=C_1\operatorname{Ai}\lb -2^{-\frac13}t\rb+C_2\operatorname{Bi}\lb -2^{-\frac13}t\rb
    \eeq
    is any  solution of the Airy equation
    $2\psi''+t\psi=0$.  The relevant monodromy corresponds precisely to non-generic points of $\mathcal S_\theta$, where the triple $\lb s_1,s_2,s_3\rb$ alone does not determine the solution uniquely:
    \beq\label{Riccatimonodromy}
    \lb s_1,s_2,s_3\rb_{\mathrm{Riccati}}=\lb \lb-1\rb^l,\lb-1\rb^l,\lb-1\rb^l\rb.
    \eeq
    These data are singular from the point of asymptotic expansions \eqref{expsII}--\eqref{expsIIG} on the complementary rays $\arg t=\pi,\pm\frac{\pi}3$. The solutions \eqref{tauRic} can however also be matched with the generic expansions \eqref{expsI}--\eqref{expsID} along the rays $\arg t=0,\pm\frac{2\pi }{3}$ in a suitable limit. 
    
    Set, for example, $\theta=2l+1$ with $l\in\mathbb N$ and $\nu_0=-\frac12+\varepsilon$, where  $\varepsilon$ will be subsequently sent to $0$. The Stokes parameters can be tuned in such a way that
    \beq
    s_1=-1+2\pi i \lb 1+\eta\rb\varepsilon+O\lb\varepsilon^2\rb,
    \quad s_2=-1-2\pi i \varepsilon+O\lb\varepsilon^2\rb,\quad
    s_3=-1+2\pi i \lb 1+\eta^{-1}\rb\varepsilon+O\lb\varepsilon^2\rb,
    \eeq
    where $\eta\in\mathbb C^*$ can be arbitrary. This implies that $e^{2\pi i \rho_0}=2\pi i \eta\varepsilon+O\lb\varepsilon^2\rb$. The Barnes $G$-function prefactor appearing in \eqref{expsI} combined with $e^{2\pi i n\rho_0}$ then gives
    \beq
    \begin{gathered}
    G\lb1+\nu_0+n+\frac{\theta}{2}\rb G\lb1+\nu_0+n-\frac{\theta}{2}\rb e^{2\pi i n\rho_0}=\hat{G}\lb n-l-1+\varepsilon\rb G\lb 1+n+l+\varepsilon\rb
    G\lb 2-n+l-\varepsilon\rb e^{2\pi i n\rho_0}= \\
    =\pm \varepsilon^{l+1} G\lb 1+n+l+\varepsilon\rb
    G\lb 2-n+l-\varepsilon\rb \lb 2\pi i \eta\rb^n\left[1+O\lb\varepsilon\rb\right].
    \end{gathered}
    \eeq
    The factor $\varepsilon^{l+1}$ may be absorbed by the overall normalization of the tau function and can therefore be discarded.  The combination $	G\lb 1+\varepsilon +n+l\rb G\lb 2-\varepsilon-n+l\rb$ vanishes as $\varepsilon \to 0$ for $n<-l$ and $n> l+1$. 	As a consequence, the sum over $n$ in \eqref{expsI} becomes restricted to a finite range of values, $n=-l,\ldots,l+1$. The parameter $\eta$ characterizes the direction along which the singular point $\lb s_1,s_2,s_3\rb=\lb -1,-1,-1\rb$ on the cubic monodromy surface $\mathcal S_1$ is approached, and is related to the ratio $C_1/C_2$ in the Hankel determinant solution. Analogous calculations can also be carried out for $\theta\in 2\mathbb N$.

    The above considerations can be used to express the asymptotic series for $\mathcal D\lb s\,|\,\nu,\theta=l\rb$ for $l\in\mathbb N$ and $\nu=-\frac{l}{2},\ldots ,\frac{l}{2}$ in terms of asymptotic series of Airy functions.

    	\begin{eg}
    Choose in \eqref{psiAiry} $C_2=-iC_1$. The corresponding linear combination of Airy functions can also be written as
    \beq
    \psi_H\lb t\rb=\operatorname{Ai}\lb -2^{-\frac13}t\rb-i\operatorname{Bi}\lb -2^{-\frac13}t\rb =e^{\frac{\pi i}{6}}\sqrt{2^{-\frac13}t\bigl/3}\;H^{(1)}_{\frac13}\lb\sqrt 2\,t^{\frac32}\bigr/3\rb.
    \eeq
    Here $H^{(1)}_{\frac13}\lb z\rb$ is the Hankel function of the 1st kind, having the following asymptotic expansion as  $z\to+\infty$:
    \beq
    H^{(1)}_{\frac13}\lb z\rb\simeq\sqrt{\frac{2}{\pi z}} e^{iz-\frac{5\pi i}{12}}\sum_{k=0}^\infty \frac{\lb \frac16\rb_k 
    \lb \frac56\rb_k}{k!} \lb-\frac i{2z}\rb^k.
    \eeq
    Then one can identify
    \beq
    \mathcal D\lb 4\sqrt 2\,e^{\frac{i\pi}{2}}t^{\frac32}\,\bigl|\,\tfrac l2,l\rb =D_l\, 
    \operatorname{det}\left \|\psi_H^{(j+k)}\lb t\rb\right\|_{j,k=0,\ldots l-1},\qquad \theta=l,\quad l\in\mathbb N^*,
    \eeq
    where the constant prefactor $D_l$ is independent of $t$. In particular, 
    \beq
    \mathcal D\lb s\,\bigl|\,\tfrac 12,1\rb =\frac{s^{-\frac16}e^{\frac{s}{12}}}{\sqrt{2\pi}}\sum_{k=0}^\infty \frac{\lb \frac16\rb_k 
    \lb \frac56\rb_k  \lb 6s^{-1}\rb^k}{k!}.
    \eeq
   	\end{eg}
    
    \section{Airy kernel and Tracy-Widom distribution}
    Recall the following well-known transformation \cite[Eqs. 32.7.3--32.7.4]{DLMF}:
    \begin{lemma}
    	Let $q\lb t\rb$ be a solution of the Painlevé II equation \eqref{PIIstandard} with $\theta=\frac12$, i.e. the ``homogeneous'' equation ${\ds\frac{d^2q}{dt^2}=2q^3+tq}$. Then the function $Q\lb T\rb$ defined by
    	\beq\label{folding1}
    	Q\lb -2^{\frac13}t\rb=2^{-\frac13}\frac{q'\lb t\rb}{q\lb t\rb}
    	\eeq
    	satisfies \PIIeq with $\theta=0$, i.e. $\ds\frac{d^2Q}{dT^2}=2Q^3+TQ+\ds\frac12$.
    \end{lemma}
    \pf Direct calculation.\epf
    
    The homogeneous \PIIeq is solved by both $q$ and $\tilde q=-q$. Recall that the  monodromy associated to $\tilde q$ is $\tilde{\mathbf{s}}=-\mathbf{s}$. The above transformation maps both $q$ and $\tilde q$ to the same $Q$. As before, let us denote the  tau function associated to $\tilde q\lb t\rb$ by $\tilde\tau\lb t\rb$ and consider
    \beq
    \frac{d}{dt}\ln \lb \tau\tilde \tau\rb= q'^2-\lb q^2+\frac t2\rb^2.
    \eeq
    Comparing this expression with the \PIIeq tau function $\boldsymbol \tau\lb T\rb$ associated to  $Q\lb T\rb$, we find that the transformation \eqref{folding1} may be equivalently rewritten as
    \beq
    \boldsymbol\tau\lb -2^{\frac13}t\rb=e^{\frac{t^3}{12}}\tau\lb t\rb\tilde \tau\lb t\rb.
    \eeq
     Thus any \PIIeq  tau function $\boldsymbol \tau\lb T\rb$ with $\theta=0$ can be expressed as a product of two \PIIeq  tau functions  with $\theta=\frac12$. The tau function on the left corresponds to monodromy data
    \beq
    \lb \mathsf{s}_1,\mathsf{s}_2,\mathsf{s}_3\rb=\lb 1-s_2s_3,1-s_3s_1,1-s_1s_2\rb.
    \eeq
    In particular, it is straightforward to check that the $\theta=0$ monodromy relation $\mathsf{s}_1\mathsf{s}_2\mathsf{s}_3-\mathsf{s}-\mathsf{s}_2-\mathsf{s}_3=-2$ is satisfied provided that $\theta=\frac12$ monodromy relation $s_1s_2s_3-s_1-s_2-s_3=0$ holds.
      Note a useful differentiation formula:
    \beq\label{derform1}
    \frac{d^2}{dt^2}\ln  \boldsymbol\tau\lb -2^{\frac13}t\rb=-q\lb t\rb^2.
    \eeq

    Let us now consider the Fredholm determinant
    \beq
    D_{\mathrm{Airy}}\lb t\rb=\operatorname{det}\lb \mathbf 1-K\bigl|_{\lb t,+\infty\rb}\rb,
    \eeq
    with 
    \beq
    K\lb x,y\rb=\lambda^2\;\frac{\operatorname{Ai}\lb x\rb \operatorname{Ai}'\lb y\rb
    -\operatorname{Ai}'\lb x\rb \operatorname{Ai}\lb y\rb}{x-y}.
    \eeq
    This determinant gives a one-parameter family of $\theta=0$ tau functions, $D_{\mathrm{Airy}}\lb t\rb= {\boldsymbol\tau}_{\mathrm{Airy}}\lb -2^{\frac13}t\rb$. The relevant monodromy data are given by
    \beq
    \lb\mathsf s_1,\mathsf s_2,\mathsf s_3\rb_{\mathrm{Airy}}=\lb 1,1-\lambda^2,1\rb.
    \eeq
    For $\lambda^2\neq 1$, the asymptotic behavior of ${\boldsymbol\tau}_{\mathrm{Airy}}\lb T\rb$ along the special rays $\arg T=0,\pm\frac{\pi}{3},\pm\frac{2\pi}{3}$ is described by the generic formulas \eqref{expsI}--\eqref{expsID}, \eqref{expsII}--\eqref{expsIIG}  or a suitable limit thereof. The asymptotics on the ray $\arg T=\pi$ (i.e. as $t\to+\infty$) is non-generic.
    
    The corresponding    pair of solutions $q$, $\tilde q$ of the homogeneous \PIIeq with ${\theta=\frac12}$  is uniquely characterized the following asymptotics as $t\to+\infty$:
    \beq
    q\lb t\rb\simeq \lambda \,\operatorname{Ai}\lb t\rb, \qquad
    \tilde q\lb t\rb\simeq -\lambda \,\operatorname{Ai}\lb t\rb.
    \eeq
    These solutions are usually referred to as Ablowitz-Segur solutions. The monodromy data associated to $q\lb t\rb$ are given by 
    \beq
    \lb s_1,s_2,s_3\rb_{\mathrm{AS}}=\lb -i\lambda,0,i\lambda\rb.
    \eeq

   The Tracy-Widom distribution \cite{TWAiry} is given by the integral
   \beq\label{FTW}
   F_{\mathrm{TW}}\lb t\rb=\exp\left\{-\int_t^{+\infty}\lb s-t\rb q_{\mathrm{HM}}^2\lb s\rb ds\right\},
   \eeq
   where $q_{\mathrm{HM}}\lb t\rb$ is the Hastings-McLeod solution of the homogeneous   \PIIeq equation, uniquely specified by the asymptotic behavior 
   $q_{\mathrm{HM}}\lb t\rb\simeq \operatorname{Ai}\lb t\rb$ as $t\to+\infty$. It can hence be viewed as a $\lambda=1$ specialization of the Ablowitz-Segur solution. Taking the logarithmic derivatives of \eqref{FTW} with respect to $t$ gives
   \beq
   \frac{d}{dt}\ln F_{\mathrm{TW}}\lb t\rb=\int_t^{+\infty} q_{\mathrm{HM}}^2\lb s\rb ds,\qquad  \frac{d^2}{dt^2}\ln F_{\mathrm{TW}}\lb t\rb=-q_{\mathrm{HM}}^2\lb t\rb.
   \eeq
   Comparing with \eqref{derform1}, we see that the distribution $F_{\mathrm{TW}}\lb t\rb$ is nothing but  the corresponding $\theta=0$ tau function
   \beq
   F_{\mathrm{TW}}\lb t\rb=\boldsymbol\tau_{\mathrm{HM}}\lb -2^{\frac13}t\rb,
   \eeq
   normalized as $\boldsymbol\tau_{\mathrm{HM}}\lb -\infty\rb=1$. The asymptotics of the Hastings-McLeod tau function $\boldsymbol\tau_{\mathrm{HM}}\lb T\rb$ is non-generic  for both $T\to-\infty$ and $T\to+\infty$.

   \section{Direct sector-by-sector check of the Zak transform}
\label{app:sectors}

Throughout this appendix, $\tau$ denotes $\tau_{\rm rect}$ with
the unit normalization fixed by \eqref{eq:DFTtau}. We collect the explicit comparison between the rectangular transseries and the ZT \eqref{eq:DFTtau}. All expressions below are understood as formal series in $z$. We write $\tau_{(n,m)}$ and $\sigma_z^{(n,m)}$ for the coefficients of $\sigma_1^n\sigma_2^m$ in $\tau$ and $\partial_z\ln\tau$, respectively.

The first transseries sectors obtained from \cite{Vega,Vega23} and directly from \eqref{eq:PIIu} are
\begin{equation}\label{eq:firstsectors}
\begin{aligned}
\Phi_{(0,0)}&=1-\frac{\alpha}{2}w^2-\Big(\frac1{16}+\frac{3\alpha^2}{8}\Big)w^4
-\alpha\Big(\frac{11}{32}+\frac{\alpha^2}{2}\Big)w^6+O\lb w^8\rb,\\
\Phi_{(1,0)}&=w\Big[1+\frac{-17+36\alpha-84\alpha^2}{96}w^2+O\lb w^4\rb\Big],\\
\Phi_{(0,1)}&=w\Big[1+\frac{17+36\alpha+84\alpha^2}{96}w^2+O\lb w^4\rb\Big],\\
\Phi_{(1,1)}&=-3w^2-\frac{21\alpha}{4}w^4+O\lb w^6\rb,\\
\Phi_{(2,0)}&=\frac{w^2}{2}\Big[1+\frac{-41+84\alpha-84\alpha^2}{48}w^2+O\lb w^4\rb\Big],\\
\Phi_{(0,2)}&=\frac{w^2}{2}\Big[1+\frac{41+84\alpha+84\alpha^2}{48}w^2+O\lb w^4\rb\Big],\\
\Phi_{(2,1)}&=\Big(1-\frac{17\alpha}{2}\Big)w^3+O\lb w^5\rb,\qquad
\Phi_{(1,2)}=\Big(1+\frac{17\alpha}{2}\Big)w^3+O\lb w^5\rb.
\end{aligned}
\end{equation}
Substituting the sector expansions into \eqref{eq:Hz} gives the rectangular expansion of $\sigma_z=\partial_z\ln\tau$. For the subleading comparisons below, the same recursively determined sectors are expanded to the required order; only their first terms are displayed in \eqref{eq:firstsectors}.

The relation between Fourier modes and rectangular sectors follows directly from \eqref{eq:DFT}:
\[
e^{2\pi i\rho}
=-\frac{\pi 8^{\theta-4\mu}e^{-i\pi(\theta-4\mu)}}
{\Gamma(1-2\mu)\Gamma(\theta-2\mu)}\,\sigma_1,
\qquad \nu=\frac\theta2-2\mu,
\qquad \mu=\sigma_1\sigma_2.
\]
Hence a positive Fourier mode $k\geq0$ has the form
\[
\sigma_1^k\times\big(\text{power series in }\mu\big),
\]
and the term proportional to $\mu^p$ is a rectangular sector
\[
\sigma_1^{k+p}\sigma_2^p,
\qquad\text{i.e.}\qquad \lb n,m\rb=\lb k+p,p\rb.
\]
For a negative mode $k=-r<0$, the block contains
$G(1-r-2\mu)$, which has a zero of order $r$ at $\mu=0$. This cancels the explicit $\sigma_1^{-r}$ and the remaining expansion produces
\[
\sigma_1^p\sigma_2^{p+r},
\qquad\text{i.e.}\qquad \lb n,m\rb=\lb p,p+r\rb.
\]
Thus, up to the $\lb 2,2\rb$ sector, the Fourier modes organize the rectangular sectors as
\begin{equation}\label{eq:mode-sectors}
\begin{array}{r|c}
\text{Fourier mode} & \text{rectangular sectors}\\ \hline
0\qquad\quad &(0,0),(1,1),(2,2),\ldots\\
1\qquad\quad &(1,0),(2,1),\ldots\\
2\qquad\quad &(2,0),\ldots\\
-1\qquad\quad&(0,1),(1,2),\ldots\\
-2\qquad\quad&(0,2),\ldots
\end{array}
\end{equation}
Accordingly, the check is performed by an ordinary expansion in $\mu=\sigma_1\sigma_2$. For example, the $k=1$ mode starts with the sector $\lb 1,0\rb$ and its first correction in $\mu$ is the sector $\lb 2,1\rb$; the $k=2$ mode starts with $\lb 2,0\rb$. For $k=-1$, the zero of $G\lb -2\mu\rb$ supplies the factor of $\mu$ needed to cancel the explicit negative power of $\sigma_1$, so the first sector is $\lb 0,1\rb$ and the next is $\lb 1,2\rb$. Thus one simply expands each Fourier mode in $\mu$, replaces $\mu^m$ by $\sigma_1^m\sigma_2^m$, and compares the resulting coefficient in $\partial_z\ln\tau$ with the coefficient obtained by substituting \eqref{eq:firstsectors} into \eqref{eq:Hz}.

\paragraph{Perturbative sector.}
The $k=0$, $\mu^0$ term gives
\begin{equation}\label{eq:C00}
\sigma_z^{(0,0)}\big|_{\rm ZT}
=-\theta\sqrt z-\frac{\theta^2}{4z}
+\frac{\theta+4\theta^3}{32z^{\frac52}}
-\frac{\theta^2\lb 8\theta^2+7\rb}{64z^4}+O\lb z^{-\frac{11}{2}}\rb.
\end{equation}
Direct substitution of $\Phi_{(0,0)}$ into \eqref{eq:Hz} gives exactly the same expression. The leading terms agree with the perturbative exponential
and power of $z$, while the $z^{-\frac52}$ and $z^{-4}$ terms reproduce $\mathcal D_1\lb \theta/2,\theta\rb$ and $\mathcal D_2\lb \theta/2,\theta\rb$ from \eqref{Dkcoefs}.

\paragraph{One-instanton sectors.}
The leading $k=1$ term gives the $\lb 1,0\rb$ sector. Using $G\lb x+1\rb=\Gamma\lb x\rb G\lb x\rb$, its contribution relative to the perturbative block begins as
\[
\frac{\tau_{(1,0)}}{\tau_{(0,0)}}=
\frac{\theta}{16}\,z^{-\frac32(1+\theta)}e^{-\frac43z^{\frac32}}
\lb 1+O\lb z^{-\frac32}\rb\rb,
\]
and therefore
\[
\sigma_z^{(1,0)}\big|_{\rm ZT}
=-\frac{\theta}{8}\,z^{-1-\frac{3\theta}{2}}e^{-\frac43z^{\frac32}}
\lb 1+O\lb z^{-\frac32}\rb\rb.
\]
For the $k=-1$ mode, the Barnes factor $G\lb -2\mu\rb=-2\mu+O\lb \mu^2\rb$ makes the first nonzero term the sector $\lb 0,1\rb$:
\[
\frac{\tau_{(0,1)}}{\tau_{(0,0)}}=
-\frac12\,z^{-\frac32(1-\theta)}e^{\frac43z^{\frac32}}
\lb 1+O\lb z^{-\frac32}\rb\rb,
\]
so that
\[
\sigma_z^{(0,1)}\big|_{\rm ZT}
=-z^{-1+\frac{3\theta}{2}}e^{\frac43z^{\frac32}}
\lb 1+O\lb z^{-\frac32}\rb\rb.
\]
Collecting the coefficients of $\sigma_1$ and $\sigma_2$ directly in \eqref{eq:Hz} gives the same two expressions. In particular, the normalization in \eqref{eq:DFT} reproduces the rectangular convention $u^{(1,0)}_0=u^{(0,1)}_0=1$.

\paragraph{The resonant diagonal sector $\lb 1,1\rb$.}
The coefficient of $\mu$ in the zero Fourier mode gives, at the level of $\ln\tau$,
\[
\ln\tau\big|_{\sigma_1\sigma_2}
=\frac83z^{\frac32}+3\theta\ln z+\dots
\]
Differentiating with respect to $z$ gives
\begin{equation}\label{eq:C11}
\sigma_z^{(1,1)}\big|_{\rm ZT}
=4\sqrt z+\frac{3\theta}{z}
-\frac{1+21\theta^2}{8z^{\frac52}}+O\lb z^{-4}\rb,
\end{equation}
in agreement with the coefficient of $\sigma_1\sigma_2$ obtained directly from \eqref{eq:Hz}. The $z^{\frac32}$ and $\log z$ terms are the tau function manifestation of the resonance encoded by the $\mu$-dependent power in \eqref{eq:transseries}.

\paragraph{Two-instanton sectors.}
For the sectors $(2,0)$ and $(0,2)$, the leading contributions to $\ln\tau$ come from
\ben
-\ds\frac{\tau_{(1,0)}^2}{2\tau_{(0,0)}^2},\qquad
 -\ds\frac{\tau_{(0,1)}^2}{2\tau_{(0,0)}^2},
 \ebn
respectively. The $k=2$ and $k=-2$ modes first contribute three powers of $z$ later. Differentiation gives
\begin{equation}\label{eq:C20}
\begin{aligned}
\sigma_z^{(2,0)}\big|_{\rm ZT}
&=\frac{\theta^2}{128}\,z^{-\frac52-3\theta}e^{-\frac83z^{\frac32}}
\lb 1+O\lb z^{-\frac32}\rb\rb,\\
\sigma_z^{(0,2)}\big|_{\rm ZT}
&=-\frac12\,z^{-\frac52+3\theta}e^{\frac83z^{\frac32}}
\lb 1+O\lb z^{-\frac32}\rb\rb,
\end{aligned}
\end{equation}
and these are again precisely the coefficients obtained from the rectangular transseries through \eqref{eq:Hz}.

\paragraph{Resonant off-diagonal sectors.}
The next term in the $k=1$ expansion is $\lb 2,1\rb$, while the next term in the $k=-1$ expansion is $\lb 1,2\rb$. Direct expansion gives
\begin{align}
\sigma_z^{(2,1)}\big|_{\rm ZT}
&=z^{-1-\frac{3\theta}{2}}e^{-\frac43z^{\frac32}}
\left[\frac14-\frac{3\theta}{4}\ln z+O\lb z^{-\frac32}\ln z\rb\right],\label{eq:C21}\\
\sigma_z^{(1,2)}\big|_{\rm ZT}
&=z^{-1+\frac{3\theta}{2}}e^{\frac43z^{\frac32}}
\left[6\ln z+O\lb z^{-\frac32}\ln z\rb \right].\label{eq:C12}
\end{align}
The same logarithms and coefficients follow from \eqref{eq:Hz}. They arise from expanding the resummed factors $z^{\pm6\mu}$ in \eqref{eq:transseries}, so this comparison directly checks the resonant part of the ZT reorganization rather than only the nonresonant instanton sectors.

\paragraph{The sector $(2,2)$.}
Finally, the $\mu^2$ term of the zero Fourier mode gives
\begin{equation}\label{eq:C22}
\sigma_z^{(2,2)}\big|_{\rm ZT}
=-\frac6z+\frac{51\theta}{4z^{\frac52}}+O\lb z^{-4}\rb,
\end{equation}
again in agreement with the direct rectangular-transseries calculation.

Thus every sector with $0\leq n,m\leq2$ agrees to the displayed orders. The subleading comparisons agree with
\[
\mathcal D_1\lb \nu,\theta\rb=\frac{\nu\lb 68\nu^2-9\theta^2+2\rb}{6},
\]
together with the value of $\mathcal D_2\lb \theta/2,\theta\rb$, in agreement with \eqref{Dkcoefs}. These are the checks of \eqref{eq:DFTtau} and \eqref{eq:DFT} quoted in Subsection~\ref{subsec:fourier}.

\end{document}